\documentclass[traditabstract]{aa}
\usepackage[varg]{txfonts}
\usepackage{rotating}

\usepackage{lscape}
\usepackage{color}
\usepackage{graphicx}

\definecolor{red}{rgb}{1,0,0}

\begin{document}

\title{Circumstellar discs in Galactic centre clusters: 
Disc-bearing B-type stars in the Quintuplet and Arches clusters
\thanks{Based on data obtained at the ESO VLT under programme IDs
085.D-0446, 089.D-0121 (PI: Stolte), 081.D-0572 (PI: Brandner), 087.D-0720, 089.D-0430 (PI: Olzcak), 
071.C-0344 (PI: Eisenhauer), 60.A-9026 (NAOS/CONICA science verification), 
as well as Hubble Space Telescope observations 
under programmes 11671 (PI:Ghez).}}

\author{A. Stolte\inst{1}, B. Hu{\ss}mann\inst{1}, C. Olczak\inst{2}, 
W. Brandner\inst{3}, M. Habibi\inst{1}, A. M. Ghez\inst{4,5}, M. R. Morris\inst{4}, 
J. R. Lu\inst{6}, W. I. Clarkson\inst{7}, J. Anderson\inst{8}}

\institute{Argelander Institut f\"ur Astronomie, Universit\"at Bonn,
Auf dem H\"ugel 71, 53121 Bonn, Germany
\and
Astronomisches Recheninstitut, Universit\"at Heidelberg, M\"onchhofstr. 12-14, 
69120 Heidelberg, Germany
\and
Max-Planck-Institut f\"ur Astronomie, K\"onigstuhl 17, 69117 Heidelberg,
Germany
\and
Division of Astronomy and Astrophysics, UCLA, Los Angeles, CA 90095-1547, USA
\and
Institute of Geophysics and Planetary Physics, UCLA, Los Angeles, CA 90095, USA
\and
Institute for Astronomy, University of Hawai'i, 2680 Woodlawn Drive, Honolulu, HI 96822, USA
\and
Department of Natural Sciences, University of Michigan-Dearborn, 125 Science Building,
4901 Evergreen Road, Dearborn, MI 48128, USA
\and
Space Telescope Science Institute, 3700 San Martin Drive, Baltimore, MD 21218, USA
} 

\abstract{We investigate the circumstellar disc fraction as determined
from $L$-band excess observations of the young, massive Arches and Quintuplet
clusters residing in the central molecular zone of the Milky Way.
The Quintuplet cluster was searched for $L$-band excess sources for the first time. 
We find a total of 26 excess sources in the Quintuplet cluster,
and 21 sources with $L$-band excesses in the Arches cluster, 
of which 13 are new detections.
With the aid of proper motion membership samples, the disc fraction
of the Quintuplet cluster could be derived for the first time 
to be $4.0 \pm 0.7$\%. There is no evidence for a radially varying disc
fraction in this cluster. In the case of the Arches cluster, a disc
fraction of $9.2 \pm 1.2$\% approximately out to the cluster's predicted 
tidal radius, $r < 1.5$ pc, is observed. 
This excess fraction is consistent with our previously found 
disc fraction in the cluster in the radial range $0.3 < r < 0.8$ pc.
In both clusters, the host star mass range covers late A- to early B-type 
stars, $2 < M < 15\,M_\odot$, as derived from $J$-band photospheric magnitudes.
We discuss the unexpected finding of dusty circumstellar discs in these UV 
intense environments in the context of primordial disc survival and 
formation scenarios of secondary discs. We consider the possibility that 
the $L$-band excess sources in the Arches and Quintuplet clusters could be 
the high-mass counterparts to T Tauri pre-transitional discs. 
As such a scenario requires a long pre-transitional disc lifetime in 
a UV intense environment, we suggest that mass transfer discs in binary 
systems are a likely formation mechanism for the B-star discs observed in 
these starburst clusters.}


\titlerunning{Circumstellar discs in Galactic center clusters}
\authorrunning{A. Stolte et al.}

\maketitle 

\section{Introduction}
\label{intro}

In view of the short lifetimes of primordial circumstellar discs
around B-type stars in dense environments, 
the detection of circumstellar discs in the UV-rich environment
of the Galactic centre Arches cluster came as a surprise (Stolte et al.~2010).
This detection raised the question of whether discs can also be found
in the more evolved Quintuplet cluster, and whether these discs 
can have their origin in massive primordial discs sustained over the 
clusters' lifetimes of several million years (Myr).
In particular, the nature of the disc sources in the Arches cluster
has remained unsolved (see the discussion in Stolte et al.~2010).
In the past five years, we have extended our disc search to larger
radii in the Arches cluster and have encompassed the Quintuplet cluster.
With the aim of shedding light on the nature of the $L$-band excess sources 
found in both starburst clusters, we compare their physical 
properties to pre-transitional discs and discuss secondary mass 
transfer discs as a possible origin of the circumstellar material.

\subsection{Circumstellar disc survival}
\label{survsec}



The survival of primordial circumstellar discs in young star clusters
is known to be a steep function of cluster age (Haisch et al.~2001,
Hern\'andez et al.~2005). For discs around low-mass stars, $M\sim 1\,M_\odot$,
planet formation theories suggest that dust agglomeration 
causes a period of grain growth, while at the same time the dense gas 
of the primordial disc is evaporated by the central star
(Cieza et al.~2012, Williams \& Cieza 2011, Owen et al.~2011).
In consequence, the thermal excess of evolved discs is 
dominated by increasingly longer wavelength emission, while the 
near-infrared contribution decreases and vanishes with time 
(e.g. Espaillat et al.~2012). The disc survival timescale in 
young stellar clusters is observed to be 3-10 Myr for 
intermediate-mass stars, $2 < M < 10\,M_\odot$ (Hern\'andez et al.~2005),
albeit with lower disc fractions at any given age as compared to
clusters dominated by their lower-mass T Tauri counterparts 
(see Stolte et al.~(2010) for a detailed discussion). 
Hern\'andez et al.~(2005) suggest that settling in the disc might depend
on the stellar mass, such that the stellar radiation could penetrate discs 
around higher mass stars more rapidly, which might accelerate the destruction
by UV radiation from the central star even further.
This mass-dependent process would 
decrease the rate of survival of the hot inner disc 
as observed through near-infrared excess emission around high-mass stars.
For Herbig Ae/Be stars in nearby OB associations, Hern\'andez et al.~find
intermediate-mass disc fractions of only 2-5\% at ages 3-10 Myr.
In particular in the dense environment of starburst clusters
such as the Arches and Quintuplet clusters near the Galactic centre,
star-disc interactions (Olczak et al.~2012) and the external UV radiation 
field (Anderson et al.~2013, Fatuzzo \& Adams 2008, Adams et al.~2004, 
Scally \& Clarke 2001, Richling \& Yorke 2000, Johnstone et al.~1998) 
might contribute to an even more rapid depletion of circumstellar material.

Until recently, circumstellar discs in young star clusters were mostly 
identified by their
$K$- and $L$-band near-infrared excess emission. This definition 
distinguishes primordial discs with dense, hot inner rims from 
disc-less stars or stars with evolved transitional discs.
The near- to mid-infrared capabilities of the recent Spitzer space mission
enabled the detection and definition of more subtle classes of discs.
Transitional discs were traditionally defined as objects lacking
near-infrared emission, yet displaying strong mid- or far-infrared
excesses (see Espaillat et al.~2012 for a summary, and references 
therein). With larger sample sizes and improved spatial resolution,
several of the proposed transition objects reveal significant
substructure in their SEDs, with small but significant near-infrared
excess contributions as a tracer for optically thick inner discs. 
In contrast to primordial discs, their near-infrared emission is accompanied 
by mid-infrared dips, suggestive of disc gaps. 
These objects are defined as \emph{pre-transitional} discs by Espaillat et al.~(2012), 
and can also be identified with the near-infrared bright subclass of the so-called 
\emph{warm debris discs} detected with the AKARI mission by Fujiwara et al.~(2013).
Although debris discs are typically found around stars with older ages of more than 10 Myr,
the ages of these near-infrared bright discs are not known.
While transitional discs are by definition absent from the classical $L$-band excess 
searches for primordial circumstellar discs, pre-transitional discs could contribute 
to the disc fractions observed in young, massive star clusters, if they are common 
around main-sequence B-type as well as T Tauri stars.

\subsection{The Arches and Quintuplet clusters}

The Arches and Quintuplet clusters are young, massive star clusters
located at a projected distance of $\sim 30$ pc from the Galactic 
centre (GC). Both clusters are host to a rich population of more 
than 100 massive O- and B-type stars (Liermann et al.~2009, Martins et al.~2008).
From the extrapolation of the observed stellar mass function,
they have estimated photometric masses of about $10^4\,M_\odot$
(Habibi et al.~2013, Hu{\ss}mann et al.~2012).
According to dynamical simulations, the underlying stellar mass is
suggested to be as high as $4\times 10^4\,M_\odot$ for the Arches cluster
(Harfst et al.~2010).
Spectroscopic age dating of the evolved population of Wolf-Rayet (WR) and 
giant or supergiant stars suggests an age of $2.5 \pm 0.5$ Myr (Najarro et al.
2004), with a possible upper age limit for the Arches cluster of 4 Myr obtained 
from supergiant member stars. For the Quintuplet cluster
the situation is not as clear. Earlier studies have suggested 
an age of $4 \pm 1$ Myr from spectral fitting to the evolved
population (Figer et al.~1999). A newer Very Large Telescope (VLT) SINFONI spectral analysis
suggests ages as young as 2-3 Myr for the WN stars (Liermann et al.~2012),
but these objects in particular may be affected by binary mass 
transfer evolution and the corresponding rejuvenation (Schneider et al.~2014). 
For the purposes of this paper, we adopt an age of $2.5 \pm 0.5$ Myr
for the Arches cluster, and $4 \pm 1$ Myr for the Quintuplet, as 
these ages are consistent with the isochrones we adopt for stellar mass
derivations (see Hu{\ss}mann et al.~2012, Habibi et al.~2013).

\subsection{Discs in the Arches cluster}

In the Arches cluster, a population of disc-bearing stars was found
from near-infrared excesses using high-resolution Keck/NIRC2 adaptive
optics observations in the dense cluster core (Stolte et al.~2010). 
From $L'$-band excesses, we derived the fraction
of disc-bearing B-type main-sequence stars to be $6 \pm 2$\% at radii
$r < 0.8$ pc from the cluster centre. A radial increase in the 
disc fraction from 3\% for $r < 0.1$pc to 10\% for $0.3 < r < 0.8$pc
suggested that circumstellar discs are prone to destruction by 
UV radiation or gravitational interactions in the dense cluster 
environment (e.g. Olczak et al.~2012). Of 24 detected excess sources,
all 21 sources with reliable proper motion measurements proved to be
genuine members of the Arches cluster. The detection of CO 2.3$\mu$m
bandhead emission in VLT/SINFONI $K$-band spectra available for 
three sources provided additional evidence that the $L$-band excess 
originates from hot circumstellar material in a disc geometry. 
For the Arches sources presented in Stolte et al.~(2010), we estimated 
a host star mass range of $3-10\,M_\odot$ based on $H$-band luminosities
which showed little to no infrared excess emission.
One of the remaining mysteries was the apparently old age of 2.5 Myr
for circumstellar material around B-type stars. Previous observational
studies of Herbig Ae/Be stars
suggested that photoevaporation destroys primordial discs around 
high-mass stars within less than 1 Myr (Hillenbrand et al.~1998,
Hern\'andez et al.~2005, Alonso-Albi et al.~2009, Gorti et al.~2009). 
A short survival timescale of less than 
1-3 Myr in UV-rich clusters is confirmed by numerical simulations
of disc survival (Anderson et al.~2013, and references therein).
When FUV, EUV, and X-ray luminosities are taken into account, 
stars with $M < 3 M_\odot$ have simulated disc lifetimes
of several $10^6$ years. The survival timescale plummets for higher-mass
stars and already reaches less than $10^6$ years for a $10\,M_\odot$ 
central star (Gorti et al.~2009, see their Fig.~12). Thus, the survival of 
primordial discs around B-type stars, even at a low rate, was unexpected.
The arguments leading to early disc destruction should be even 
more  valid  for the older Quintuplet cluster. 

Here, we investigate VLT high-resolution adaptive optics $K_sL'$ 
observations in combination with HST/WFC3 $JH$ photometry to extend
the cluster area covered in the Arches out to $r = 1.5$ pc,
close to the cluster's predicted tidal radius (Habibi et al.~2013),
and to detect disc candidates in the more evolved Quintuplet cluster
for the first time. Proper motion membership from two-epoch adaptive optics 
$K_s$ data is employed to confirm the cluster origin and hence the youth 
of the discovered $L$-band excess sources. 

Towards the end of this contribution, 
we provide a detailed discussion of the origin of the 
$L$-band emission (Sect.~\ref{discussion}), and compare the 
expected location of the hot dust with the dust sublimation 
and inner disc radii. We compare our findings to studies of 
transition discs with near- to mid-infrared emission recently
discovered by the Spitzer and AKARI surveys (e.g. Fujiwara et al.~2013,
Maaskant et al.~2013, 
Espaillat et al.~2010, 2011, Muzerolle et al.~2010, Texeira et al.~2010, 
Furlan et al.~2009) to investigate the evolutionary state
of the disc candidates in the Arches and Quintuplet clusters.
The problem of the disc lifetime is addressed, and secondary disc formation 
will be proposed as a possible scenario to explain the Arches and Quintuplet
near-infrared excess sources and their apparent expanded lifetime
compared to primordial discs around Herbig Be stars.

In Sect.~\ref{obssec}, we present the VLT and HST observations,
and the photometric and astrometric analysis is summarised in 
Sect.~\ref{photsec}. The disc fraction of the Quintuplet cluster
is derived in Sect.~\ref{quindiscsec}, and new Arches disc sources
are presented in Sect.~\ref{archdiscsec}. The physical properties 
of the $L$-band excess sources are discussed in Sect.~\ref{discussion},
and we summarise our findings in Sect.~\ref{summary}.

\section{Observations}
\label{obssec}

\subsection{VLT/NAOS-CONICA}

For the membership campaign of the Arches and Quintuplet clusters,
multi-epoch $K_s$ imaging with time baselines of 3 to 5 years was 
obtained with the Very Large Telescope (VLT) adaptive optics
system NAOS and its near- to mid-infrared camera CONICA (hereafter NACO, 
Lenzen et al.~2003, Rousset et al.~2003) during the time period 2002-2012.
Complementary $L'$-band images were observed in June and August 2012.
A complete list of all NACO data sets analysed in this paper is 
provided in Table \ref{obstab}.

\begin{figure*}
\includegraphics[width=8cm]{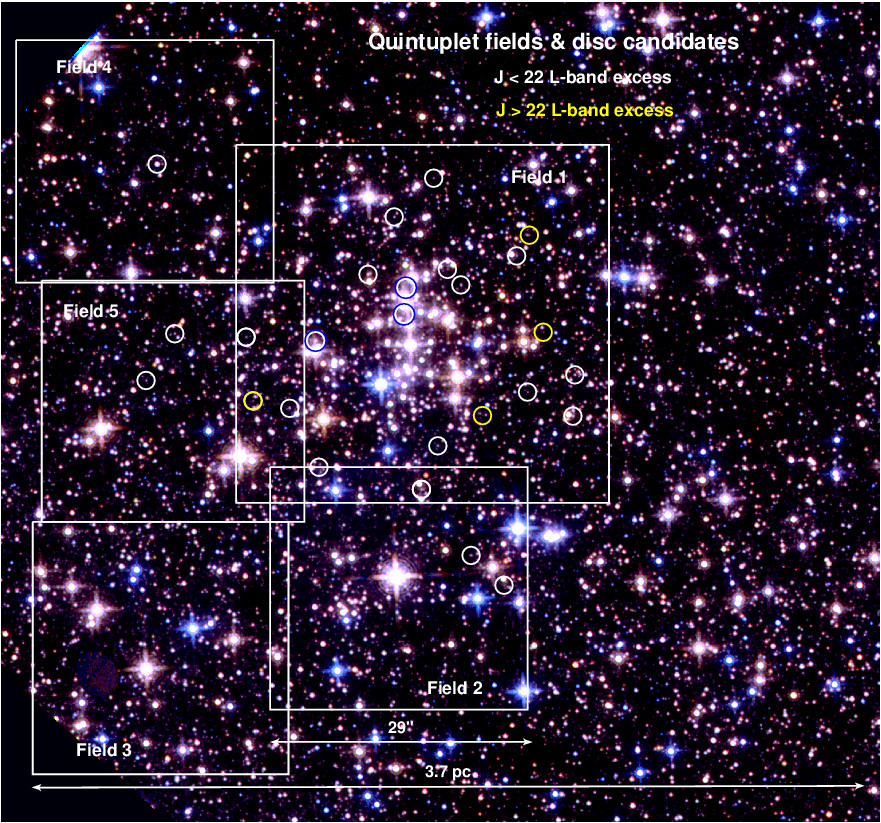}
\includegraphics[width=8.6cm]{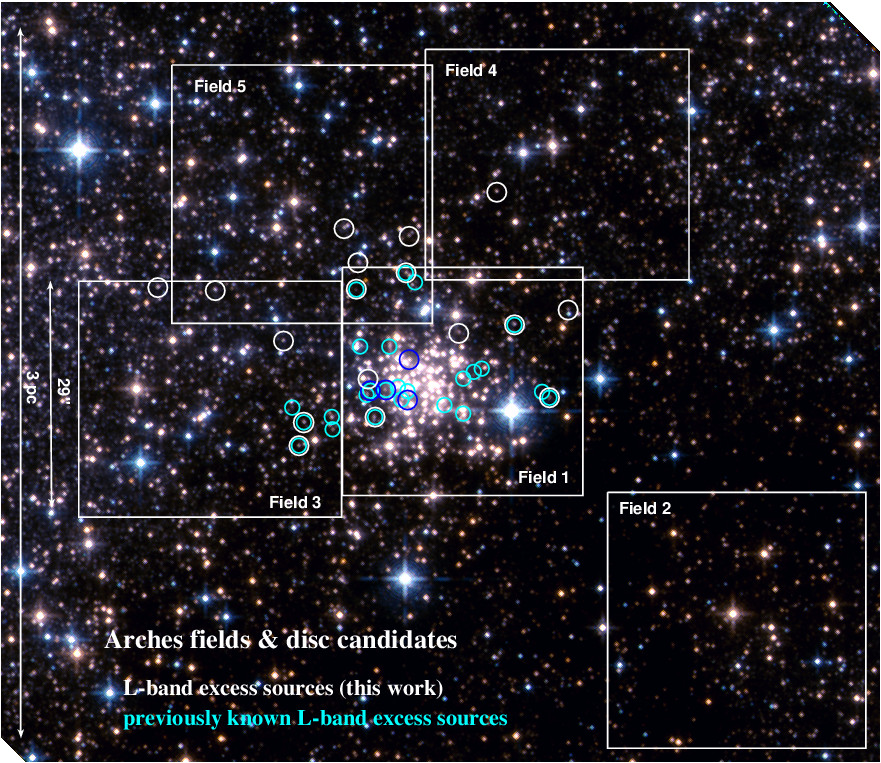}
\caption{\label{mosaic} WFC3 JH colour composite of the Quintuplet cluster (left panel) 
and Arches cluster (right panel) covered with both HST/WFC3 and NACO observations. 
The NACO fields are shown as boxes. $L$-band excess sources detected in both clusters
are overlaid as circles. Previously detected excess sources in the Arches are shown 
as smaller cyan circles. Excess sources in the cluster centres are circumscribed in blue 
for clarity.}
\end{figure*}

The $K_s$ images were obtained with the S27 camera with a pixel scale 
of 27.1 mas/pixel, covering a $27^{''} \times 27^{''}$ field of view. 
Five fields were identified
in each cluster with suitable natural guide stars operating mostly
with the near-infrared wavefront sensor. As a consequence of the 
high foreground extinction of $A_V \sim 25$ mag, optical guide stars 
are rarely available along the Galactic centre line of sight, 
such that the unique NIR sensing capability of NAOS was extensively
exploited to obtain the wide area coverage of both clusters.
The Arches and Quintuplet mosaics cover maximum distances of 
$48"$ and $60"$ from the cluster centres, corresponding to 1.8 pc 
and 2.3 pc at a distance of 8.0 kpc, respectively. Arches field 2 
reaches larger distances out to $63"$ (Fig.~\ref{mosaic}), but is not 
part of the coherent mosaic of this cluster.
For the Quintuplet cluster with evolved stars as bright as
$K_s =7.3$ mag, the N20C80 dichroic could be used,
which distributed 80\% of the light to the science camera and 
only 20\% to the wavefront sensor for natural guiding.
As the brightest sources in the Arches cluster are substantially fainter,
$K_s = 10.4$ mag, the N90C10 dichroic
had to be employed with only 10\% of the light diverted to CONICA.
The Arches $K_s$ data are correspondingly shallower than 
the Quintuplet data sets. Detection limits are provided in Table~\ref{phottab}.

The first epoch data were optimised for deep photometry using
individual detector integration times (DITs) of up to 20s. 
In all newer data sets DITs were kept 
short to avoid saturation of the brighter stars to establish
the cluster reference frame for astrometric proper motion 
measurements. Because of the different setup, these DITs ranged
from 2s for the N20C80 observations to 10s for N90C10 imaging.

The $K_s$ observations were complemented with single-epoch 
$L'$ imaging obtained in 2012. 
In the case of $L'$, NACO offers a $JHK$ dichroic, which passes all
near-infrared light to the wavefront-sensor, and allows the full
$L$-band channel to be diverted to the science detector.
The limiting factor in $L'$ 
is the sky brightness, such that the detector saturates rapidly
even in very short integration times. To avoid saturation,
the exposure time was set to 0.175s - 0.2s (the shortest feasible
integration times with the CONICA detector) using the uncorrelated
readout mode. Between 150 and 170 individual DITs were coadded
to a total integration time of $\sim 30$s per science image.
The L27 camera was used with a pixel scale of 27.1 mas/pix to 
provide the same spatial coverage as for the $K_s$-band observations.
The data were obtained in dither mode for both filters with 
dither offsets between 30 and 70 pixels ($0.8"$ to $2"$)
to allow for sky subtraction and the removal of hot pixels,
which are of particular concern in infrared detectors.
The dither offset was chosen to be less than 1/10 of the field
size in the astrometry-oriented epochs (2008 and onwards)
to minimise the impact of optical distortions and to optimise 
the relative astrometric performance between proper motion epochs. 

In order to monitor sky variations, separate sets of sky images
were observed in $K_s$ after, and in $L'$ interleaved with, the 
science images. Sky fields are observed in open-loop 
mode without adaptive optics correction
with NACO, such that residual starlight is spread out
across the detector and leads to a biased sky level 
in the combined master sky. As it is difficult to find star-free 
fields in the vicinity of the clusters because of the high stellar 
density along the GC line of sight, all sky fields contained some 
residual star light. Given these complications, three 
different procedures were attempted especially for the $L'$-band
data, where the sky is the most limiting factor on sensitivity
(Sect.~\ref{nacoredsec}).

\subsection{HST/WFC3}

Hubble Space Telescope (HST) images were obtained with the 
wide-field camera WFC3 in the near-infrared channel in the 
medium-band WFC3 F127M and F153M filters, corresponding 
approximately to the $J$ and $H$ broadband filters in the 
ground-based near-infrared system, in August, 2010, under 
programme ID 11671 (PI: Ghez). The integration times were 
600s and 350s in F127M and F153M per individual frame,
and a total of 12 and 21 single images lead to combined 
image integration times of 120 min in each filter (see Table \ref{hsttab}).
WFC3 offers a field of view of $2.2' \times 2.0'$, which was 
also the approximate size of the drizzled image, such that
each cluster is covered with a single pointing (Fig.~\ref{mosaic}).
Images were dithered with small positioning offsets between 
0.6 and 10 pixels ($0.08{''} - 1.2{''}$) 
to allow for image reconstruction and bad pixel removal.
While the reduced, distortion-corrected and pre-combined Arches 
images were obtained from the MAST archive, the Quintuplet images 
were obtained before the final version of the distortion solution 
was integrated into the standard WFC3 pipeline.
These images were  processed using the {\sl multidrizzle} algorithm 
(Koekemoer et al.~2002, Fruchter et al.~2009) with the most recent 
distortion solution applied.
Prior to image combination, the image shifts were adjusted with the pyraf 
{\sl tweakshifts} routine provided for WFC3 pipeline reduction 
(see the WFC3 Data Reduction Handbook for details)\footnote{%
http://www.stsci.edu/hst/HST\_overview/documents/multidrizzle/ 
ch56.html}.

\section{Analysis}
\label{photsec}

\subsection{VLT/NACO}
\label{nacoredsec}

\subsubsection{Data reduction}

The data reduction for the VLT/NACO $K_s$ and $L'$ images
was carried out with a custom-made reduction pipeline based
on python/pyraf and IDL routines. The basic steps in $K_s$
included dark subtraction, flat fielding, the removal of 
cosmic ray hits and hot pixels, and the subtraction of a 
master sky. 
Twilight flat fields were observed with decreasing brightness, 
such that the pixel sensitivity could be obtained as the 
slope of the brightness variation in each pixel. 
Bad pixels were identified during the combination of the 
master flat field as pixels where the sensitivity deviated
significantly from the mean. In addition to this universal
bad pixel mask for each data set, individual bad pixel masks
were created using the iraf task {\sl cosmicrays} to identify
positive, spatially confined flux peaks. Where possible,
the sky image for each science set was derived from the adjacent 
sky exposures at an off-cluster position without adaptive optics
correction. As these images frequently contained residual 
star flux, the best sky subtraction was in some data sets achieved
by median-combining both dithered sky and science images to create a
flat thermal background image free of residual stellar light.
The  choice of the sky subtraction procedure depended on the 
individual dataset. 

In $L'$, no master darks are provided by ESO, hence the basic
reduction steps consisted of flat fielding and sky subtraction.
The thermal background is the dominant source of uncertainty
at $3.8\mu$m. 
For most of the $L'$ data sets several attempts 
at sky subtraction had to be made: i) one master sky was median-combined 
from all science and interleaved sky images, ii) one master sky
was median-combined from the sky images only, and iii) individual skies
were created from the images closest in time to the considered
science frame. After each of these master skies was subtracted
from the science images, the final image with the 
least background noise and the highest contrast and sensitivity 
was selected for photometric analysis. 

Both $K_s$ and $L'$ images were combined into one deep exposure
with the drizzle task (Fruchter \& Hook 2002). The routines
{\sl precor}, {\sl crosscorr}, and {\sl shiftfind} were employed to derive the 
positional offsets. For these adaptive optics data, the {\sl precor} task, which 
allows masking of background patterns, aided the identification 
of the true spatial shifts from the cross-correlation image.
Finally, the individual bad pixel masks were applied to each 
image during the image combination process. We note here that 
no distortion correction was applied during drizzling. Despite 
continued efforts to characterise the optical distortions in NACO, 
the distortion correction still bears large uncertainties,
especially for the median-field camera S27 (see e.g. Trippe et al.~2008,
Fritz et al.~2010). 
We interpret this as an effect of anisoplanatism, which depends on 
the nightly conditions and causes additional point source image 
distortions. As these distortions depend on the isoplanatic angle
and the adaptive optics correction during the time of the observations, the 
natural guide star distance from the field centre, and the brightness
of the NGS, they add a strong random component to each source 
position, such that no uniform instrument distortion solution
could be derived (see Sect.~2.3 in Habibi 2014 for details)\footnote{The 
PhD thesis may be requested from mhabibi@mpe.mpg.de.}. 
During the NACO observations, the influence of optical
distortions on the relative astrometric uncertainty is minimised by 
conducting a small-scale dither pattern to avoid large image shifts, and 
by using the same observational setup and positioning in each field during 
each epoch (see also Ghez et al.~2008, Yelda et al.~2010).

\subsubsection{Photometry and astrometry}
\label{photastsec}

The crowded nature of the GC cluster fields required PSF 
fitting to perform the photometry and astrometry on each image.
At a field size of $27{''}$, the NACO images are severely 
influenced by anisoplanatism. As shown in Hu{\ss}mann et al.~(2012),
a constant PSF across the field yields reliable astrometric measurements
only in the area near the natural guide star, which is typically (but not always)
located near the field centre. While we restricted the analysis
of the stellar mass function of the Quintuplet cluster in this 
previous paper to a radius of $r < 500$ pixels or $13{''}$, 
the purpose of the present disc study is to cover the cluster areas
as completely as possible. Therefore, the IRAF daophot package (Stetson 
et al.~1992) was employed to obtain PSF fitting photometry
and astrometric positions of all stars. Quadratically varying 
PSF functions were used to minimise the spatially varying effects
of source elongation due to anisoplanatism on the derived astrometry and 
photometry as much as possible.

In the deep Quintuplet central field of the 2003 observations
obtained exclusively with single-frame integration times of 20 seconds,
a quadratic variation provided unsatisfactory results.
In particular in the deep exposure, stars within $\sim 5{''}$ from 
the edge of the drizzled image were rejected during the PSF fitting
procedure as a consequence of their irregular shape.
Therefore, the long-exposure combined image (see Table \ref{obstab})
was split into four quadrants for which individual PSFs were created
and subtracted. After the quadrants were separated, a linearly
varying function was sufficient to obtain a good fitting performance
across each quadrant. In the overlap regions, the standard deviations 
in x and y residuals between astrometric measurements were below 
0.2 pixels, which is the expected variable PSF fitting accuracy 
in these performance-limited data sets.
The individual quadrant photometric lists were then 
recombined using the offset shifts applied earlier when splitting 
the images. 

The $K_s$ photometry of the Arches cluster was calibrated with 
respect to Espinoza et al.~(2009). 
Fields 3-5 were calibrated
from overlap areas with the central field and calibrations were 
cross-checked in each overlap region. Field 2 has no overlap with 
any of the previous observations and the remainder of the Arches fields,
and was calibrated against the UKIDSS $K_s$ Galactic Plane Survey 
(Lucas et al.~2008, spatial resolution $1{''}$).
The Quintuplet $K_s$ photometry was calibrated
with respect to the UKIDSS GPS survey as well, although in a two-step 
procedure, which helped overcome the resolution differences.
First, VLT/ISAAC observations\footnote{These data were taken under 
Proposal ID: 67.C-0591, PI: Stolte, and are used only for calibration
purposes in this analysis.} taken under excellent conditions
were calibrated over a 2.4 arcminute
field of view with respect to UKIDSS $K_s$. No colour terms were 
found, and a zeropoint offset was applied. The NACO $K_s$ high-resolution 
photometry with a spatial resolution of $\sim 0.1{''}$ 
was calibrated with respect to the seeing-limited ISAAC
$K_s$ photometry with a spatial resolution of $\sim 0.4{''}$
in a second step. Details can be found in Hu{\ss}mann (2014).

The $L'$ photometry was zeropointed with respect to  3.8$\mu$m
IRAC photometry obtained with the Spitzer GLIMPSE survey
(Benjamin et al.~2003, Churchwell et al.~2009).
Because of the large number of detections missing from the 
source catalogues, photometry on the IRAC 3.8$\mu$m images was
rederived using daophot PSF fitting. The large area coverage
of the images allowed for a robust calibration against the 
downloaded Spitzer catalogues. For the cluster fields, however, 
the low resolution of the Spitzer data implied that only a few sources 
(4-6) were available as references, even in the PSF fitting photometry 
source lists. In the $J-H$, $K_s-L'$ colour-colour diagrams, the
uncertain $L$-band calibration manifested itself as offsets from 
the reddening path. We therefore checked the $J-H$, $H-K_s$ 
colour-colour diagrams for consistency with the reddening path
(Nishiyama et al.~2009). Any remaining offsets in $L'$ were
adjusted in the $JHK_sL'$ colour-colour diagram such that 
the $K_s-L'$ colour was also consistent with the reddening path.

\subsubsection{Photometric and astrometric uncertainties}

Photometric and astrometric uncertainties were derived by 
independent repeated measurements of the photometry of each 
star. We combined three subsets of each individual
image stack to perform these measurements. In one data set, 
the number of frames and the image quality were compromised,
such that only two subsets could reasonably be obtained.
To ensure that the data quality of each subset image is 
comparable to that in the deep image,
care was taken that each auxiliary image created from a single
subset contained frames over the full range of spatial resolutions
and photometric sensitivities as contributed to the deep image.
For this purpose, 
the range in performance was measured as the FWHM of the PSF 
on each individual image, and each subset list contained images
over the entire range of FWHM values.
Daophot was then run with the same PSF fitting parameters as 
in the respective deep image on each auxiliary image to produce 
repeated measurements of 
the photometry. Each star that entered the final source 
catalogues was required to be detected in at least two 
auxiliary images in addition to the deep science frame.
The photometric and astrometric uncertainties were then
calculated either as the standard error given as the standard 
deviation from the mean divided by sqrt(3)
if a star was detected in all three auxiliary images, or as the 
deviation from the mean of the two measurements if the star
was only detected in two auxiliary frames. This was mostly
the case for faint stars, but could also be caused by incomplete
area coverage of the subset images in comparison to the complete
deep image. In the data set where only two auxiliary images could 
be obtained, the uncertainties are always given as the deviation
from the mean, and all stars are again required to be detected 
in both auxiliary images.

Photometric uncertainties are shown in Fig.~\ref{kserr} for the 
Quintuplet and in Fig.~\ref{kserr_arch} for the Arches cluster. 
All uncertainties are accordingly
quoted in the final source catalogue along with the magnitude
and positional measurements of the deep science image.\footnote{%
All source catalogues will be available in electronic 
format from the CDS and in the online version of the journal.}

\subsubsection{Geometric transformations}
\label{astromsec}

For   both the Arches and Quintuplet clusters, we derived 
membership information for all fields out to  a radius of 
$\sim 1.5$ pc. For this purpose, we have combined
at least two epochs of $K_s$ NACO observations to construct
the proper motion diagram for each cluster and field.

Proper motions are derived via the matching of two epochs 
of imaging observations separated by a time baseline of
3 to 5 years. The positions of each star have to be 
transformed to one epoch that serves as the reference epoch
for the respective field. The geometric transformation
between both epochs was derived for bright stars with
$K_s < 17.5$ mag. The IRAF task {\sl geomap} provided residual
rotation, image shifts, and relative distortions by fitting
a second-order polynomial to the x and y positional 
differences for all stars in each field. In general, 
the earlier epochs provided deeper images with a higher
spatial resolution as a result of the degradation in 
adaptive optics performance
of the NAOS system over time. Hence, the first epoch was
used as the reference epoch in all cases, and the positions
measured in later epoch images were converted to the first
epoch using the {\sl geomap} geometric solutions. 
After a first transformation,
cluster candidate stars were selected from the proper motion
plane as stars located close to the origin in the cluster
reference frame (i.e.~with zero relative motion with respect
to the mean cluster motion). These cluster member candidates
were iteratively used to generate the final geometric 
transformation solution. This procedure ensures that the 
mix between rapidly moving field stars and slowly moving
cluster members with respect to the cluster reference frame
does not bias the transformation. It should be noted that 
the method implies minimisation of the relative motions 
between cluster stars, such that the cluster population 
appears more confined after the second iteration, which 
improves the distinction between cluster members and field 
stars. The uncertainties in the 
proper motion measurements include the individual positional
uncertainties in x or y of each star as described above, added 
in quadrature to the residual rms uncertainties 
of the geometric transformation. 

The proper motion uncertainties are shown in Fig.~\ref{pmerr}.
A distinct increase in the median uncertainty can be seen in 
all data sets towards fainter magnitudes, as expected. 
As the limit where reliable membership measurements can be obtained
depends on the data set, we adopt limiting magnitudes of
$K_s < 17.5$ in all outer Quintuplet fields and $K_s < 18$ mag 
in the Quintuplet central field, where both $K_s$ epochs are equally deep, 
as indicated in Fig.~\ref{pmerr}.
In the central field, we also include in the source catalogue
an indication of membership for fainter stars, 
in which case we mark the membership information as uncertain. 
In contrast to Hu{\ss}mann et al.~(2012), the
source catalogue of the central field (Field 1, F1) in the search for 
disc candidates presented here includes 
the full $40{''}\times 40{''}$ dithered field of view.
\\

A realistic estimate of the uncertainties in the proper motion 
measurement of each star can also be obtained from the comparison
of the derived proper motion values with our earlier NACO-NIRC2 member selection
in the Arches cluster (Stolte et al.~2010). Substantial differences between 
the motion samples might introduce uncertainties in the main-sequence member 
selection, and hence in the disc fractions.

In Fig.~\ref{compare_ast}, the astrometry presented here is compared to
our earlier results obtained with Keck/NIRC2 and VLT/NACO over a 
time baseline of 4 to 5 years (Stolte et al.~2010). 
Proper motions in Stolte et al.~(2010) were derived from Keck/NIRC2 
astrometry with respect to the NACO 2002 central field also used as 
reference for Field 1 in this study.
The area that could be matched covers the central 13$''$ of the 
cluster, and is dominated by the crowding-limited cluster core.
The characteristic deviations in proper motions both in the 
right ascension and declination directions are $\pm 0.5$ mas/yr
for the majority of cluster members. This uncertainty is very 
similar to the absolute motion uncertainties in Arches Field 1 as 
derived from multiple astrometric measurements (Fig.~\ref{pmerr},
top right panel), suggesting that the repeated PSF fitting
on the shallower auxiliary subsets yields realistic motion 
uncertainties. The relatively large 
differences originate from the fact that all earlier observations 
with Keck/NIRC2 and VLT/NACO were taken under excellent 
conditions, resulting in higher spatial resolutions of 60-80 
milliarcseconds closer to the Keck and VLT diffraction limits than 
the more recent VLT data sets covering the full cluster area 
presented here. 
The comparison is restricted to the area of the earlier data sets
in the cluster centre, and hence proper motion deviations are
induced by crowding. 
No systematic bias is observed in both proper motion directions, 
confirming the robustness of our transformation method. 

As can be seen from Fig.~\ref{compare_ast}, the membership selection 
is barely affected by the proper motion uncertainties. Stars previously
selected as cluster members in Stolte et al.~(2010) are shown as red
asterisks, while stars selected as members in the new NACO data set
are shown as blue diamonds. Out of 442 stars matched between 
both samples, 309 were chosen to be members from the NIRC2-NACO 
astrometry, and 332 are chosen to be members from the NACO-NACO 
astrometry presented here, with 299 stars being common members in 
both samples. Only ten stars chosen as members in our previous
study are not recovered as members here, and six of these have 
substantial proper motion deviations, suggesting that their motions
are uncertain. An additional 33 stars are detected as members
in this work, probably because   the rms
scatter in the motion plot is larger and the corresponding 
member selection needs to be wider than chosen before
(1.50 mas/yr as compared to 1.30 mas/year in Stolte et al.~2010). 
Hence, a total of 43 stars or 10\% of the comparison sample 
have deviating membership designations, which represents the 
uncertainty in the derived main-sequence reference samples 
due to the proper motion membership selection alone.

As a consequence of the field size limitations in our earlier 
observations, this comparison only covers the crowding-limited
cluster core. We expect less deviation in the individual motions
for stars at larger radii, including both stars in the outer
fields of the Arches cluster as well as in the Quintuplet cluster
where stellar densities are lower. Therefore, a maximum deviation
in the main-sequence number count of 10\% is expected in the 
Arches cluster core, while the deviation will be even smaller
for all other main-sequence reference  samples.

\subsection{Membership derivation}
\label{memsec}

The resulting proper motion diagrams of the central Quintuplet and Arches 
fields are presented in Figs.~\ref{quinpm} and \ref{archpm}, while the outer 
fields are presented in Figs.~\ref{quinpm_app} and \ref{archpm_app}. 
A confined cluster population is observed at the origin, as expected
after transforming to the cluster reference frame. The field star
population is elongated along the orientation of the 
Galactic plane (dashed line in Fig.~\ref{quinpm}), indicative of 
a wide range of orbital parameters for field stars in the GC
(see also Stolte et al.~2008 for a discussion). The histograms 
of motion in the Galactic longitude and latitude directions are 
shown in Fig.~\ref{pmhists} for the central Arches and Quintuplet fields. 
The peak of cluster member candidates (hereafter cluster members) 
stands out clearly
in longitudal motion (motion along the plane), followed by 
the extended tail of field stars. In the latitude direction, 
the peak is dominated by the positional uncertainties.
The gaussian fits of the latitudinal motion in the central fields
provides the 2$\sigma$ selection criterion for cluster members.
As the outer fields of the Arches cluster are dominated by field 
stars, the cluster peak is not pronounced and the same selection 
criterion as for the central Arches field 
is used ($\mu < 1.5$ mas/yr). For the Quintuplet outer Fields 2 to 5,
membership propabilities are available from our mass function campaign
(Hu{\ss}mann 2014). In these fields exclusively, we use the membership
probabilities to discern cluster members from field stars (see
Hu{\ss}mann 2014 for details). 

There is one major difference in the derivation of membership probabilities
 compared to earlier membership studies in nearby clusters. 
The combination of the large cluster distance
and the small time baseline did not allow the use of the classical membership
likelihood as the cluster member selection criterion.
Instead, the criterion for a star to be considered a cluster member was constrained
using Monte Carlo simulations of the distribution of field and cluster stars
in the proper motion plane (see Sects. 4.2.2.1 to 4.2.2.3 in Hu{\ss}mann (2014) 
for details). 
In these models, two artificial Gaussian distributions were populated with stars
to represent the cluster and the field, respectively. The properties of these
distributions were modelled after the observed proper motion measurements.
Each artificial star was assigned a proper motion uncertainty drawn randomly
from the observed distribution of astrometric errors, and as in the real 
observational data, these uncertainties were used to weight the 
probability for each star to belong to the cluster or the field.
The simulated cluster and field distributions were then fitted with an 
expectation-maximisation algorithm as described in Hu{\ss}mann (2014).
For these simulated cluster/field distributions, the recovery fractions
of cluster stars were compared to the inserted number of simulated cluster
stars. Because of the overlap between the cluster and field ellipses, 
the distinction cannot be perfectly made. A trade-off probability value 
was derived for which the maximum number of cluster stars is recovered, 
while the minimum number of field interlopers contaminates the sample.
This probability represents the likelihood of finding each star at a 
certain location in the proper motion plane, and is weighted by each star's
respective proper motion uncertainty (Eq.~4.11 in Hu{\ss}mann 2014).
These membership indicators are included in Table \ref{jhktab_quin} for reference. 
Extensive simulations of Field 2, which has the largest number of cluster stars 
outside the central field and hence facilitates statistical modelling, 
suggested that cluster stars are most efficiently separated from 
field stars using a formal probability threshold of 0.4. Stars above
this threshold are likely cluster members, while stars below this limit
are likely field stars.
\\

As will be discussed further below, neither the 2$\sigma$ membership selection 
nor the probability method allow  a perfect distinction
between cluster and field stars. 
Some stars will unavoidably move on orbits similar to the cluster 
motion, even if their motions are extreme compared to circular orbits
of stars and clouds in the inner bulge (Stolte et al.~2008). 
Some of these stars will be bulge stars with elongated or high-velocity orbits 
located at a similar distance as the clusters. This population is 
dominated by bulge giants and red clump stars and has the  
red colours of evolved stars with substantial extinction.
In addition, a number of foreground stars move on 
characteristic Galactic disc orbits with lateral velocities 
in the range of $\sim 200$ km/s along the Galactic centre line of sight, 
similar to the cluster orbital velocity (Stolte et al.~2008, Clarkson et al.~2012). 
Galactic disc stars significantly nearer to the Sun can be discerned from cluster 
stars as their proper motion will appear larger than the cluster velocities.
However, Galactic disc stars at larger distances are not easily 
distinguished on the basis of their proper motion alone.
The Galactic disc population is concentrated along the spiral arms,
and is predominantly composed of main-sequence stars with 
blue colours and a lower foreground extinction. Both the 
bulge and Galactic disc contaminants are distinct in the colour-magnitude
diagram on the basis of their colours, which are different 
from the reddened cluster main sequence. The colour information 
therefore provides an additional tool with which to improve the membership
selection and obtain an almost genuine cluster member sample,
and will be employed to derive the main-sequence reference
samples for the disc fractions in Sects. \ref{quindiscsec} and 
\ref{archdiscsec}.

\subsection{HST/WFC3}
\label{hstphot}

As adaptive optics observations suffer from marginal
adaptive optics corrections at shorter wavelengths, we employed the WFC3
$J$ and $H$ images to derive $JHK_s$ photometry of each source.
In the case of the HST/WFC3 observations, either the pre-reduced 
images (Quintuplet) or the combined drizzled image (Arches) were
downloaded from the HST MAST archive. The drizzled images provided
the basis for the photometric analysis.  The observational properties
are summarised in Table \ref{hsttab}.

\subsection{Photometry}

After applying the most recent distortion solution, starfinder PSF 
fitting with a constant, empirically extracted PSF across the field 
was applied to the Quintuplet images. 
For the Arches, the reduced $JH$ images were, at the time 
of processing, not optimised for astrometry, and 
daophot photometry with a quadratically varying penny or moffat
function provided the lowest photometric residuals, and was hence
used on both the $J$ and $H$ images.

Photometric uncertainties were established for all datasets
as described above. Auxiliary images were either created or 
consecutive data sets were used to obtain independent measurements, 
and the PSF fitting was repeated in the same way as for the deep drizzled
images. As in the case of the NACO data, the uncertainties 
were calculated as the standard deviation or the deviation 
from the mean (Figs.~\ref{jherr} and \ref{jherr_arch}).

The $JH$ photometry lists were combined into a single catalogue, 
which was calibrated with respect to UKIDSS (Lawrence et al.~2007). 
The UKIDSS resolution of $1{''}$ implies an improvement of a factor of two 
compared to the 2MASS catalogues, which is particularly crucial 
in these crowding-limited cluster fields. 
We used a version of data release 6 (DR6) of the UKIDSS Galactic Plane Survey 
(GPS, Lucas et al.~2008), which we corrected with respect to 2MASS
for foreground extinction effects (see also Hodgkin et al.~2009 for 
details on the UKIDSS calibration). This correction proved necessary 
in the GC fields, as the automatic pipeline zeropointing is based
on Schlegel maps (Schlegel et al.~1998), which suffer from low resolution 
and provide extinction values that are a factor of 3 too high in the GC region.
Only sources detected in both $J$ and $H$ are included in the final 
catalogue to allow for colour term correction. Colour terms were
derived directly with respect to the corrected UKIDSS catalogue
in the case of the Arches cluster. 
As in the case of the Quintuplet $K_s$ calibration (Sect.~\ref{photastsec}), 
the Quintuplet WFC3 photometry 
was calibrated with respect to ISAAC $J_sHK_s$ photometry, and the colour
terms were derived accordingly. The ISAAC photometry was calibrated over
the full ISAAC field of view of 2.5 arcminutes with respect to UKIDSS,
which facilitated the selection of isolated calibration stars.
The NACO photometry was then calibrated in a second step with respect
to this ISAAC photometry (see also Hu{\ss}mann (2014) for details).

The colour conversions are derived to be

\begin{eqnarray}
J_{s,ISAAC} - J_{WFC3} & = & ZPT_{J,WFC3} + \nonumber \\
                    &   & 0.09\, (\pm 0.02) \cdot (J_{WFC3}-H_{WFC3}) \\
H_{s,ISAAC} - H_{WFC3} & = & ZPT_{H,WFC3} - \nonumber \\
                    &   & 0.36\, (\pm 0.04) \cdot (J_{WFC3}-H_{WFC3}) 
\end{eqnarray}in the Quintuplet field with respect to ISAAC $J_sH$, and to be 

\begin{eqnarray}
J_{UKIDSS} - J_{WFC3} & = & -0.23\, (\pm 0.11) + \nonumber \\
                   &   & 0.12\, (\pm 0.06) \cdot (J_{WFC3}-H_{WFC3}) \\
H_{UKIDSS} - H_{WFC3} & = & 0.85\, (\pm 0.16) - \nonumber \\
                   &   & 0.30\, (\pm 0.06) \cdot (J_{WFC3}-H_{WFC3}) 
\end{eqnarray}in the Arches field with respect to UKIDSS $JH$. The colour terms 
of the WFC3 F127M and F153M filters found with respect to
the ground-based ISAAC and UKIDSS $JH$ filters are
consistent within their uncertainties.
In the Quintuplet field, because of the different calibration procedures, the 
offset term was not fit independently of the instrumental zeropoints 
$ZPT_{J,WFC3} = 1.31 \pm 0.06$ and $ZPT_{H,WFC3} = 1.96 \pm 0.07$, 
such that they are not directly comparable to the absolute offsets derived 
in the Arches field after a preliminary calibration offset had been applied.
After colour terms were removed, the Arches $JH$ photometry was adjusted
to match the photometry by Espinoza et al.~(2009) in the cluster centre, 
as in the case of the $K_s$ observations (Sect.~\ref{photastsec}).

Finally, the combined $JH$ WFC3 catalogue with calibrated, colour-term
corrected magnitudes was transformed to each $K_s$ reference epoch 
image and matched with the $K_s$ and $K_sL'$ photometry lists separately.
Source counts and sensitivity limits for these combined catalogues
are provided in Table \ref{phottab}. The sensitivity limits are derived
from the peak of the luminosity functions (Figs.~\ref{lfs} and \ref{lfs_arch})
as a first indication of the completeness of each data set in each filter
after matching $JH$ with $K_s$, and $L'$ with $JHK_s$ in the case of the $L'$
luminosity function exclusively. A rigorous incompleteness treatment, 
including the effects of catalogue matching, is presented in the next
section.

\subsection{Incompleteness simulations}
\label{incsec}

There are two major limitations to the sample of excess sources and detected
main-sequence members of the Arches and Quintuplet clusters. Stars with 
main-sequence colours are mostly limited by detection in the bluest filter,
WFC3 $J$. This limitation arises from the high foreground extinction towards
both clusters and the GC in general.
In addition, the lower resolution of HST/WFC3 of 230 mas
compared to the NACO $K_s$-band resolution of typically 80-120 mas 
(see Table \ref{obstab}) prohibited the detection of $J$ and $H$ 
counterparts for $K_s$ sources in areas with high stellar densities.
Especially in the crowding-limited core of the Arches cluster, 
the loss of one neighbour in close pairs is common. Sources with 
$L$-band excess, on the other hand, are additionally limited by the 
$L$-band completeness limit at $L' \sim 14.5$ mag. As all $L$-band
sources are also matched with the $JH$ source list to reveal the excess
in the two-colour plane, the detection of excess sources is limited 
by the constraints in both $J$ and $L$. 

With the aim of quantifying the losses and deriving completeness-corrected
disc fractions, we have performed artificial star experiments.
For both clusters, artificial stars were inserted in the WFC3 $J$ and $H$
images, and photometry was performed with starfinder (Quintuplet) or
daophot (Arches) as on the original images. The $J$ and $H$ catalogues
of recovered artificial stars were matched to account for matching losses.
For the NACO $K_S$ epochs, the less sensitive and hence more limiting 
proper motion epoch was used to derive recovery fractions. Artificial stars
were inserted in the same physical location in each filter. The magnitudes
and colours of each artificial star were chosen to represent a typical
main-sequence star in each field, with a range of $4.2 < J-K_s < 4.8$ mag 
in the Quintuplet and $4.3 < J-K_s < 6.2$ mag in the Arches. 
The $K_s -L'$ colour was chosen to be 2.4 mag in the Quintuplet and 
2.5 mag in the Arches for all excess source simulations, in accordance
with the observed characteristic colour of the excess sources in each cluster. 
As for the real photometry, the $K_s$ and $JH$ artificial source lists were 
combined with the same matching parameters. In the real source lists, 
a limit of $K_s < 17.5$ mag was imposed to allow for proper motion member
selection, and the same selection limit is applied for the calculation 
of all completeness fractions. The $JHK_s$ artificial star lists 
are employed to calculate the completeness fraction of main-sequence stars.
The $JHK_s$ catalogues are then combined with the $L'$ artificial star
lists to derive the completeness fraction of the excess samples.

The $JH$ images are mainly limited by crowding effects, as WFC3 provides
uniform photometric performance across the field. The NACO $K_s$ and $L'$ 
epochs and fields, on the other hand, are influenced by the adaptive 
optics correction (Strehl ratio, sensitivity, and anisoplanatism) as well
as fluctuations in the thermal background especially in $L'$. These effects 
cause each field to display different completeness curves.
In Fig.~\ref{incfit}, the completeness fractions are fitted with 
fourth-order polynomials. In the Quintuplet simulations, the drop in completeness
with fainter magnitudes
is significantly steeper than in the more crowding-limited Arches 
simulations. Two separate polynomials were fitted in the shallow parts
of the incompleteness curves at brighter magnitudes and the steep parts
of the curves towards faint magnitudes in the Quintuplet simulations.
Polynomial fitting provides the advantage that
a completeness correction can be calculated for each star's observed magnitude
without binning into magnitude bins. The completeness-corrected star
counts are used in Sects.~\ref{quindiscsec} and \ref{archdiscsec} to
obtain corrected excess samples and main-sequence reference samples in 
each field.

\section{Results}

\subsection{Disc candidates in the Quintuplet cluster}
\label{quindiscsec}

A robust estimate of the cluster disc fraction, especially at cluster ages
above 1-2 Myr, can be obtained from the fraction of $L$-band excess sources 
with respect to the main-sequence reference sample (Haisch et al.~2001).
To derive $L$-band excess fractions in each field and in the entire cluster, 
we combine the $L'$ photometry with the $JHK_s$ source lists.
The $J-H, K_s-L'$ colour-colour diagram of the central Quintuplet field 
is presented in Fig.~\ref{ccds}, while the outer fields are included in the 
Appendix (Fig.~\ref{ccds_app}).
Excess candidates are selected if their colours are redder than 3 $\sigma$
with respect to the reddening vector, where $\sigma$ is calculated as the standard deviation
of the $K_s-L'$ colour
of all cluster members with $K_s < 17.5$ mag. In the central field, the deeper sensitivity 
allowed for a member selection down to stars with $K_s \sim 20$ mag, such that only a 
proper motion accuracy limit was imposed in Field 1 exclusively. 
As Wolf-Rayet stars also display infrared excesses due to their strong envelope
emission, only stars fainter than $K_s > 12$ mag are considered to be disc candidates. 
The WR candidates are shown in blue in all colour-colour and colour-magnitude diagrams.
The astrometric accuracy of these sources is frequently compromised by non-linearity in the 
PSF core, limiting the value of the membership criterion, yet they display colours and 
brightnesses consistent with Wolf-Rayet stars belonging to the Quintuplet population.
Although the combined $JHK_sL'$ catalogue contains the most reliable photometric sources,
not all sources measured in all four filters have proper motion membership information.
A few sources are too faint in the less sensitive second epoch observations or happen to 
fall close to equally bright or brighter neighbours, such that the astrometric uncertainties
are too large for these sources to derive membership information (black sources in 
the colour-colour diagrams). In most Quintuplet fields, there is characteristically
one non-cluster member in the excess sample, and one source where the membership situation
is unclear (see Figs.~\ref{ccds} and \ref{ccds_app}). 
This is particularly striking in Field 3, where no cluster member excess source
is found, yet one source without membership and one non-member show $L'$ excess emission.
A similar situation is observed in Fields 1 and 5, while all excess sources observed in 
Field 2 are cluster members. Only one faint excess source is observed in Field 4
owing to the shallow sensitivity of the $L'$ photometry in this field. The formally derived
low excess fraction in Field 4 is therefore not representative for the remainder of the cluster.
With just two epochs of proper motion measurements, the motion of each star has to be 
derived as the difference in the positions between both epochs, and no linear fit 
to the motion can be obtained. Likewise, the likelihood that each star belongs to 
the cluster or the field does not provide an absolute distinction between cluster
members and non-members. More accurate motions, obtained from multiple proper motion
epochs, are required to definitively conclude whether the apparent non-member excess 
sources might belong to the cluster as well. It is therefore too early to discuss
their possible origin and the nature of their excess, and we exclude them from 
the member sample in the following analysis.
  
A total of 24 cluster members with $L'$ excess emission is found distributed 
across three of the Quintuplet fields (Figs.~\ref{mosaic} and \ref{ksmosaic}).
Two sources are detected in the overlap regions of two fields, providing 
independent confirmations of their $K_s-L'$ excesses.
Two additional excess sources do not have
membership information, such that a maximum of 26 excess sources is currently 
observed in the Quintuplet cluster. The positions and photometry of these sources
are provided in Table \ref{disctab}, and a summary of the observed number counts
is included in Table \ref{extab}.
Most $L$-band excess sources in the Quintuplet are detected in the central field, 
at radii $r < 1$ pc from the cluster centre. The central field harbours 16 excess
sources with $J < 22.0$ mag, while a total of 18 excess members plus one star with 
unknown membership are detected down to $J = 22.8$ mag. 
In contrast to the central field, the outer fields only feature  between 0 and 4 
members with $L'$ excess, which suggests that the outer regions of the 
Quintuplet cluster harbour very few disc candidates. 
The limiting factor in the 
detection of $L$-band excess sources is the $L'$ sensitivity itself. The strong 
thermal background emission at 3.8$\mu$m, which is a combination of sky brightness and 
dust emission from optical elements such as telescope mirrors, 
leads to substantial noise in each $L$-band image.
In the outer fields, all detected $L'$ excess sources are brighter than $L'=15.0$ mag.
All $L$-band excess sources are included in the source catalogue, yet for the derivation
of the global excess fraction, only sources brighter than $L=15.0$ mag and $J = 22.0$ mag
are considered. 

In order to derive the $L$-band excess fraction and hence the potential fraction of 
circumstellar discs in the Quintuplet cluster, a main-sequence reference sample has to 
be defined. This sample has to contain membership information to distinguish cluster 
stars from field interlopers, and it has to cover the colour range expected for main-sequence  sources belonging to the Quintuplet. A problem arises from the fact that 
disc sources are easier to detect in $L'$ than main-sequence stars because of 
their infrared brightness enhancement. Hence, the $JHK_s$ source lists are used to 
derive the reference sample instead of the $JHK_sL'$ combined catalogue
(see Stolte et al.~2010). We use the $J$-band brightness as an indicator of the 
stellar photospheric luminosity and the stellar mass, as the excess from circumstellar 
disc emission is expected to be weak at bluer wavelengths. The limiting $J$-band 
magnitude in the excess sample can therefore be used to define the main-sequence
reference sample in each field.
The $J, J-K_s$ colour-magnitude diagrams (Figs.~\ref{jjk_cmds} and \ref{jjk_cmds_app}) 
reveal
that almost all disc candidates with $L' < 15$ mag are brighter than $J=22.0$ mag, and only
three sources between $22.0 < J < 22.8$ mag are found in the central field. 
The deeper member/non-member distinction in this field allowed us to include these sources, 
imposing the same $L' < 15$ mag limit as in the outer fields, in the full sample of disc 
candidates. The main sequence was selected down to $J=22.8$ mag accordingly.
The characteristic main-sequence colour of $J-K_s \sim 4.5$ mag implies that a 
limiting magnitude of $K_s = 17.5$ mag used for the membership samples in the outer
fields corresponds to $J=22.0$ mag as well. We therefore use $J=22.0$ mag as the 
faint limit to define the main-sequence samples in all outer fields. 

The  membership source lists derived from proper motions cannot exclude blue disc
stars or red bulge giants which happen to move with velocities similar to the clusters
(see Hu{\ss}mann et al.~2012 for a detailed discussion).
Hence, we apply an additional colour selection to remove these contaminants. While 
the blue limit of the cluster main sequence stands out clearly in the CMDs, the red
limit of the cluster population is more difficult to discern. The red limit was determined
from the red clump population evident in each CMD at $J \sim 20$ mag, and progressing
towards redder colours and fainter magnitudes along a distinctive path (the reddening vector).
The red limit is chosen such that the cluster main sequence is completely included in the 
selection while red clump stars are rejected as efficiently as possible. As a consequence 
of the varying extinction across the Quintuplet cluster field, the colour selection had 
to be adapted to the main-sequence colour spread observed in each CMD. The colour selections
for main-sequence stars are included in Table \ref{extab}, and are shown in Figs.~\ref{jjk_cmds}
and \ref{jjk_cmds_app}
as dashed vertical lines. In summary, the following constraints define the main-sequence
reference samples: i) $J < 22.0$ mag, ii) $3.7-4.0 < J-K_s < 4.5-5.5$ mag (see Table \ref{extab}), 
and iii) source is a proper motion member. 
To estimate the unavoidable biases imposed by the main-sequence colour selection, the 
red boundary was shifted by $\pm 0.2$ mag to derive upper and lower limits of the excess 
fractions in each field individually. These uncertainties are included in Table \ref{extab}.
Both the main-sequence samples and the excess source counts are corrected for incompleteness
using individual completeness values for each star as outlined in Sect.~\ref{incsec}.
In Table \ref{extab}, we include the observed and incompleteness corrected disc fractions
for each field. In the two bottom rows, the cluster disc fractions are calculated from 
the full area coverage of all fields. Table \ref{extab} also provides a lower and an 
upper limit on the true excess fraction, $f_{ex} = n_{ex}/n_{ms}$, where $n_{ex}$ is the 
number of excess sources and $n_{ms}$ the number of stars in the main-sequence reference sample.
The upper limit originates from including each of the three fields with excess source populations. 
Here, the main-sequence reference samples are derived from Fields 1, 2, and 5 exclusively, 
and $f_{ex}$ is correspondingly large as a result of the small main-sequence number count $n_{ms}$.
A lower limit to the total disc fraction is derived when including all fields in the 
main-sequence reference sample. In this case, Fields 3 and 4, which feature 0 and 1  excess cluster
members with $K_s > 12$ mag, are also included, such that the cluster main-sequence population 
can be considered more complete. The real situation is slightly more difficult to assess, 
as the varying extinction  particularly affects the detection of faint cluster stars in Field 4, 
where the $L$-band sensitivity is also compromised. Here, the maximum number of presently known
Quintuplet members is used as the main-sequence reference sample, increasing $n_{ms}$ and 
minimising $f_{ex}$. For the combined excess fraction in the Quintuplet cluster, a $J$-band
limit of 22 mag is imposed in all fields (F1-F5) on the main-sequence reference samples.

The values for all fields and all main-sequence reference samples are in the range 3-5\%
(again with the exception of Field 4). Combining all excess source members, we find
an upper limit to the incompleteness-corrected excess fraction of 4.1\% and a lower limit 
of 3.7\% for the full cluster area. Including the three excess sources without membership 
information in the excess number count increases the upper limit to 4.3\%. 
Employing these upper and lower limits as uncertainties, we conclude that the excess fraction 
in the Quintuplet cluster at its present age of 4 Myr is $4.0\pm 0.3$\%. 


   Although the Quintuplet cluster has a lower density than the core of the Arches cluster, 
   a maximum uncertainty in the disc fraction can be derived by assuming the same 10\%
   membership selection uncertainty as found in the Arches core (Sect.~\ref{astromsec}). 
   A number count error of $\pm 10$\% in either the Field 1+2+5 main-sequence sample 
   or in the full main-sequence sample changes the respective combined excess fraction by $\pm 0.4$\%.
   If we additionally assume the propagated uncertainty in incompleteness-corrected 
   number counts is 10\% as well, the corresponding uncertainty in the corrected disc fraction 
   is found to be $\pm 0.5$\%. Taking into account the uncertainty imposed by the colour 
   selection (the higher value of $\sigma_{f_{ex},low}$ and $\sigma_{f_{ex},high}$ in Table 
   \ref{disctab}), and assuming that both uncertainties are independent, 
   the disc fraction in the Quintuplet cluster is found to be $4.0\pm 0.7$\%
   (Table \ref{disctab}).

The stellar masses of the host stars of these circumstellar disc candidates can be estimated
from their $J$-band brightness, which is presumed to represent the stellar photosphere.
This assumption is validated by the 
fact that most of our sample sources show little to no near-IR 
excess emission in the $2.2\mu$m $K_s$ band. The mean extinction towards each field
was assumed in the conversion of $J$-band magnitudes into stellar masses.
Our imposed $J$-band limit of 22.0 mag corresponds to a lower mass limit of $2.2\,M_\odot$
for both a 4 and 5 Myr Geneva main-sequence isochrone (Lejeune \& Schaerer 2001).
The brightest source with mid-infrared excess emission has $J=19.6$ mag, which 
corresponds to a stellar mass of $14.2\,M_\odot$ for the 4 Myr isochrone.
Effective temperatures covered by these stellar masses in the Geneva models range
from 10500 to 29300 K. Following the recent compilation of Currie et al.~2010 (see their 
Table 7) for main-sequence stars with solar metallicity, these temperatures suggest
that the disc candidates in the Quintuplet cluster are found around B9.5V to B0V stars. 

Currently, it is unclear whether the fraction of excess sources increases towards
lower-mass stars; the sample statistics are not sufficient to provide excess fractions
in various mass bins. From these data alone, we can therefore not conclude that stars
with lower or higher masses are more likely to exhibit $L$-band excess emission.
Given  the UV-intense environment of this massive cluster, one might not expect to find 
disc candidates at an age of 4 Myr at all.
The implications of this finding and possible explanations for an elevated disc fraction, 
especially in a more evolved, massive young cluster, are discussed in Sect.~\ref{discussion}.

Despite the large range of radial distances from the cluster centre covered by the different 
pointings, the excess fractions derived individually for Fields 1, 2, and 5 are identical 
within their uncertainties. Hence, we find no indication for a radial variation of the 
disc fraction in the Quintuplet cluster. This finding contrasts with our earlier detection 
of a significant variation in the disc fraction of the Arches cluster, where a radial increase
from 3\% to 10\% was observed (Stolte et al.~2010).
The radial variation in the Arches cluster might be caused by disc destruction in 
the dense cluster core, where UV radiation from neighbouring sources and dynamic
interactions could affect disc survival (Olczak et al.~2012). 
These disc destruction mechanisms might not be effective at the lower central density 
of the Quintuplet cluster, such that disc sources are equally distributed at all radii.
Alternatively, the generally lower disc fractions in the Quintuplet cluster of $\sim4$\%
might mask the radial dependence, as radial variations of $\pm 0.5$\% would remain undetected 
given the main-sequence selection uncertainties. 
We therefore conclude that, within these uncertainties, the disc fraction in the Quintuplet
cluster shows no radial variation.

\subsection{Disc candidates in the Arches cluster}
\label{archdiscsec}


In the Arches cluster, a total of 28 candidates with 3$\sigma$ $L'$ excess emission is found,
of which 20 are proper motion members of the cluster, 6 are non-members, and for 2 excess 
sources in the central field the membership is unknown (Figs.~\ref{ccds_arch} and \ref{ccds_arch_app}).
Three excess sources detected at the edge of Field 5 are also observed in either Field 1 or Field 3.
Among the proper motion members, one excess source qualifies 
as a Wolf-Rayet candidate with $K_s = 11.2$ mag and is excluded from the final disc sample.
The sample of circumstellar disc candidates identified as likely cluster members therefore 
contains 19 sources (21 including the two sources with unknown membership). 
The excess source list can be found in Table \ref{disctab_arch}, and the number counts 
in each field are summarised in Table \ref{extab_arch}.
As in the Quintuplet sample selection, the main-sequence reference sample was chosen 
around cluster members with characteristic main-sequence colours given the extinction in 
each field (see Figs.~\ref{jjk_cmds_arch} and \ref{jjk_cmds_arch_app}). 
Given the similarity in the $J$-band photometries,
and the main-sequence colour range, a limiting magnitude of $J < 22$ mag is applied
to obtain a complete main-sequence sample. In the Arches, there are two sources in the 
central Field 1 and one source in Field 5 with $J > 22$ mag, yet the majority of excess 
sources are significantly brighter than this limit 
(Figs.~\ref{jjk_cmds_arch} and \ref{jjk_cmds_arch_app}). 

\begin{figure*}
\includegraphics[width=8cm]{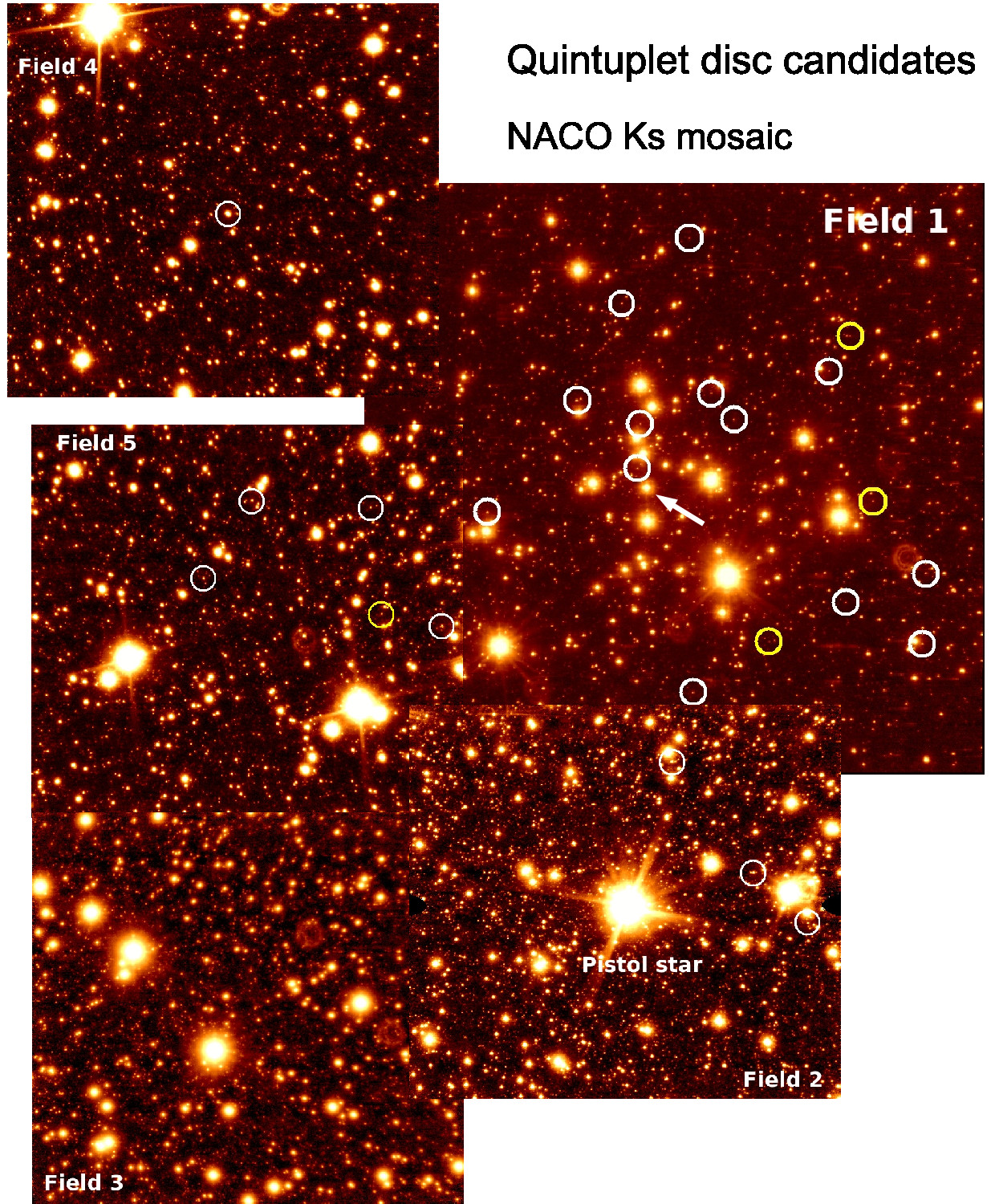}
\includegraphics[width=12cm]{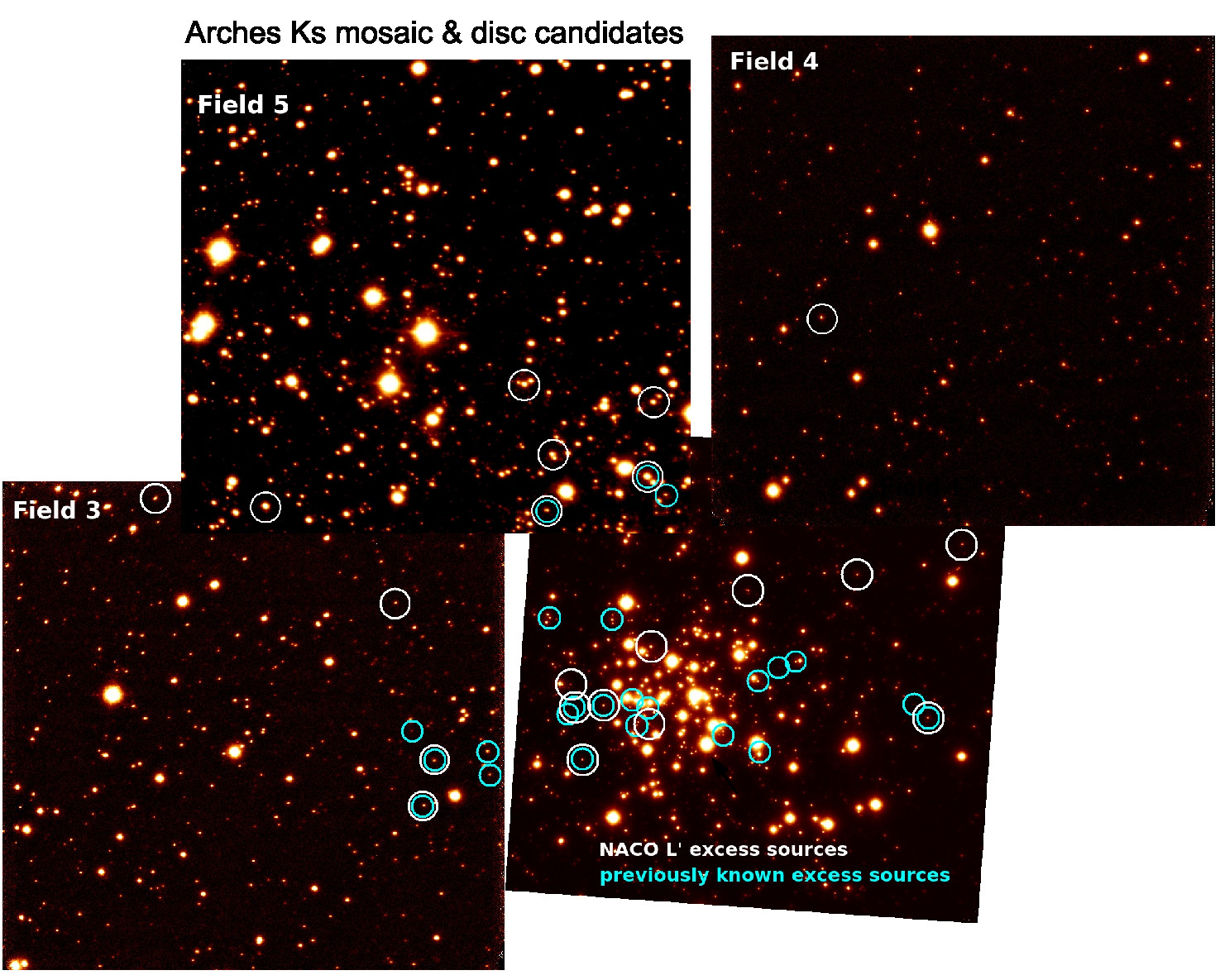}
\caption{\label{ksmosaic} NACO $K_s$ mosaic of fields in the Quintuplet (left)
and Arches (right) clusters. The same fields are approximately covered in $L'$.
In the Arches cluster, Field 2 is not shown and is located to the lower
right of the cluster (see Fig.~\ref{mosaic}). Disc candidates are 
indicated in both clusters as white circles. Yellow circles in the Quintuplet mark
disc candidates fainter than the completeness limit of $J=22$ mag in the left panel.
Previously known disc candidates in the 
Arches detected in our earlier higher-resolution Keck/NIRC2 observations
are included for comparison as smaller cyan circles in the right panel (Stolte et al.~2010).
The star serving as coordinate reference for all catalogues is indicated by the arrow.}
\end{figure*}

\begin{figure*}
\centering
\includegraphics[width=16cm]{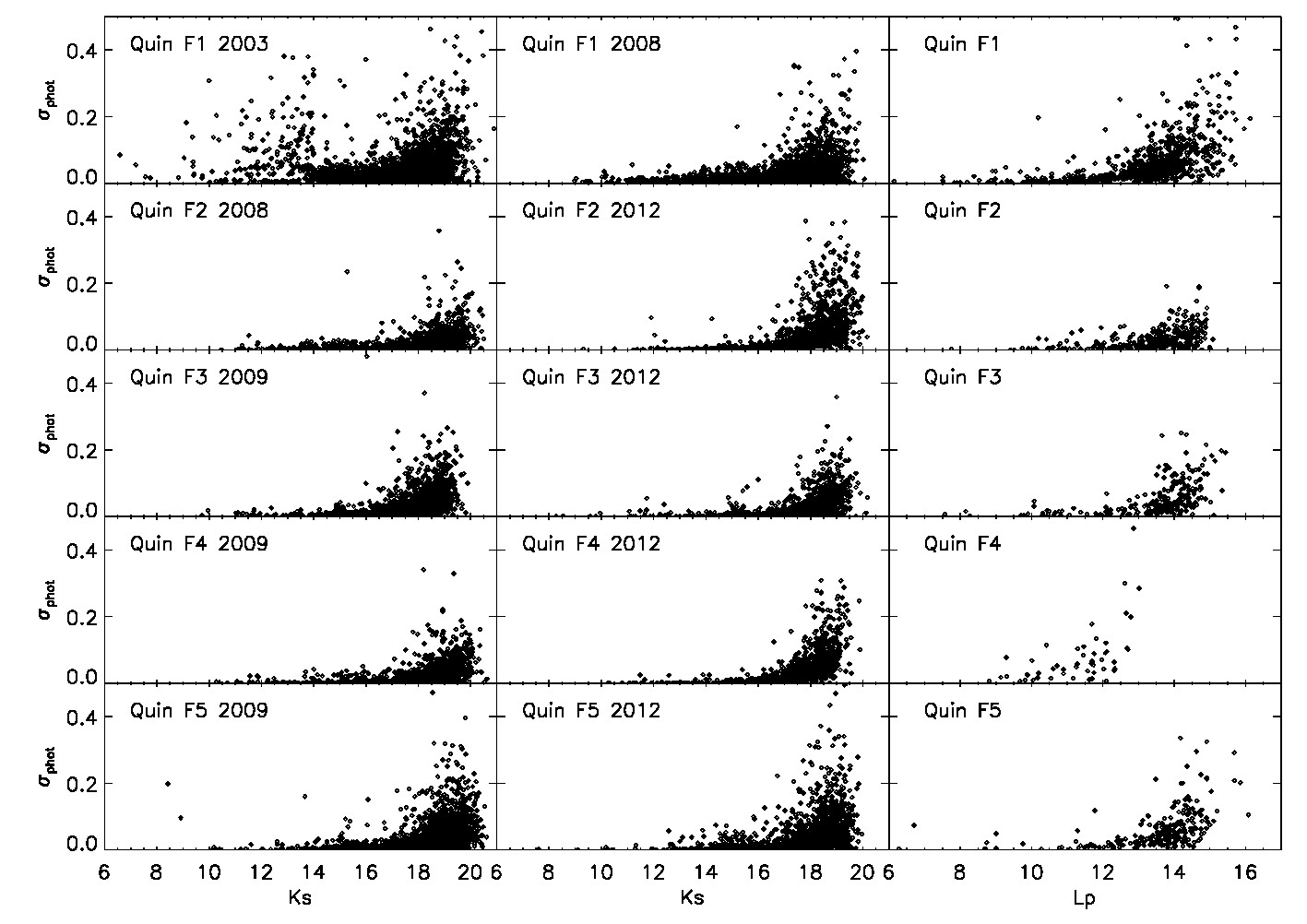}
\caption{\label{kserr} Photometric uncertainties of the NACO $K_s$ and $L'$ observations
of the Quintuplet cluster for Fields 1 through 5 (top to bottom).
The left panels show the first epoch and the middle panels the second epoch
$K_s$ uncertainties for the two epochs from which proper motions were derived. 
The $K_s$ photometry in Field 1
was combined from short and long exposures, with a saturation transition at 
$K_s = 14$ mag. The transition is marked by the prominent improvement in photometric 
performance in the long exposure data. 
$L'$ photometric uncertainties (right panels) indicate the 
large differences in performance in the $L'$ observations due to thermal background
variations. While Fields 2, 3, and 5 show comparable sensitivities, Field 4 is 
compromised by background fluctuations. Field 1 features significantly deeper
photometry, such that the $L'$ selection was truncated at $L' < 15$ mag to 
obtain the excess fraction consistently across the entire cluster area.}
\end{figure*}

\begin{figure*}
\centering
\includegraphics[width=16cm]{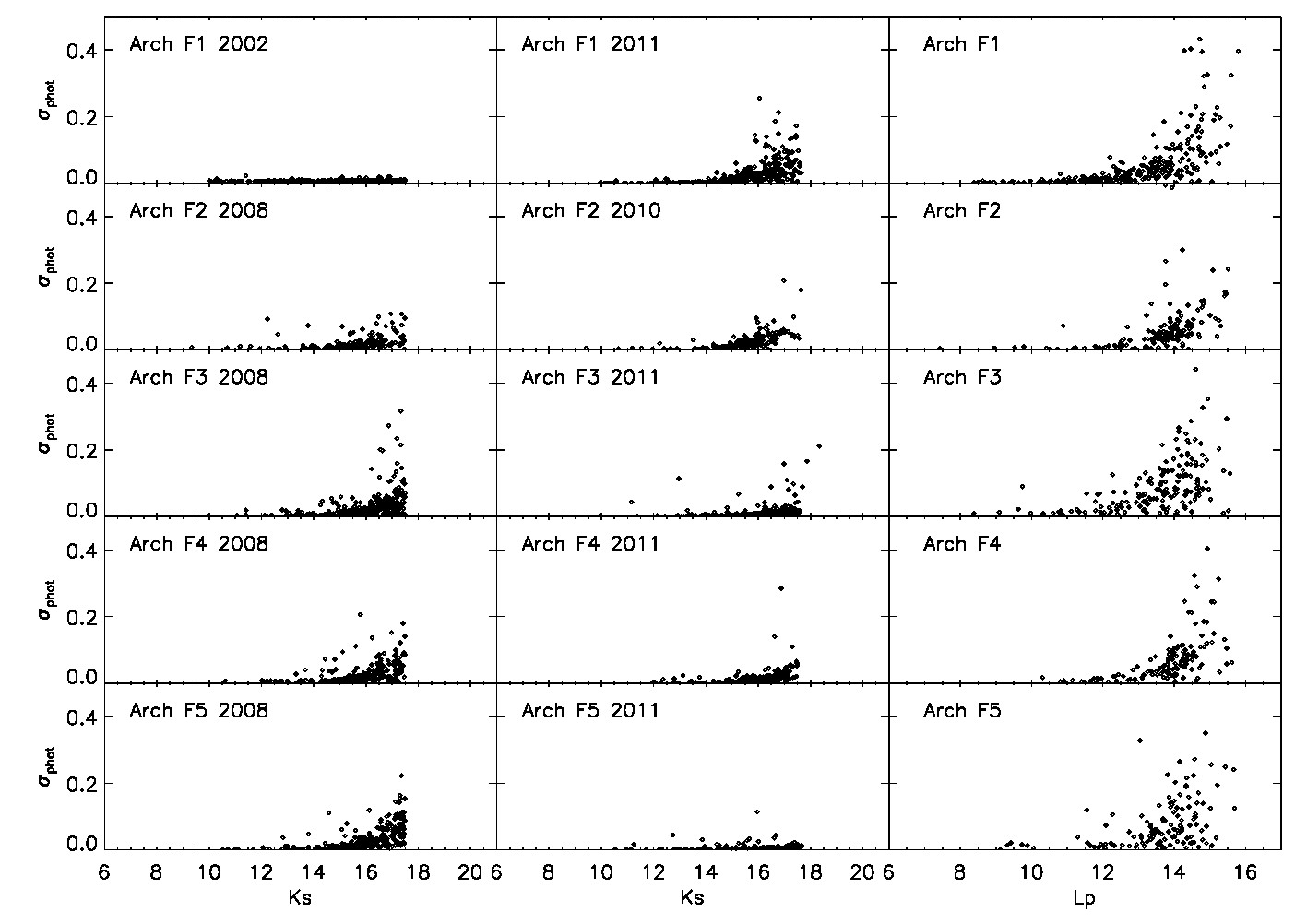}
\caption{\label{kserr_arch} 
Photometric uncertainties of the NACO $K_s$ and $L'$ observations of the
Arches cluster for Fields 1 through 5 (top to bottom).
The left panels show the first epoch and the middle panels the second epoch
$K_s$ uncertainty for the two epochs from which proper motions were derived. 
The photometric uncertainties of the first epoch are daophot PSF fitting 
uncertainties, and are not derived from multiple measurements. 
As in the case of the Quintuplet cluster, $L'$ photometric uncertainties 
(right panels) indicate the differences in performance in the $L'$ observations 
due to thermal background variations. The sensitivity limits show more consistency
than in the Quintuplet, with a detection threshold of $L' \sim 14.5-15.0$ mag.
The larger scatter in Fields 3 and 5 is caused by background fluctuations.}
\end{figure*}

\begin{figure*}
\centering
\includegraphics[width=8.4cm]{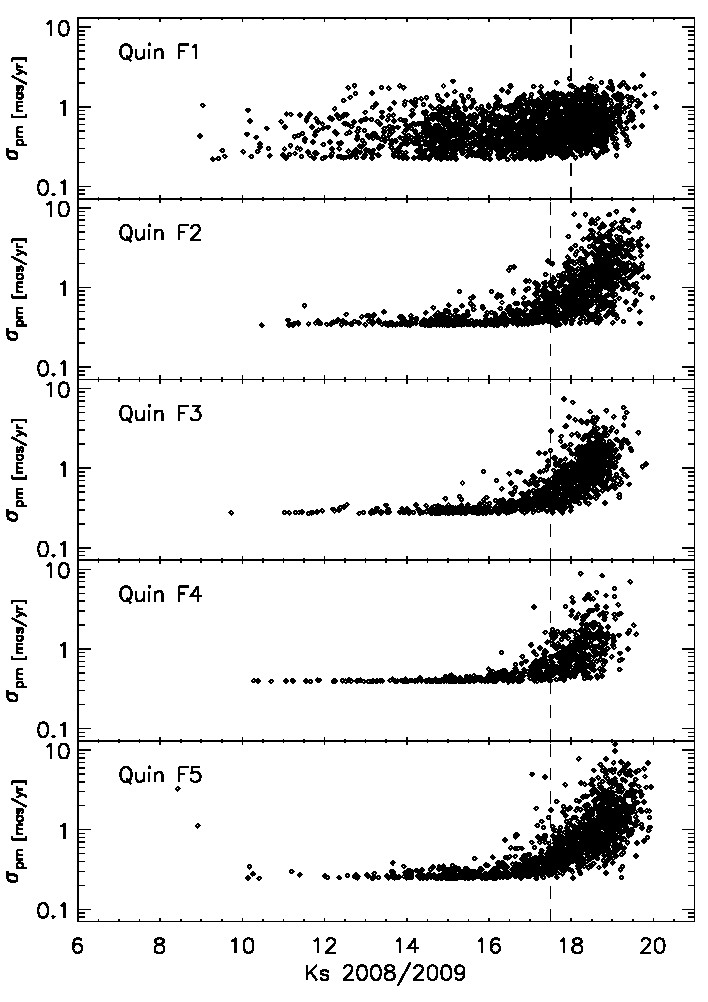}
\includegraphics[width=8.4cm]{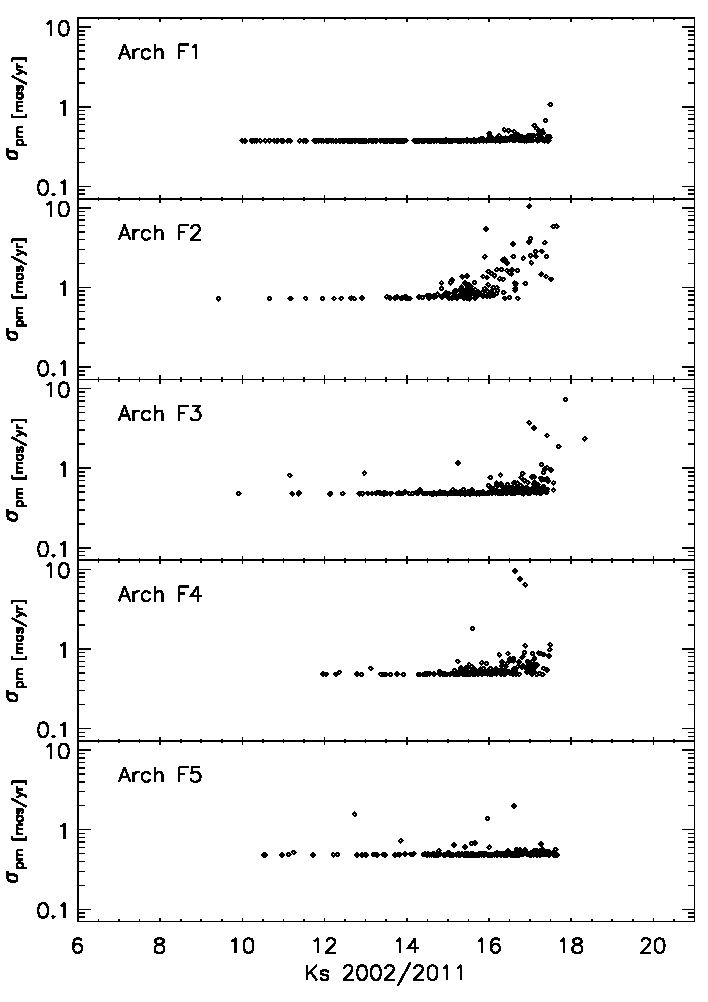}
\caption{\label{pmerr} Proper motion uncertainties of stars in the Quintuplet (left) 
and Arches (right) clusters  for Fields 1 through 5 (top to bottom) in each 
cluster.
The proper motion uncertainties include the astrometric uncertainties from 
each individual epoch in addition to the standard deviation in the geometric 
transformation from epoch 2 to epoch 1.
The deeper $K_s$ observations in the central Quintuplet field provide good astrometric
accuracies down to $K_s \sim 19$ mag, while Fields 2-5 are truncated at $K_s=17.5$ mag (dashed lines),
beyond which astrometric accuracies degrade. The wide spread in Quintuplet Field 1 is caused
by crowding. For the purposes of membership derivation, only sources with $K_s < 18$ mag
(dashed line) are used in Field 1 for compatibility with the outer fields.
In most Arches fields (right panels), the astrometric accuracies are small down 
to $K_s \sim 18$ mag,
with the exception of Field 2, where proper motion derivations are limited to 
$K_s < 17$ mag.}
\end{figure*}

\begin{figure*}
\centering
\includegraphics[width=8.4cm]{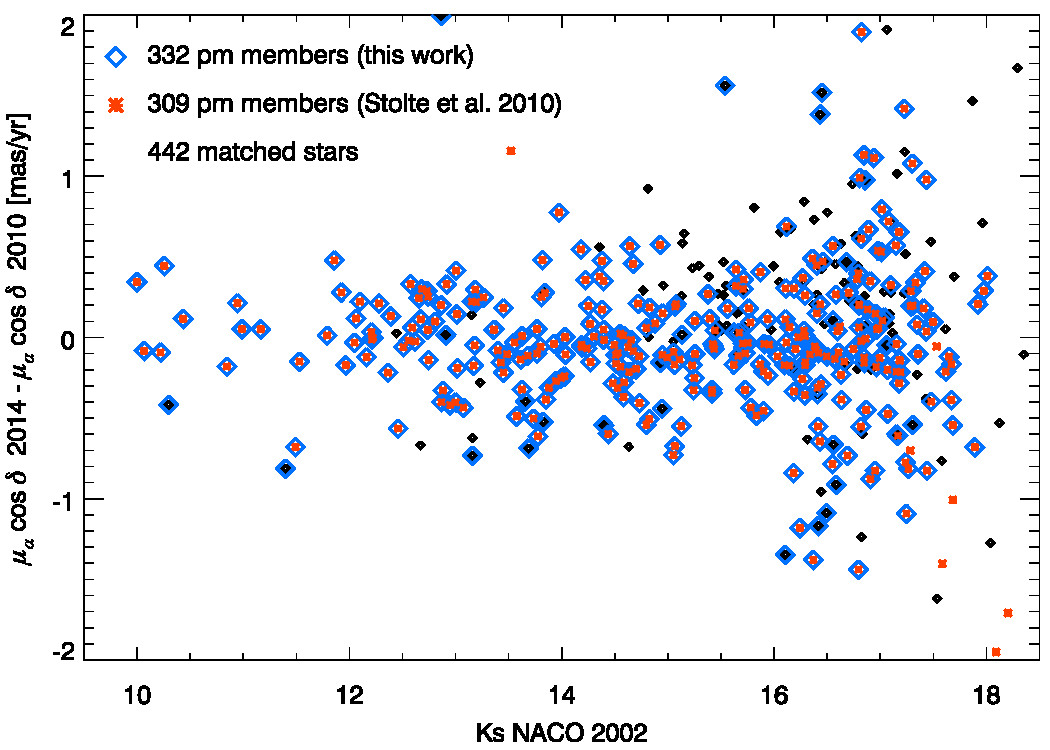}
\includegraphics[width=8.4cm]{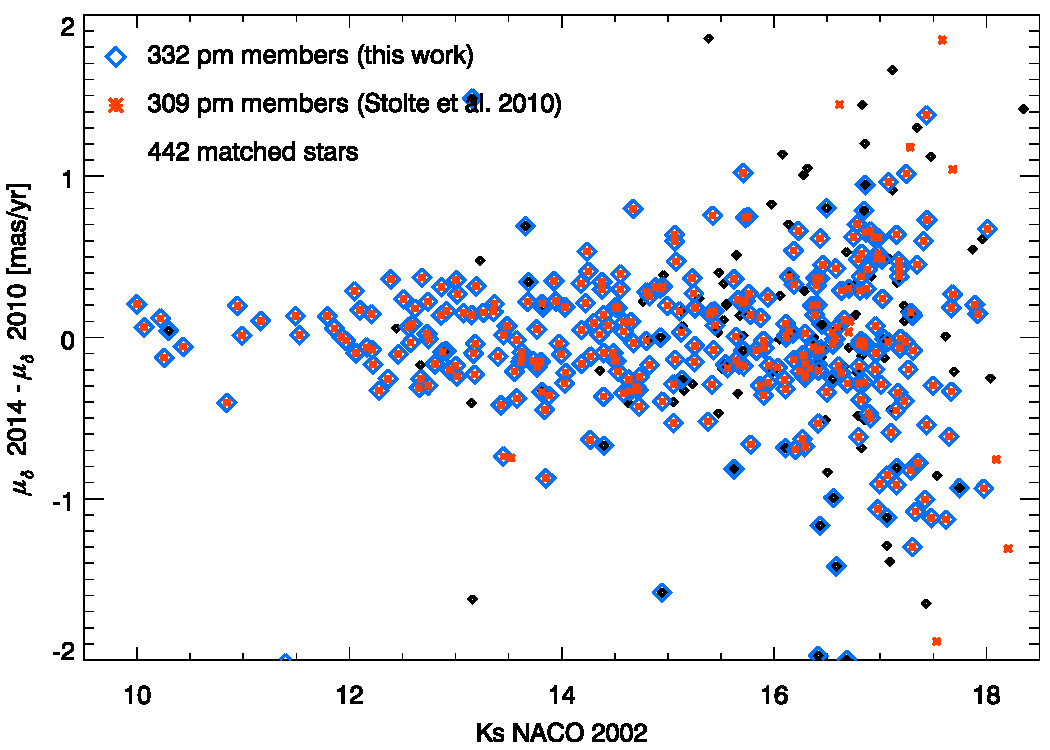}
\caption{\label{compare_ast} Comparison of proper motion values 
derived in our 2010 disc paper of the central Arches field and in 
this study. Differences in the proper motion are shown individually 
in the right ascension (left) and declination (right) directions.
Note the similarity between the membership selection used to derive
the central disc fraction in Stolte et al.~(2010, red asterisks)
and in this work (blue diamonds). Non-members are shown as black dots.
The majority of stars in the common sample
shows proper motion differences below 0.5 mas/yr, almost identical
to the relative astrometric uncertainties in Arches Field 1 shown in 
Fig.~\ref{pmerr}. Faint stars ($K_s < 16$ mag) show larger motion 
differences, as expected from their higher astrometric uncertainties.}
\end{figure*}

\begin{figure*}
\centering
\includegraphics[width=9cm]{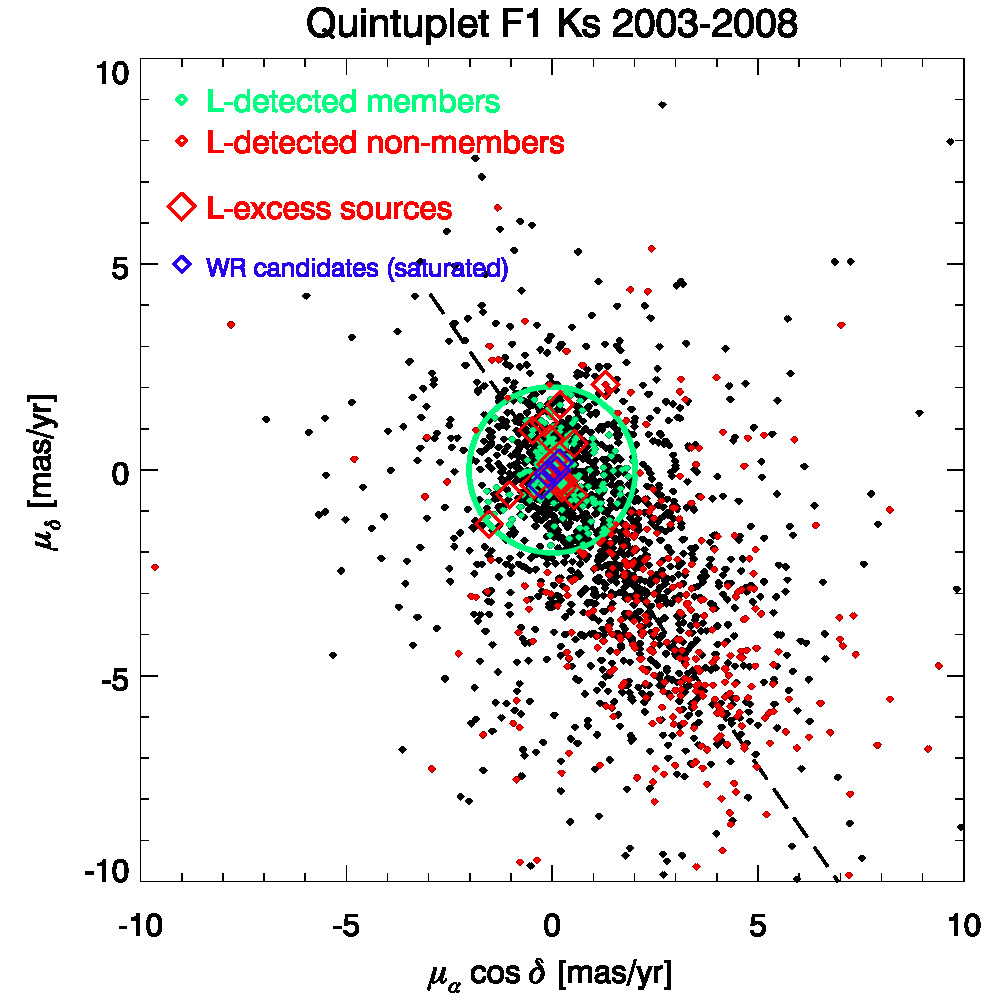}
\includegraphics[width=9cm]{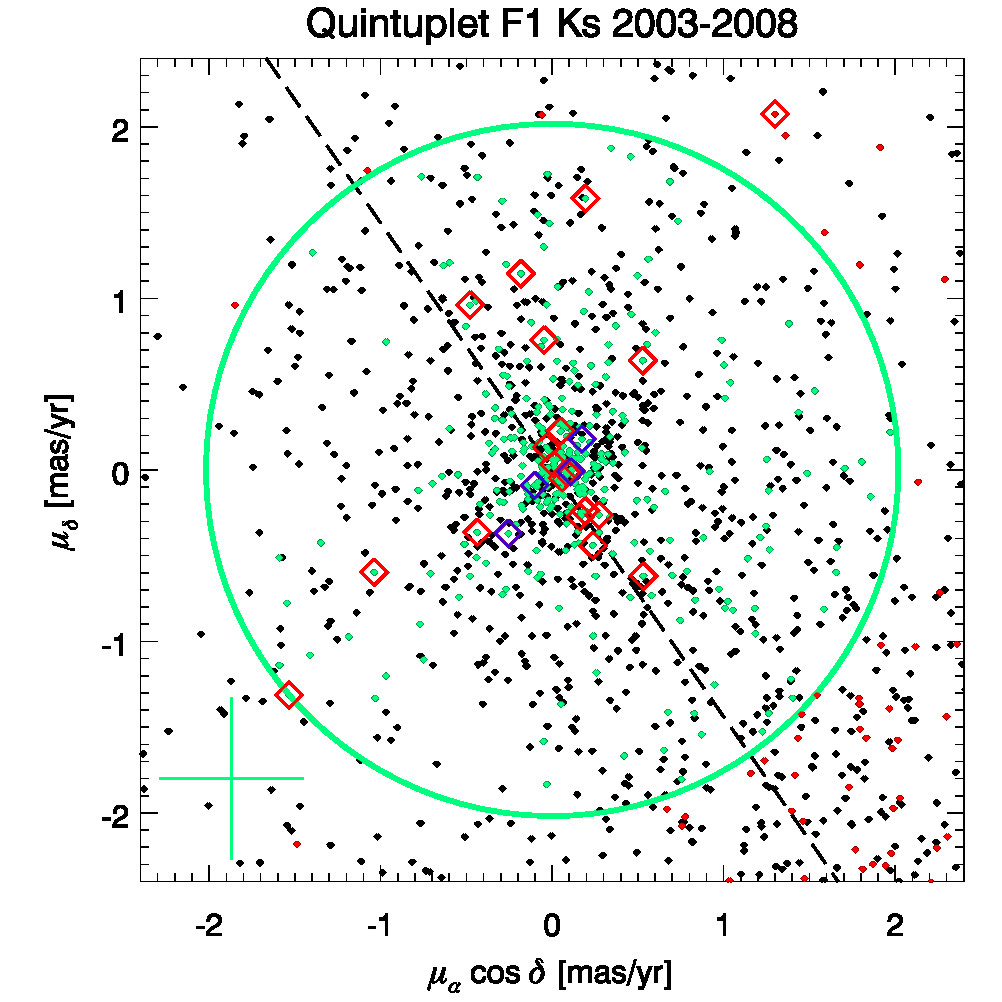}
\caption{\label{quinpm} Proper motion diagram of the central field of the Quintuplet cluster.
The circle around the origin indicates the selection of cluster members. 
$L$-band detected
proper motion members are shown in green, while non-members are shown in red. 
Black sources are not detected in $L'$. Those sources are cluster members
if located inside the green circle. $L'$ excess sources are marked as red diamonds,
while Wolf-Rayet candidates are shown in blue (see Sect.~\ref{quindiscsec}).
The {\sl right panel} displays a zoom on the cluster selection and the median
astrometric uncertainty for $L$-band detected members in the lower left corner.}
\end{figure*}

\begin{figure*}
\centering
\includegraphics[width=9cm]{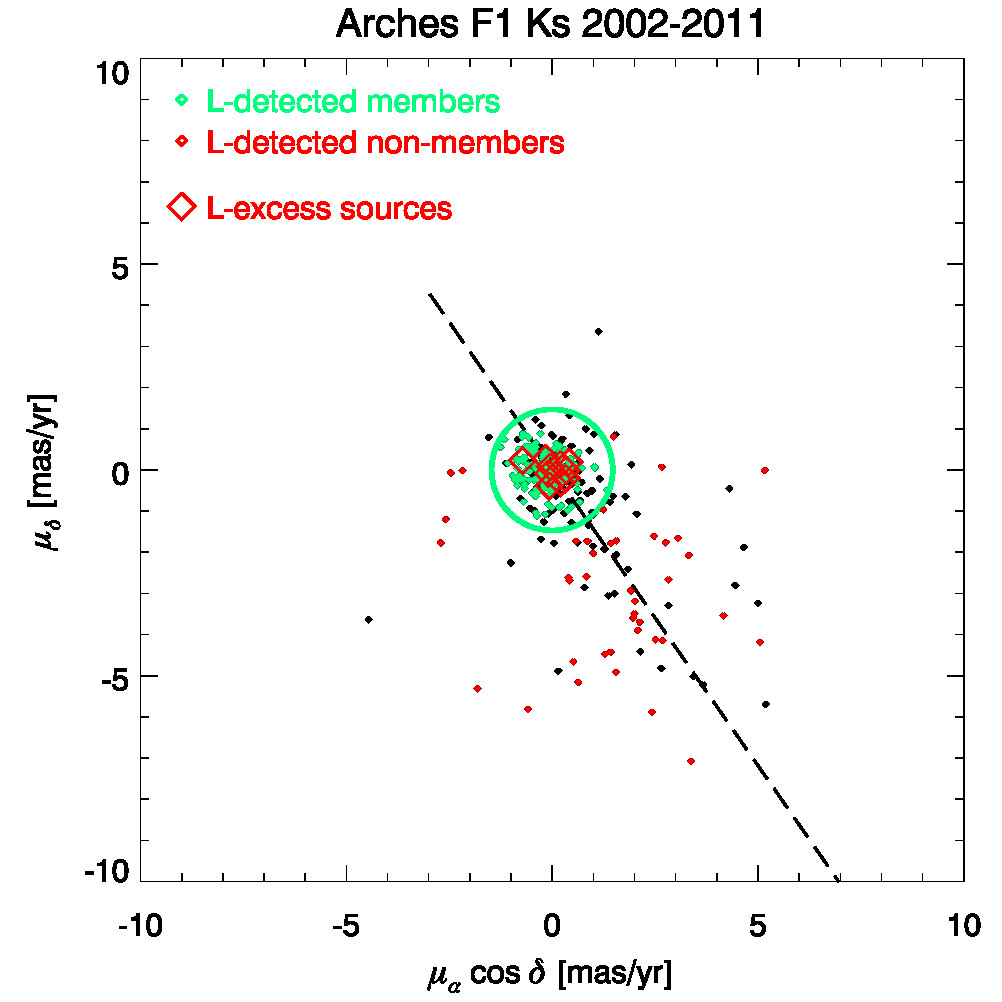}
\includegraphics[width=9cm]{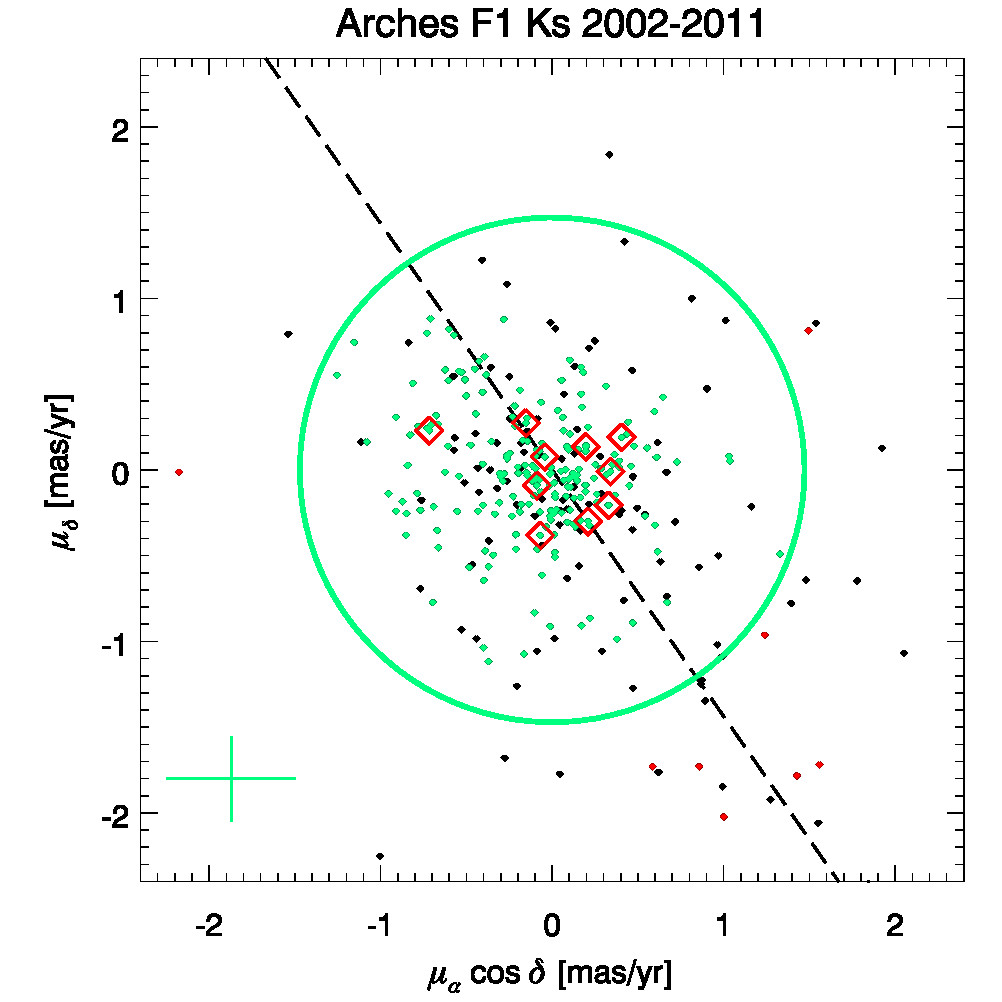}
\caption{\label{archpm} Proper motion diagram of the central field of the Arches cluster.
The circle denotes the selection limit for cluster members.
Members with $L'$ detections are highlighted in green, while 
$L'$-detected non-members are shown in red. As already found in
Stolte et al.~(2010), all excess sources 
recovered in the cluster centre are proper motion members of the Arches.
The {\sl right panel} displays a zoom on the cluster selection and the median
astrometric uncertainty for $L$-band detected members in the lower left corner.}
\end{figure*}

\begin{figure*}
\includegraphics[width=8cm]{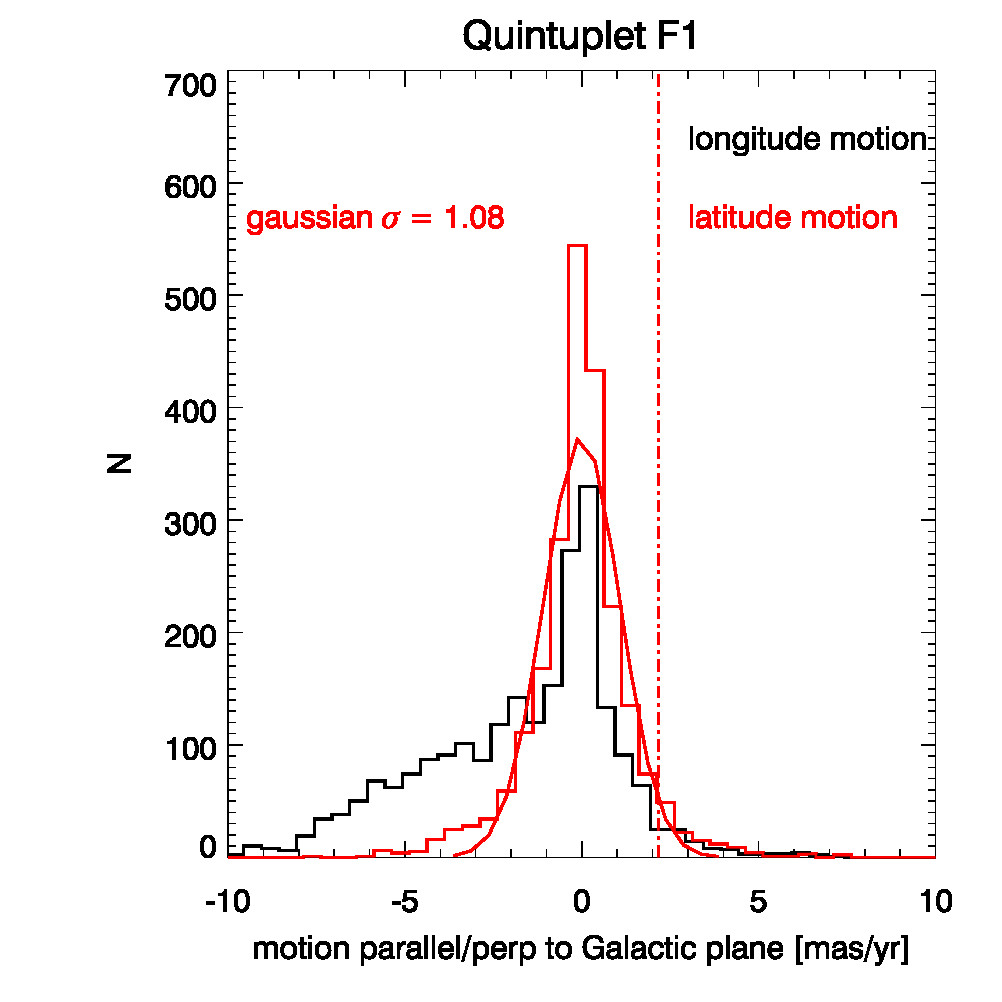}
\includegraphics[width=8cm]{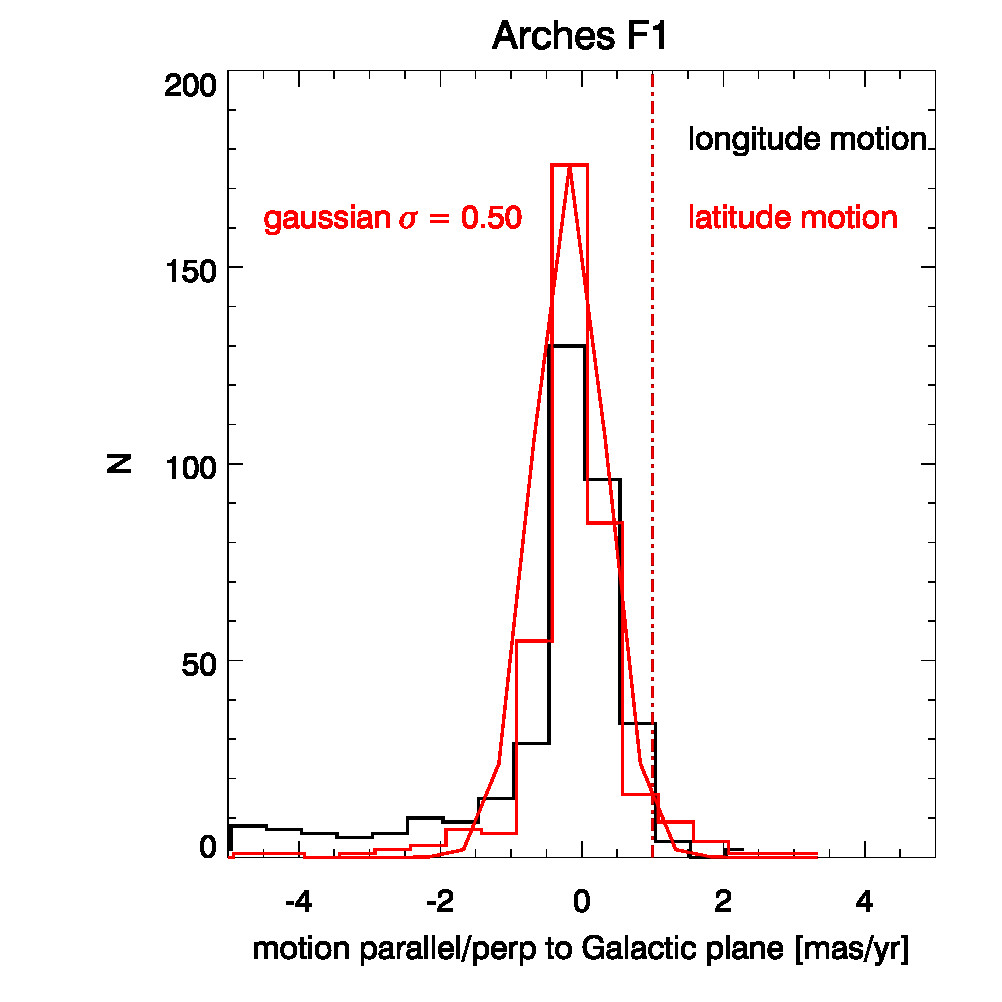}
\caption{\label{pmhists} 
Proper motion histogram of stars in the central fields of the 
Quintuplet (left) and Arches (right) clusters.
The motion parallel to the Galactic plane (black) shows a pronounced
tail of field sources, as indicated in the proper motion diagram
(Figs.~\ref{quinpm} and \ref{archpm}), while the motion perpendicular 
to the plane (red histograms) is dominated by the velocity dispersion. The Gaussian
fit to the latitude motion (red solid lines) in Field 1 provides the 3$\sigma$ criterion 
for selecting cluster members in both central fields. In the case of the 
Arches cluster, the same membership criterion is imposed in all outer fields.}
\end{figure*}

\begin{figure*}
\centering
\includegraphics[width=16cm]{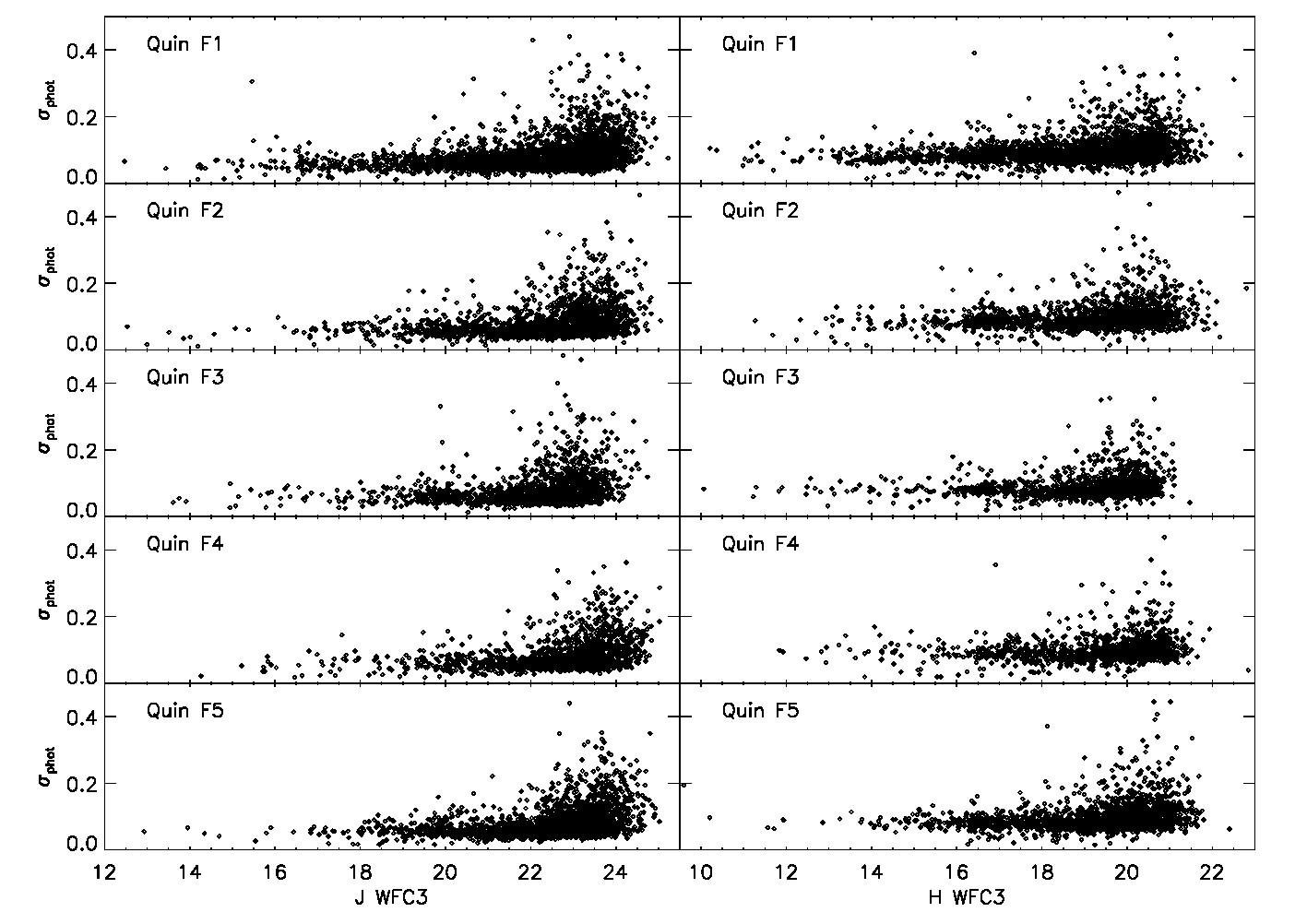}
\caption{\label{jherr} Photometric uncertainties of the WFC3 $J$ and $H$ photometry
for each Quintuplet field after matching with NACO $K_s$.
The apparent differences in the detection limits reflect the different $K_s$ 
sensitivities of the first epoch NACO observations.}
\end{figure*}

\begin{figure*}
\centering
\includegraphics[width=18.4cm]{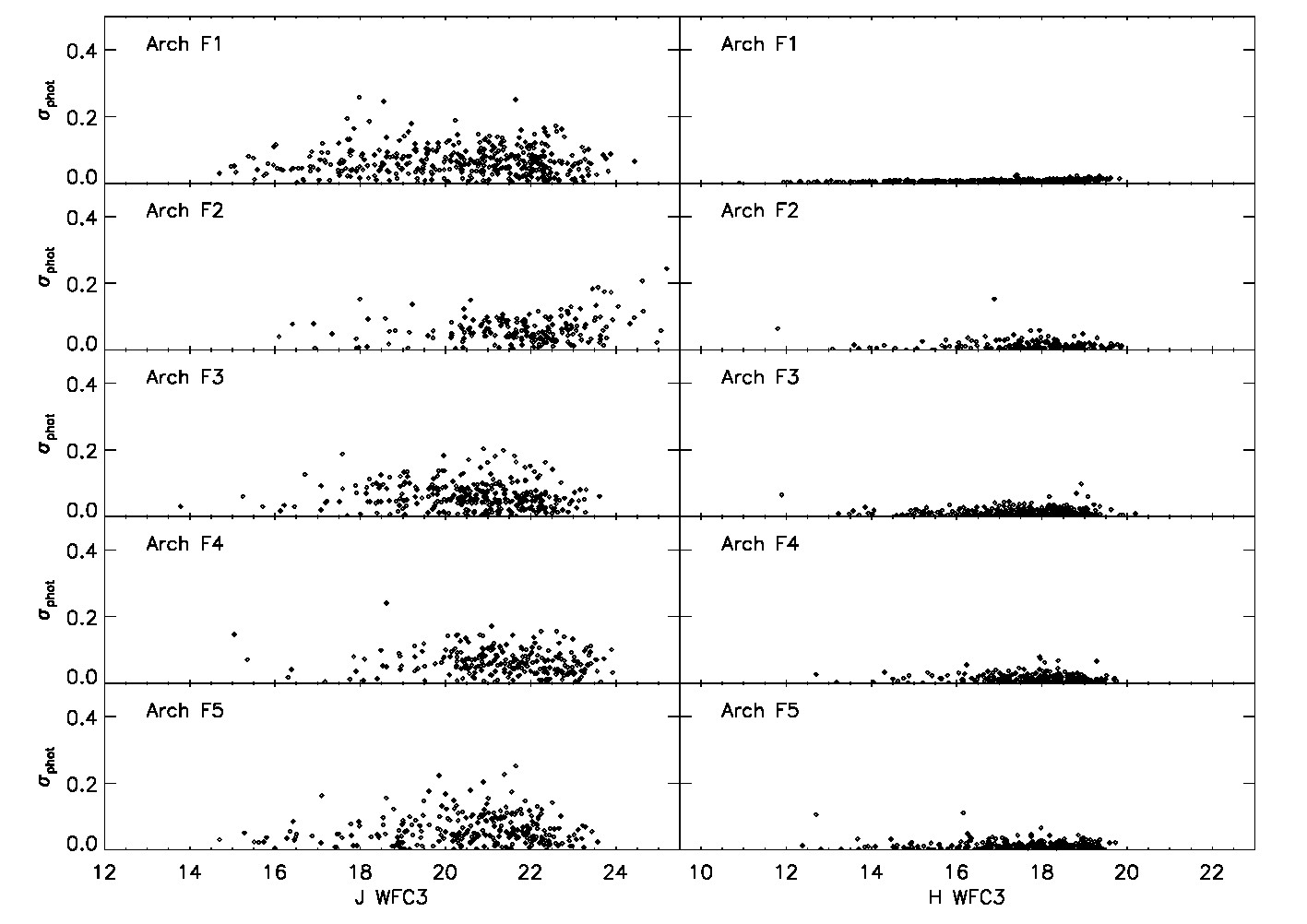}
\caption{\label{jherr_arch} Photometric uncertainties of the WFC3 $J$ and $H$ photometry
for each Arches field after matching with NACO $K_s$.
The apparent differences in the detection limits reflect the different $K_s$ 
sensitivities of the NACO observations in each field which determine the depth 
of the $J, J-K_s$ CMDs.}
\end{figure*}

\begin{figure*}
\centering
\includegraphics[width=18cm]{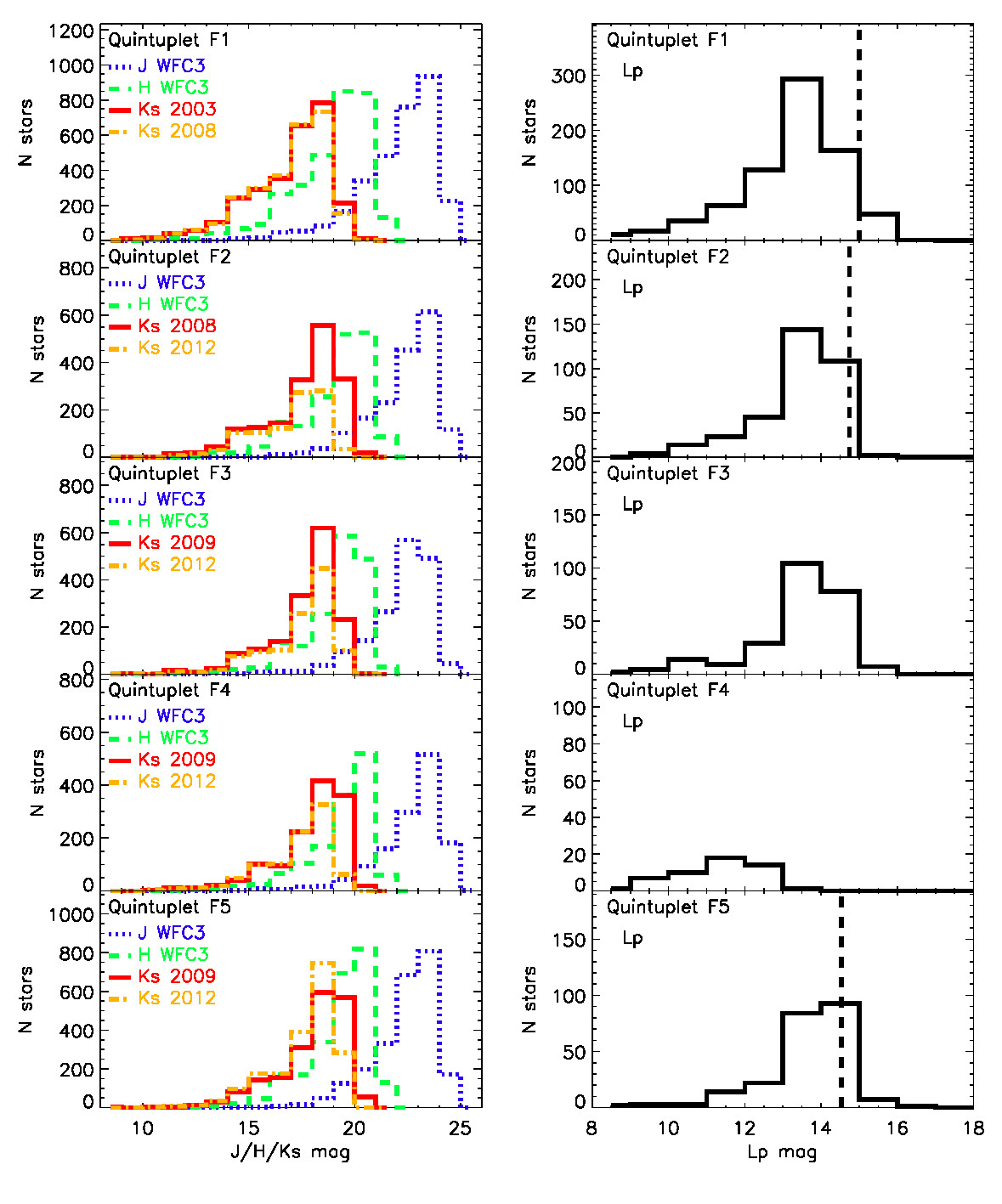}
\caption{\label{lfs} 
Quintuplet luminosity functions of all filters (left: $JHK_s$, right: $L'$). 
All $JHK_s$ luminosity functions 
are shown for the combined sample used for the main-sequence source counts.
These luminosity functions therefore indicate the limitations of the source counts in the final 
$JHK_s$ and $JHK_sL'$ catalogues relevant for the main-sequence and excess samples
here, and not the true detection limits in each filter. 
The $K_s$ luminosity functions are shown for both proper motion epochs used for membership derivation.
Likewise, the $L'$ luminosity functions are derived from the matched $JHKL$ source lists used for further
analysis. The dashed line in the $L'$ luminosity function in Field 1 indicates the truncation imposed 
when combining F1 with the outer fields. The dashed lines in the $L'$ luminosity functions in 
Fields 2 and 5 indicate the faintest $L$-excess source in each field. Except for 
Field 4, all $L'$ data sets extend towards or beyond a sensitivity of $L'=15.0$ mag.}
\end{figure*}

\begin{figure*}
\centering
\includegraphics[width=18cm]{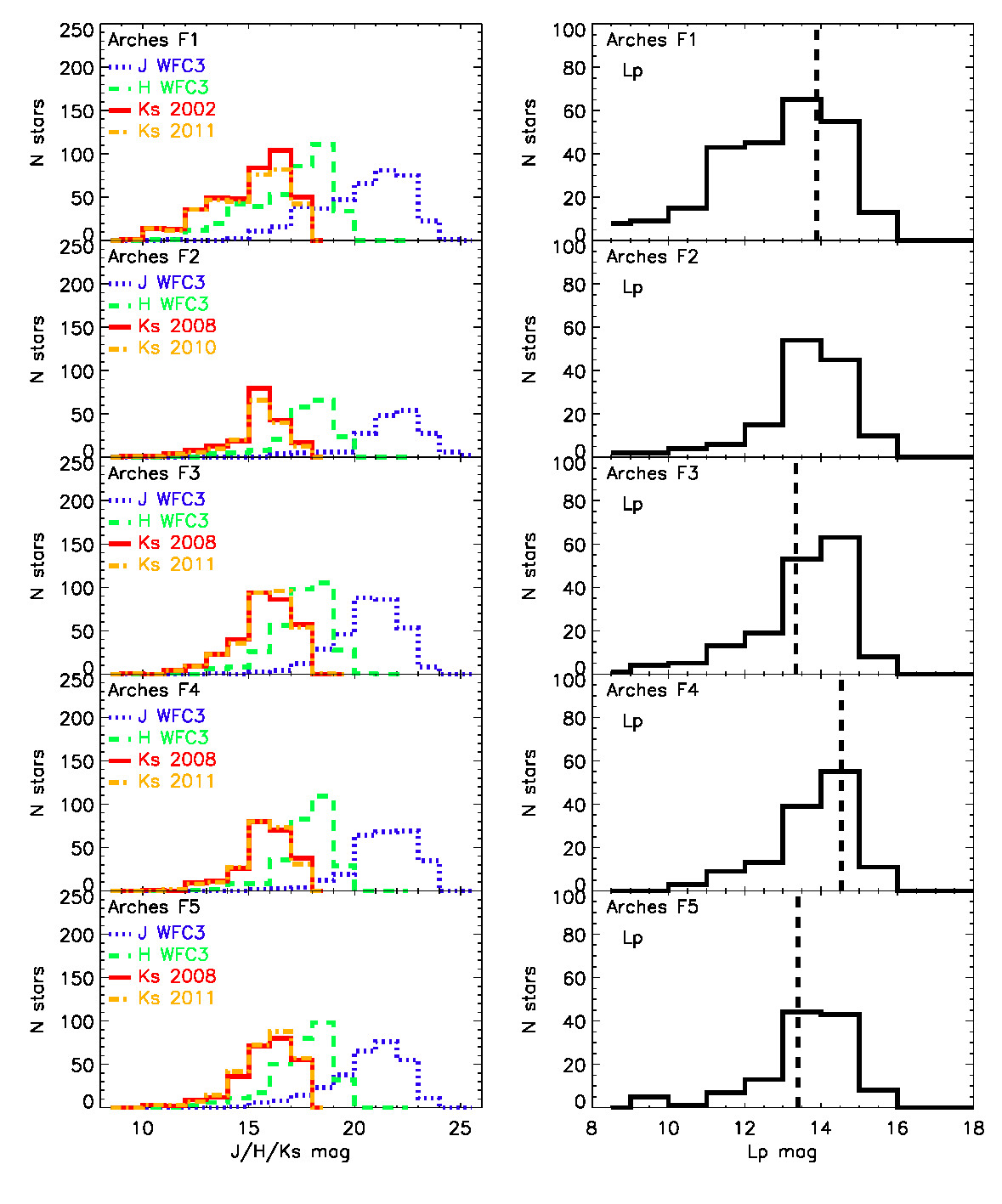}
\caption{\label{lfs_arch} Arches luminosity functions of all filters (left: $JHK_s$, right: $L'$). 
The $K_s$ luminosity functions are shown for both proper motion epochs used for membership derivation.
The detection limits in the WFC3 $JH$ observations are comparable to those obtained 
in the Quintuplet, yet the absolute number of sources matched with $K_s$ is limited
by the shallower Arches NACO photometry caused by the fainter guide stars and required
adaptive optics observing mode. The dashed lines in the $L'$ luminosity functions 
indicate the faintest $L'$-excess source in each field. In Arches Field 2, no cluster 
member with circumstellar disc emission is detected.
All $L'$ data sets show a sensitivity limit close to $L'=15.0$ mag, fainter than 
the faintest detected excess source in each field.}
\end{figure*}

\begin{figure*}
\includegraphics[width=9cm]{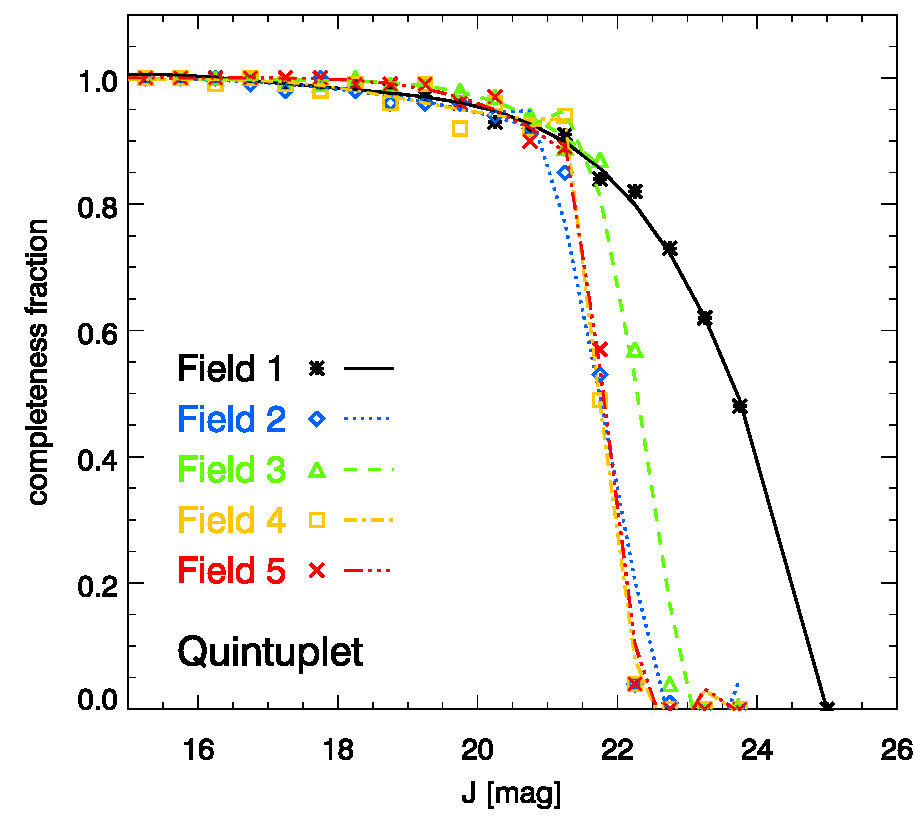} 
\includegraphics[width=9cm]{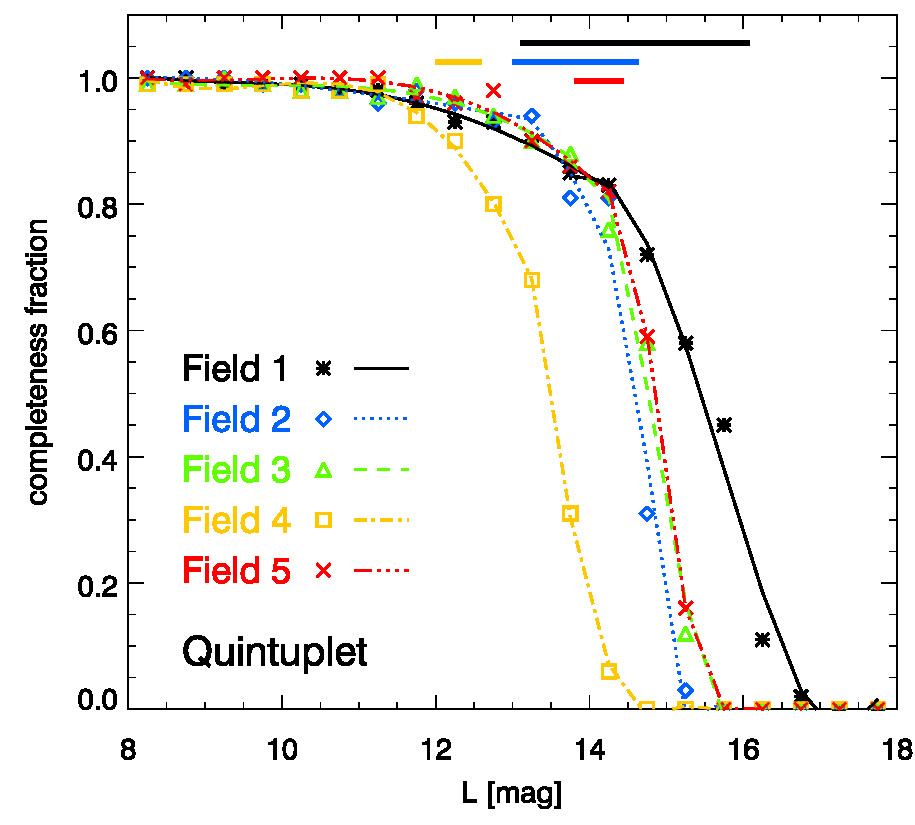} \\
\includegraphics[width=9cm]{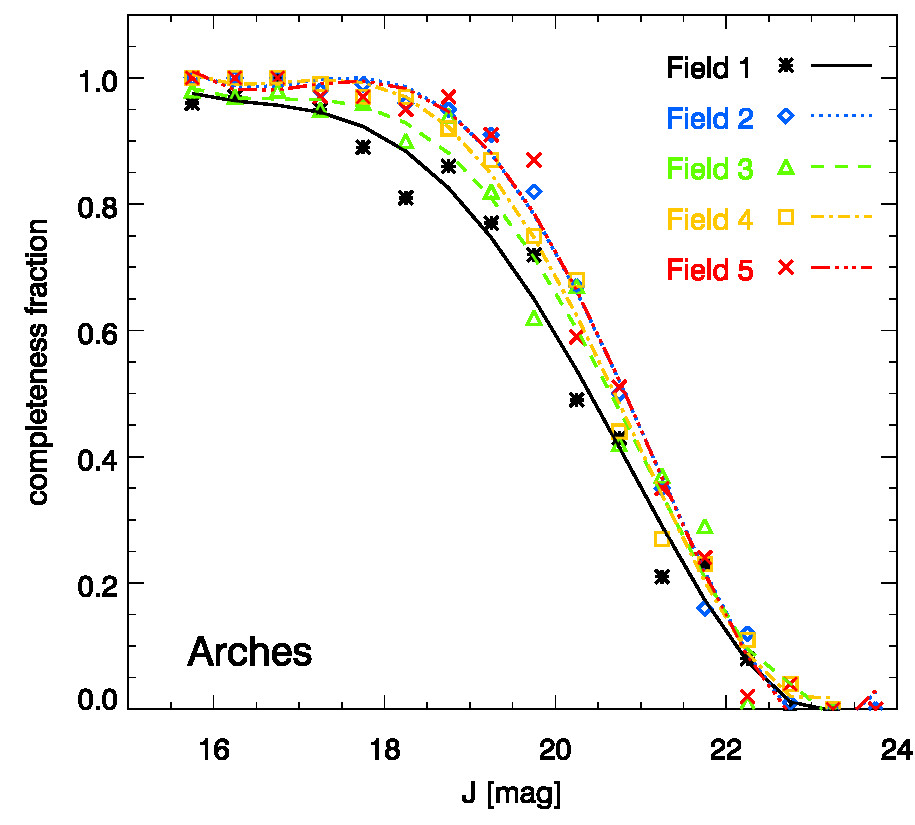}
\includegraphics[width=9cm]{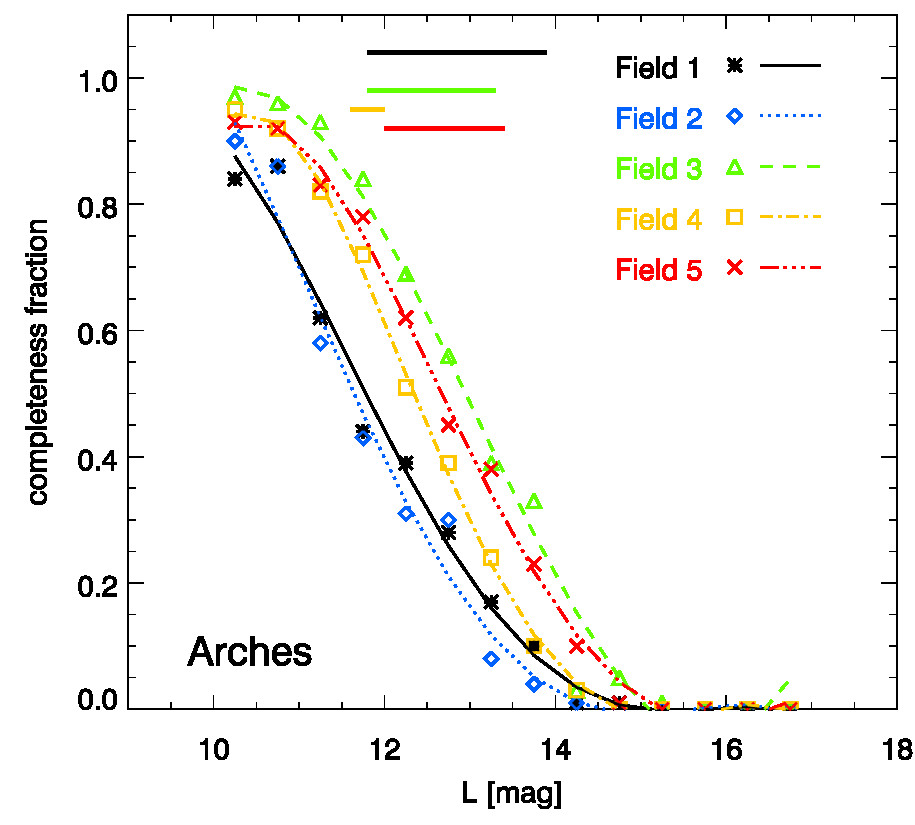}
\caption{\label{incfit}
Incompleteness fractions and fitted polynomials for the 
main-sequence reference samples (left panels) are shown vs.~the 
$J$-band magnitude. Artificial stars were matched in $JHK_s$
prior to the derivation of the displayed completeness fractions.
For the excess source populations (right panels), incompleteness
fractions and fits are shown vs.~the $L'$-band magnitude for all
artificial stars recovered in $JHK_s$ and $L'$ simultaneously. 
The top panels show the completeness fractions in the Quintuplet
cluster, while the bottom panels refer to the Arches observations.
Thick bars in the top of the $L$-band panels represent the range of
$L'$ magnitudes observed in the excess sources in each field.}
\end{figure*}

All disc fractions provided below are derived from $JHK_sL'$ excess sources and 
$JHK_s$ main-sequence stars resolved with NACO and WFC3 simultaneously. 
Compared to our high-resolution Keck/NIRC2 study in the central 0.8 pc of the Arches cluster
(Stolte et al.~2010), a more rigorous incompleteness analysis was implemented here.
For both the NACO $K_s$ as well as the WFC3 $JH$-band observations, the spatial 
resolution limits source detections especially in the crowding-limited cluster core,
and has a larger effect on $J$ and $H$ detections than in the sparser Quintuplet cluster.
Among the 12 excess sources resolved in the central field (see Fig.~\ref{ksmosaic}), 
8 were previously known from our earlier Keck/NIRC2 investigation (Stolte et al.~2010). 
While 4 new sources are detected
in the central field in the analysis presented here, 16 excess sources resolved with 
Keck/NIRC2 remain undetected in the central cluster region with WFC3. 
These sources are located
near brighter neighbouring stars in the dense cluster centre, and are not resolved
in the WFC3 images. As the same losses affect main-sequence stars near bright 
neighbours, and as our incompleteness simulations are calculated for the combination 
of NACO and WFC3, the previously known sources are not included in the following analysis.
Even for stars brighter than the $J$-band limiting magnitude of $J \sim 22$, 
significant losses are observed as a result of the combined effects of crowding 
and undersampling in the WFC3 images. This limit coincides with 
the proper-motion imposed $K_s < 17.5$ mag selection especially for 
fainter and redder stars ($J-K_s \ge 4.5$ mag), and additionally coincides with
the $L'$ detection limits of 14-15 mag for excess sources. As a consequence, the individual
correction factors can be very large with values up to $\sim 80$\% for excess sources 
(see Fig.~\ref{incfit}), where both a detection in $J$ and in $L'$ is required 
simultaneously. The requirement that excess sources be detected in all 
four filters leads to higher correction factors than for the $JHK_s$ main-sequence
reference sample, such that the disc fraction tends to increase when incompleteness 
corrections are taken into account (see Table \ref{extab_arch}). 

The numbers of excess sources and the reference main-sequence samples, including
incompleteness-corrected values, are summarised in Table \ref{extab_arch}. 
Excess fractions are derived for each full field to keep the covered areas
comparable. Three excess sources are located in the overlap regions
between Fields 1, 3, and 5, as indicated in Table \ref{disctab_arch}.
The combined disc fractions are corrected for redundancy. The correction implies
using only the unique excess and main-sequence samples, and reducing the incompleteness
corrected number counts accordingly. The total numbers in the final two rows
in Table \ref{extab_arch} are therefore not the sums of the individual field
counts in Rows 1-5.
The incompleteness-corrected excess fractions range from 2\% in Field 4, 
where only one excess source is found, to 11-13\% in Field 5 and the cluster centre. 
No excess sources are observed in Field 2, where the unusually high extinction 
limits the detection of $J < 22$ mag main-sequence cluster members to just 21, 
and the detection of excess sources is likely impeded by the enhanced foreground 
extinction. Excluding Field 2,
a total incompleteness-corrected excess fraction of 9.4\% is calculated for 
Fields 1, 3, 4, and 5, which decreases slightly to 8.9\% for the full area 
coverage including Field 2. Including the two excess sources with unknown membership 
and their respective incompleteness corrections, the excess fractions increase
to 10.0\% and 9.6\%, respectively.

%

   As shown in Sect.~\ref{astromsec}, individual astrometric uncertainties can induce
   an error in the membership selection of at most 10\%. Assuming this uncertainty
   in the main-sequence samples as a conservative estimate leads to an uncertainty of 
   $\pm 0.5$\% in the overall uncorrected disc fraction. The additional assumption 
   that such an error also imposes a 10\% uncertainty in the incompleteness-corrected
   main-sequence number counts (a 10\% variation in Col. 4 in Table \ref{extab_arch}), 
   provides an estimate of the maximum uncertainty in the 
   incompleteness-corrected disc fractions. This uncertainty in the membership 
   selection supercedes the uncertainties by the main-sequence colour selection when 
   the global disc fraction is derived. Allowing for this 10\% variation in the corrected
   main-sequence number counts and 
   taking into account the uncertainty imposed by the field coverage
   (inclusion or exclusion of Field 2, see Table \ref{extab_arch}), 
   the global disc fraction in the Arches cluster is found to be between 8.1\% and 10.4\%, 
   or $9.2\pm1.2$\%.

In our previous study, we found that the excess fraction 
increases from 3\% in the immediate cluster core ($r < 0.2$ pc), which 
is not well resolved in the NACO and WFC3 data presented here, 
to 10\% at $r > 0.3$ pc (Stolte et al.~2010, see their Fig.~11).
The fraction of 10\% is larger than the total excess fraction of 
$6\pm 2$\% found in the central area of the cluster previously (Stolte et al.~2010),
but is consistent with $f_{ex} = 10$\% found for stars at larger radii, $r > 0.3$ pc.
This consistency is expected as the NACO/WFC3 sample is dominated 
by stars outside the cluster core (see Fig.~\ref{ksmosaic}). 
With the increased area coverage, the total disc fraction in the Arches cluster 
has therefore increased from $6\pm 2$\% ($r < 0.8$ pc) to $9.2\pm 1.2$\% 
including stars with radii out to $r < 1.5$ pc. Here, the uncertainty is 
estimated from the variation in the excess fraction due to different choices
of the main-sequence reference sample and the inclusion of the two excess
sources with unknown membership (Table \ref{extab_arch}).

In addition to the area coverage, there is one striking difference with our
earlier Keck/NIRC2 study. Given the limiting spatial resolution of $\sim 200$ mas
with HST/WFC3 compared to $\sim 60$ mas with Keck/NIRC2, high incompleteness
corrections are applied to each star in the cluster core. This is evidenced in the
large incompleteness-corrected number of excess sources, $n_{ex,corr} = 66$, extrapolated 
from just 10 detected excess proper motion members in the central field. 
This correction bears a high level
of uncertainty, and might over-extrapolate the true number of excess sources located 
in the core. As excess sources are more prone to incompleteness because of their faint
$JH$ magnitudes, the corrected disc fraction of 12-13\% in Field 1 has to 
be considered an upper limit. A lower limit to the central disc fraction is obtained
when only the observed excess sources and main-sequence members, without incompleteness 
corrections, are taken into account. 
The total \emph{observed} disc fraction is then found to be $4.8 \pm 0.5$\% in the 
\emph{central field only}, which is dominated by sources in the inner 0.4 pc.
Although several central excess sources are not resolved here, this excess fraction 
compares well with the disc fraction of $6\pm 2$\% found for the 
inner cluster in the higher-resolution Keck/NIRC2 data (Stolte et al.~2010).

Consistent with the findings for the Quintuplet fields above, there are one or two excess
candidates qualifying as field interlopers in each $27{''} \times 27 {''}$ NACO field.
The lower data quality and less confined proper motion plane derived from the Arches
images, however, does not allow for a final conclusion on the apparent non-members
among the excess sample. The large scatter observed in the proper motion diagrams 
(Figs.~\ref{archpm} and \ref{archpm_app}) 
indicates that cluster stars might have scattered into the field
distribution. A third proper motion epoch would be required to fit each star's motion
with a linear fitting function, which might lead to an even more reliable and 
extended member sample among both the excess sources and the main-sequence reference
sample.

Comparing $J$-band magnitudes to a 2.5 Myr Geneva isochrone results in a mass range of
$2.1 < M < 17.5\,M_\odot$ ($9500 < T_{eff} < 32200$ K), corresponding to A1V to O9V stars
(Currie et al.~2010), similar to the mass range of disc host stars observed in the 
Quintuplet cluster. 

In summary, a fraction of $9.2\pm 1.2$\% of early A- to early B-type stars are found 
to display $L$-band excesses in the Arches cluster when individual, local 
incompleteness corrections are taken into account.

\section{Discussion} 
\label{discussion}

A fraction of 4.0\% of the Quintuplet and 4.8\% of the Arches cluster members are 
observed to display $L$-band excess emission, which is here interpreted as evidence 
of circumstellar discs. Incompleteness-corrected excess and main-sequence samples
result in
disc fractions of $4.0 \pm 0.7$\% in the Quintuplet and $9.2\pm 1.2$\% in 
the Arches cluster. The disc host stars are dominated by B-type main-sequence 
stars with a mean mass of $\sim 6\,M_\odot$ in the Quintuplet and of $\sim 7\,M_\odot$
in the Arches cluster. Disc survival in the UV radiation field of B-type 
main-sequence stars (Herbig Be stars) for time periods exceeding 2.5-4 Myr 
is unexpected. Such a long lifetime would argue against primordial discs, and 
raises the question whether the circumstellar material can have formed recently
from secondary processes. This hypothesis is discussed below in the context of 
transition discs and binary mass transfer.

In this section, we first discuss the survival of discs in the Arches and 
Quintuplet clusters  compared to young star clusters outside the GC region
(Sect.~\ref{survivalsec}).
We proceed to discuss the possible nature of the $L$-band excess sources.
In Sect.~\ref{naturesec}, we provide estimates of the physical properties of the
observed $L$-band excess population, and compare the derived size scales 
and masses to known circumstellar, pre-transitional, and transitional discs. 
The origin of the circumstellar dust emission is discussed in the context
of primordial disc survival vs.~a possible secondary origin of dusty 
discs from binary interactions (Sect.~\ref{secondsec}).

\subsection{Disc survival in young star clusters}
\label{survivalsec}

A detailed discussion of the low disc fractions observed
in the Arches cluster was presented in Stolte et al.~(2010).
The updated Haisch diagram including the data point of the 
Quintuplet cluster is shown in Fig.~\ref{discage}.
From a statistical viewpoint, the Arches cluster has a disc
fraction substantially lower than expected from nearby young 
populations, while the older age of the Quintuplet cluster renders
the disc frequency more consistent with the expected time evolution.
In dense star clusters, gravitational interactions accelerate
disc disruption, especially for sources near the cluster core.
Olczak et al.~(2012) showed that in the dense core of the 
Arches cluster, a rapid removal of 30\% of all discs is expected 
from dynamical interactions alone during its 2.5 Myr lifetime. 
As expected, disc mass loss is shown 
to be most efficient in the cluster core in their simulations, 
yet a strong dependence with host star mass is also found.
Olczak et al.~(2012) suggest that survival is most likely around 
B-type stars. Discs around lower-mass stars are more rapidly depleted 
by numerous encounters, while discs around the highest mass stars 
are most affected by close encounters with lower-mass stars during
gravitational focusing events.

In a cluster as dense as the Arches core, and presumably the Quintuplet
at a younger age, gravitational interactions would affect primordial 
and secondary discs alike. If interactions were the dominant source
of disc depletion, we would expect only relatively massive stars to 
retain their discs in the dense cluster core, including high-mass
binaries with substantial amounts of mass transfer. For stars with 
lower presumed disc masses, the chances of disc survival are higher
at larger cluster radii. Investigating the radial distribution of disc 
host star masses as derived from the $J$-band luminosity above 
(Fig.~\ref{massrad}), we do not observe a prominent radial decrease
in the host star mass. Nevertheless, the 2-3 most massive disc-bearing
stars, with $M > 12\,M_\odot$, are found in both clusters near the core 
at $r < 0.5$ pc.
This finding provides weak evidence that gravitational interactions
play a r\^ole in disc depletion, as mass segregation also causes 
the most massive stars to sink to the cluster centre. Deeper observations
detecting discs around lower mass host stars will shed more light
on the dominant disc destruction mechanism in these massive, young
Galactic centre clusters.

\subsection{The nature of the excess sources}
\label{naturesec}

At a wavelength of 3.8$\mu$m, it is assumed that 
the $L$-band emission originates from hot material at the 
inner rim of a circumstellar disc (Espaillat et al.~2011, see 
Alonso-Albi et al.~2009 especially for Herbig Be stars). 
Here, we provide an estimate of the distance of the material from 
the central star and the disc mass, and compare this to the observed 
properties of debris and transition discs detected around lower-mass stars.

\subsubsection{Distance of the dust from the star}
\label{distsec}

A first approximation of the distance of the hot dust from 
the star can be deduced assuming radiative equilibrium.
The radiation from the central star 
hitting a dust particle with radius $a$ and area $\pi a^2$
is given by
\\

$L_{abs} = \frac{L_{star}}{4\pi D^2} \cdot \pi a_{dust}^2 (1-A)$ ,
\\

where 
$L_{star} = 4\pi D^2 F_{star}(D) = 4\pi \sigma_{SB} R_{star}^2 T_{star}^4$ 
is the stellar luminosity, $R_{star}$ is the radius and
$T_{star}$ the effective temperature of the central star,  
$a_{dust}$ is the radius of a dust particle,  A is the 
albedo of the dust, 
$\sigma_{SB}$ is the Stefan-Boltzmann constant,
and $F(D)$ is the flux 
that reaches the particle at distance $D$ from the star.
A low albedo implies that the entire flux is absorbed by 
the dust, while a high albedo means total reflectivity.
\\

At the same time, the particle emits radiation at the characteristic
dust temperature $T_{dust}$:

\begin{equation}\label{ldust}
\hspace*{5mm}
L_{dustgrain} = 4\pi \sigma_{SB}\ a_{dust}^2 T_{dust}^4
 .\end{equation}
In equilibrium, the radiation absorbed and re-emitted 
by dust particles at the dust temperature has to be the same,
$L_{abs} = L_{dustgrain}$, which leads to
\\

%

$\Rightarrow$ $D^2 = \frac{1-A}{4} \frac{R_{star}^2 T_{star}^4}{T_{dust}^4}$ .
\\

Using Wien's law with the wavelength of the NACO $L'$ filter, 
$\lambda_{L'} = 3.8\mu$m, yields an emission temperature of 
$T_{dust} = 763$ K. 
However, from the observation of an $L$-band excess alone the wavelength 
of the peak emission of the dust is unknown. 

A  minimum distance of the dust to the star can be estimated from 
the maximum possible temperature in the absence of a $K$-band excess, 
as observed in more than half of the excess sources. We note that 
even in the presence of a $K$-band excess, 
the $K$-band excess has to be smaller than the $L$-band excess 
for a strong $K_s-L'$ excess to be observed, as equal excesses 
in both filters would cancel each other. With characteristic colour
uncertainties of 0.2 mag, less than 20\% of the $\sim\! 1$ mag $K_s-L'$
excess should be present at 2.2$\mu$m. The temperature of 
a black body with 20\% of the radiation at 2.2$\mu$m corresponds
to 1220K and the peak will occur between the $K_s$ and $L'$ bands 
at 2.4$\mu$m. 
A lower limit to the temperature is not as strictly defined, as we have 
no knowledge of the longer-wavelength emission. If the discs are evolved, 
consisting of substantial amounts of larger grains, a radiation maximum 
in the mid-infrared can be expected. Under the assumption that the 
observed $L$-band emission at 3.8$\mu$m represents only 
20\% of the total disc emission, the black-body temperature corresponds 
to 704K with a peak wavelength of 4.1$\mu$m, shortly beyond $L'$.

With a characteristic B2V $10\,M_\odot$ star 
for our excess sources, we assume a main-sequence temperature
of $T_{star} = 22000$K, stellar luminosity of $5770\,L_\odot$, and 
from Stefan-Boltzmann's Law a radius of $5.2\,R_\odot$.

Allowing for disc temperatures between 704K and 1220K, 
the distance of dust from the star under the black-body approximation 
is estimated to be
\\
\\
\begin{tabular}{ll}
$\Rightarrow$ & 
$ D = \sqrt{1-A}\cdot\ 3.8$ AU \hspace*{7mm} for $T_{dust} = 1220\,$K \\
 & $ D = \sqrt{1-A}\cdot 11.3$ AU \hspace*{7mm} for $T_{dust} = 704\,$K . \\
\\
\end{tabular}
\\
In the case of complete light absorption, a maximum of 
the distance of the irradiated matter from the central star
is derived: \\

Albedo A = 0 \ \ $\Rightarrow$ $D_{max} = 3.8-11.3$ AU . \\
\\
With the more moderate assumption that half of the stellar
radiation is absorbed by the dust, lower minimum and maximum 
radii are obtained,
\\

Albedo A = 0.5   $\Rightarrow$ $D = 2.7-8.0$ AU ,
\\

%
where the minimum represents the case where 20\% of the 
radiation is emitted at $K_s$ and the maximum represents the case
where 20\% of the emission is emitted at $L'$.
The simplified assumption of black-body radiation provides only
a very approximate range of likely dust radii. Detailed dust 
distribution calculations would be required to obtain a more 
realistic radial distribution of the dust disc; however, the 
single excess value at $L$-band does not provide firm observational
constraints for such models, and such an effort would be beyond
the scope of this paper.

For a B2V $10\,M_\odot$ star we expect a dust distance of 3-11 AU 
in the limiting cases that 20\% to 60\% of the dust radiation are 
emitted at $L'$, respectively.
Distances in the range of several AU 
suggest that dust in the inner disc is destroyed, consistent with 
evolution models of discs around high-mass stars. The fact
that only weak $K$-band emission is detected in most of the excess
sources also suggests that
the inner, and hotter, part of the dusty disc component is depleted.

%
%
%
%

\subsubsection{Limits to disc masses}
\label{discmass}

A lower limit to the disc mass can be obtained when assuming the disc
is optically thin at the observed wavelength. Following the derivation 
in Hartmann 2000 (pp.~113-114), we can estimate the disc mass from

\begin{equation}
\hspace*{5mm}
\nu L_\nu = 8\pi k\nu^3 \kappa_\nu \frac{M_{dust} T(D)}{c^2} \frac{2-p}{2-p-q},
\end{equation}
where $M_{dust}$ is the desired dust mass, 
$\nu$ is the frequency where the luminosity $L_\nu$ is measured, 
$\kappa_\nu$ is the dust opacity, 
$T(D)$ is the temperature in the disc at the distance $D$ from the 
star, $p$ and $q$ are the exponents of the radial density and temperature
profiles of the disc, $k$ is the Boltzmann constant, and $c$ is the speed of light.
Typical values for the exponents of the radial and temperature 
profiles are $p=q=0.75$, hence $\frac{2-p}{2-p-q} = 2.5$.
\\

For the dust opacity, following Eq.~6.13 in Hartmann (2000), we assume
\\

$\kappa_\nu = 0.1(\nu / 10^{12}\ {\rm Hz})\ {\rm cm^2\, g^{-1}} = 
              1\ {\rm m^2\, kg^{-1}} $ \\

at a frequency $\nu = 10^{14}$ Hz.
This value is consistent with the dust opacity observed by 
Indebetouw (2005), $\kappa_\lambda(4.5\mu{\rm m}) = 0.9\ {\rm m^2 kg^{-1}}$.

With $\nu L_\nu = 4\pi d^2 \nu F_\nu = 4\pi d^2 \lambda F_\lambda$, and 
 $\lambda F_\lambda$ inserted from Appendix \ref{disclum}, we find
 $\nu L_\nu = 
3.4\times 10^{26}\ \frac{\rm kg\, m^2}{\rm s^3}$.
Using the temperature limits derived in the previous section,
704K and 1220K,
the lower limit to the dust mass is calculated to be 
\\

$M_{dust} \sim 2-3 \times 10^{20}\, {\rm kg} \sim 10^{-10}\, M_\odot
\sim 10^{-7}\,M_{Jup}$, 
\\

where we have assumed that the $L$-band excess stems entirely 
from the dust emission.
If the discs are still gas-rich, the frequently used 
gas-to-dust ratio of 100:1 would imply the total disc mass to be 
higher by a factor of 100. We note here that the entire calculation is 
based solely on the brightness of a B2V star and the assumption
of an $L$-band excess of 1 magnitude above the stellar photosphere.
\\

This disc mass estimation implies several assumptions originally
derived for optically thin discs at long wavelength radiation (1.3\,mm,
Beckwith et al.~1990). Both the density and the temperature profiles of 
the disc are assumed to be power laws (Hartmann 2000, Beckwith et al.~1990).
However, the structure of our excess objects cannot be deduced from 
$L'$ observations alone.
In addition, the Planck function determining $L_\nu$ is represented by 
the Rayleigh-Jeans approximation in the long-wavelength tail, and $T(D)$
represents the temperature at the \emph{outer boundary} of the disc.
This is not true in our estimation above. Here we have assumed that 
the radiation at $L'$ is the dominating dust radiation component, 
such that Wien's law can be used to obtain the dust temperature.
As the dust is heated most intensely at the inner rim, the radius
corresponds to the inner disc radius rather than the outer boundary.
Beckwith et al.~(1990) argue that the mass estimate depends only weekly
on the outer boundary in the case of optically thin discs. In the
above calculation, we have made the assumption that all dust 
particles are located at the \emph{inner} dust sublimation radius
derived to be on the order of 10 AU in the previous section.
However,
a lower temperature $T(D)$ at the outer boundary decreases the 
disc mass estimate. If the temperature follows the assumed power law
profile $T(D) = T_0 (r/r_0)^{-0.75}$ (see also Hartmann 2000, 
Beckwith et al.~1990), $T(D)$ could be as low as 140 K 
at an outer radius of 100 AU, and the disc mass would be lower by 
a factor of 5.
\\

While this mass estimation reflects the standard derivation 
of disc masses in the literature, which are mostly obtained from
millimetre radiation, some of the uncertain assumptions above can 
be avoided. Here, we start from the total luminosity of the 
dust grains as given in Eq.~\ref{ldust}, 
$L_{dust} = N_{dust} \times L_{dustgrain}$, where $N_{dust}$ is the
total number of dust grains in the disc contributing to the total 
dust luminosity $L_{dust}$. The value of $N_{dust}$ can be expressed in terms
of the standard dust grain size distribution derived for 
interstellar dust, $n(a)\propto a^{-3.5}$ (Mathis et al.~1977) 

\begin{equation}
\hspace*{5mm}
N_{dust} = c_{dust} \int_{a_{min}}^{a_{max}} a^{-3.5} da 
,\end{equation}where $a$ is the grain size with minimum and maximum
grain radii of $a_{min}$ and $a_{max}$, and $c_{dust}$ represents
an unknown normalisation factor.

Inserting $L_{dustgrain}$ from Eq.~\ref{ldust} yields

\begin{equation}\label{ltocdust}
\hspace*{5mm}
L_{dust} = \left. c_{dust} \times 2\pi \sigma_{SB} T_{dust}^4 a_{dust}^{-0.5}\, \right|_{a_{max}}^{a_{min}}
.\end{equation}

The total disc mass can then be calculated from the mass of the 
individual dust grains by integration over the assumed size
distribution. The mass of each individual grain is given by

$m_{dustgrain} = \frac{4}{3}\pi\, a_{dust}^3\, \rho_{grain}$
\\

assuming a mean density of $\rho_{grain}$ to be constant for 
all dust grains in the disc. Integrating over all grains
assuming the same grain size distribution, $n(a)\propto a^{-3.5}$,
yields the total dust mass in the disc:
\\

$ M_{dust} = \frac{4}{3}\pi\, \rho_{grain} \times c_{dust}\, 
            \int_{a_{min}}^{a_{max}} a_{dust}^{-0.5}\, da  
$

\begin{equation}
\hspace*{14mm}
         = \frac{2}{3}\pi\, \rho_{grain} \times c_{dust}\, 
           (a_{min}^{0.5} - a_{max}^{0.5})
.\end{equation}

The normalisation factor $c_{dust}$ is derived from the known
$L$-band excess emission of the disc using Eq.~\ref{ltocdust}. 
For the order of magnitude estimate desired here, we use a 
characteristic $L$-band excess of 1 mag above the stellar photosphere
as before, and the corresponding total 
dust luminosity of $L_{dust} = 3.4\times 10^{26}\ \frac{\rm kg\, m^2}{\rm s^3}$
(Appendix~\ref{disclum}). Using the normalisation factor $c_{dust}$ to 
derive the dust mass, a minimum and maximum grain size of 
0.025 and $0.250\mu$m, and assuming a grain density of 
$\rho_{grain} \sim 2.3\, {\rm g\, cm^{-3}}$ adequate for carbon grains,
yields a total dust mass of 
\\

$M_{dust} \sim 1.5 \times 10^{18}\, {\rm kg} \sim 7.5\times 10^{-13}\, M_\odot
\sim 7.5\times 10^{-10}\,M_{Jup}$, 
\\
\\
for a temperature of 704K. Using the upper limit to the dust 
temperature, 1220K, the dust mass decreases to 
\\

$M_{dust} \sim 1.6 \times 10^{17}\, {\rm kg} \sim 8\times 10^{-14}\, M_\odot
\sim 8\times 10^{-11}\,M_{Jup}$, 
\\
\\
two to three orders of magnitude smaller than the disc mass estimated
above. The lower mass estimate is a direct consequence of the 
assumed steep grain size distribution, $n(a)\propto a^{-3.5}$.
This distribution, which is used as the standard size distribution 
in protostellar discs, implies that most of the dust particles are 
in the small spatial regime, $a_{dust} << \lambda_{obs}$, and are small
compared to the wavelength of the observation. While this might be 
a good approximation for early, primordial dust discs, the particles
in evolved discs are expected to have undergone grain growth.
At the same time, the grain size distribution in mass transfer 
discs is entirely unknown.
Even if the discs are of primordial origin, a flatter grain size 
distribution with more particles at larger size scales
or a maximum grain size larger than the classical value of 
$a_{max} = 0.250\mu$m (Mathis et al.~1977) 
would give rise to a higher dust mass in the disc.

In this estimate, we again had to make the assumption that all dust
contributing to the total dust luminosity in $L$-band and hence
the observed infrared excess emission was distributed near 
the dust sublimation radius at a dust temperature between 704
and 1220 K.

These disc masses are strict lower limits for two reasons.
\\

1. We only derive the mass from the contribution of warm dust
that is heated by the stellar flux to temperatures around 800 K.
This dust is located at a distance of $\sim 10$ AU, and 
is likely  only a small fraction of the total dust mass in the disc.
Any colder dust at larger radii is not accounted for.
\\

2. We have assumed an optically thin disc in the first estimate, 
which might be 
reasonable for transition discs, where dust and gas are already 
clumpy and partially depleted. However, the inner rim of the disc,
where the $L$-band excess is emitted, is hot and likely partially
optically thick. The optically thick portion 
of the disc cannot be penetrated and is not included in the mass
estimate. 
\\

Both aspects of these considerations lead to an underestimate of the disc mass.
We therefore conclude that the observed excess originates from 
circumstellar discs at least as massive as $5\times 10^{-10}\, M_{Jup}$,
and likely as massive as $10^{-7}\, M_{Jup}$.
If the discs contained a large fraction of gas
close to the primordial gas-to-dust ratio of 100:1, the total disc mass
could be as large as $10^{-5}\,M_{Jup}$. 
These low masses support the suggestion that we are indeed observing either 
mass transfer or transition discs with large $L$-band excesses,
as discussed in Sects. \ref{fujisec} and \ref{secondsec}.

\subsubsection{Comparison to pre-transitional discs}
\label{fujisec}

The discs in the Arches and Quintuplet clusters are detected because of
their $K_s-L'$ colour excess. The prerequisite for such a colour excess 
is that $L$-band emission exceeds both photospheric levels as well as 
any excess emission in $K$-band. While primordial discs around very 
young Herbig Be stars ($<1$ Myr) feature prominent near-infrared excesses
to wavelengths as short as 2$\mu$m, a weak or absent $K$-band excess in
combination with a substantial $L$-band excess indicates a later stage of 
disc evolution. For high-mass stars, this implies the formation of an inner 
hole leading to a larger radius for the hot, inner rim observed at 2-4$\mu$m
(e.g. Alonso-Albi et al.~2009). As discussed in Sect.~\ref{survsec}, a 
survival timescale of more than 2 Myr is also not expected for primordial
discs around B-type stars because of their strong UV radiation field.
Transitional or debris discs, on the other hand, are already depleted in 
$L$-band emission, such that the excess sources in the Arches and Quintuplet
clusters display properties between primordial and transitional discs. 
If the observed discs in the Arches and Quintuplet clusters at ages 2.5-4 Myr 
around B-type stars have evolved from primordial discs,
they could be the counterparts to the recently found 
near-infrared bright evolved F star discs (Fujiwara et al.~2013),
yet around higher-mass host stars. In order to conclude whether the 
observed $L$-band emission might originate from pre-transitional discs,
the inner (sublimation) disc radius and the amount of the $L$-band excess
emission are compared to the physical properties of the pre-transitional
discs in the sample of Fujiwara et al.~(2013) displaying $L$-band excesses.
\\

Recently, Fujiwara et al.~(2013) presented a survey of 18 AKARI
sources in the transition phase from protoplanetary to debris objects. 
They defined their sample from nearby ($d < 120$ pc) spectrally classified 
main-sequence A-M stars with AKARI $18\mu$m detections, which were cross-correlated
with the 2MASS $JHK_s$ data base. WISE 3.4 and 4.6$\mu$m photometry
was obtained to cover the mid-infrared region of the SEDs. 
In their sample, four sources
show clear indications of mid-infrared excess down to 3.4$\mu$m ($L$-band)
(namely, HD165014, HD166191, HD167905, HD176137; compare to 
Fig.~3 in Fujiwara et al.~2013). The mid-IR excess emission is measured 
in their sample around low-mass G and F main-sequence stars,
and they find inner disc radii of 0.7-1.4 AU with maximum 
temperatures of 400-500 K for the spectrum of a $1.4\,M_\odot$ star.

Following their procedure, we can use the luminosity of our B2V 
template star as an example to calculate the inner disc radius, 

\begin{equation}
R_{in} = \sqrt{\frac{L_*}{16\pi \sigma_{SB}}} \ T_{in}^{-2}
,\end{equation}where $L_*$ is the stellar luminosity, $T_{in}$ the inner disc
temperature, and $\sigma_{SB}$ the Boltzmann constant (see Eq.~3
in Fujiwara et al.~2013). With an absolute luminosity
of $5770\, L_\odot$ for a B2V star, and an inner disc temperature
between 704 and 1220 K (Sect.~\ref{distsec}), 
the inner disc radius of a $10\,M_\odot$ $L$-band excess source is 
expected to be in the range $4-12$ AU, such that the inner disc
of a B-type star is depleted to substantially larger radii than
the FG discs, as expected. The same authors also provide the ratio
of the dust to stellar luminosity, $L_{dust}/L_*$, from which the 
expected $L$-band excess for these transition discs can be derived.
For a $1.4\,M_\odot$ main-sequence star, a stellar luminosity of
$L_{bol} = 3.47\,L_\odot$ is assumed. As the ages of the transition discs 
in the  Fujiwara et al. (2013) sample are not known, the $L$-band magnitude is 
adopted from the 2.5 Myr Geneva isochrone used above to derive the 
masses of our disc host stars. For a $1.4\,M_\odot$ star, these
isochrones suggest $M_L = 2.54$ mag, such that we find for the stellar flux 
\\

$F_{1.4,L} = 10^{-2.54/2.5} \cdot 2\times 10^{-11}\, {\rm W/m^2} = 
            1.9 \times 10^{-12}\, {\rm W/m^2}$.
\\

Converting to the standard distance of 10pc to obtain the absolute
luminosity, $L_{1.4,L} = 2.3 \times 10^{24}$ W, and using 
$L_{bol} = 3.47\,L_\odot$, we derive $L_{1.4,L} = 6 \times 10^{-3}\,L_\odot = 
1.7 \times 10^{-3}\,L_{1.4,*}$ as the expected photospheric luminosity of 
the disc host stars in a standard $L$-band filter.
From the dust-luminosity-to-star ratio provided in  Table 13 of 
Fujiwara et al.~(2013), the excess emission fraction from the discs for 
the four strong emission sources can be estimated. With
$L_{dust}/L_* = 6 \times 10^{-3}, 9 \times 10^{-3}, 2 \times 10^{-2}, 5 \times 10^{-2}$, 
and using $L_{1.4,L} = 1.7 \times 10^{-3}\,L_{1.4,*}$, the excess factor of the
luminosity above the stellar photosphere in $L$-band caused by the
dust contribution is calculated to be 
$\frac{L_{dust}/L_*}{L_{1.4,L}/L_*} = 3.5, 5.3, 11.8, 29.4$,
where again $L_{dust}/L_*$ is the dust luminosity in terms of 
the total luminosity of a $1.4\,M_\odot$ host star, 
and $L_{1.4,L}/L_*$ is the photospheric $L$-band contribution.
These relative fractions correspond to $L$-band excesses of
$\Delta L_{dust} = 1.4, 1.8, 2.7, 3.7$ mag. 
These values are comparable to, and even slightly larger than, the $L$-band 
excesses measured in the Arches and Quintuplet excess sources. 


In summary, under the assumption that the central stars are B2V stars,
the hot dust emission likely occurs at large radii of $3-12$ AU. 
As expected, the inner dust rim is further from the central star 
as in the case of F-type stars with dust sublimation radii of 0.7-1.4 AU.
The L-band excess emission derived from the dust luminosity of these
lower-mass stars, however, is consistent with the observed L-band 
excesses in the dusty Arches and Quintuplet sources. The question 
remains whether primordial discs could have survived for a sufficiently 
long time to evolve into pre-transitional discs, or whether the discs  
formed later as a result of mass transfer. 
If the host star masses were lower than estimated, 
towards the low end of our $J$-band magnitude range ($3\, M_\odot$),
a survival of primordial discs as observed around F-stars 
yet with dust at larger radii would be more likely.

\subsection{A secondary disc origin}
\label{secondsec}

Espaillat et al.~(2012) recently suggested a physically motivated
distinction between primordial, pre-transitional, and transitional
discs. Following their definition, a disc is only transitional when 
a  near-infrared excess is no longer present, which is interpreted
as transitional discs being discs with inner holes (optically thin 
inner discs). A pre-transitional disc, on the other hand, is composed 
of an optically thick inner disc at a few AU from the central star
leading to near-infrared excesses, 
a lack of mid-infrared emission corresponding to a radial gap, 
and strong far-infrared excess emission. As our sources show weak
$K_s$ excess, yet substantial $L'$ excesses, they would formally 
qualify as pre-transitional discs from their near-infrared properties
alone. Without knowledge of their far-infrared spectral energy distributions,
other possibilities have to be considered as well.

In addition to the detection of the (pre-)transitional discs in Fujiwara et al.~(2013)
discussed above, Mo\'or et al. (2011) detected an unexpectedly massive CO gas 
component in young debris discs with ages up to 30 Myr. For their two A1V and A3IV-V
stars having strong CO emission, line profile modelling suggests an inner radius
for the dust disc of $\sim\! 60$ AU with a dust temperature of 60-80K. This radius
is even more extended than the $L$-band emission region estimated for the Arches
and Quintuplet sources of $\sim\! 10$ AU above. 
Mo\'or et al.~(2011) argue that UV radiation
from both the central star and the interstellar radiation field is capable of 
destroying the molecular gas component on very short timescales, as shielding 
becomes inefficient within 500 years at every location in their modelled transition disc. 
They therefore argue that the gas discs are of secondary origin.
In $K$-band spectra available for three of the Arches excess sources, strong 
CO bandheads at 2.3$\mu$m are detected, indicating that the Arches discs
contain a gaseous component as well (Stolte et al.~2010).
At the higher central stellar masses \emph{and} the stronger ambient radiation 
field of our B-type cluster stars, rapid gas depletion is even more expected 
(Hollenbach et al.~1994, Alexander et al.~2006, Cesaroni et al.~2007),
such that a secondary origin of the CO emission observed in the three 
spectroscopically identified Arches excess sources 
might provide an explanation of the Arches and Quintuplet excess sources alike.

Two  scenarios have been discussed to explain secondary discs around lower-mass stars.
Both are related to the B[e] and Be phenomena. In single Be stars, equatorial mass 
loss due to rapid rotation close to the critical rotation velocity removes angular
momentum from the surface of the star (Granada et al.~2013). Although absolute numbers
of expected rapid rotators with optically thick discs are not known, theoretical models
predict a fraction of a few per cent at cluster ages of a few Myr (Granada et al.~2013, 
see their Fig.~2), which would be 
consistent with the disc fractions observed in the Arches and Quintuplet clusters.
The simulations naturally predict denser, more extended, and more massive discs around
higher-mass stars, which could explain why our sample is dominated by mid- to late
B-type stars. 
For rapidly rotating stars in the mass range $2-9\,M_\odot$, Granada et al.~(2013)
find modelled disc extents of 10-30 AU and 
disc masses of $10^{-8} - 10^{-11}\,M_\odot$ reaching the same order of magnitude
as our lower disc mass limit ($M_{gas+dust} \sim 10^{-8}\,M_\odot$)  
estimated for the Arches and Quintuplet excess emission sources (Sect.~\ref{discmass}).

However, the predicted number ratios are sensitive to the initial conditions, and
these models generate classical Be stars only towards the end of the main-sequence phase, 
which takes more than 10 Myr for all stars considered in their models ($M_{max} = 9\,M_\odot$).
As a consequence of the initial rotation profile, an early adjustment phase 
prohibits the generation of critically rotating stellar surfaces during the early
main-sequence phase (see Granada et al.~2013 for details), as would be required to explain 
the very young B-star disc population in the GC clusters. 

An additional caveat arises
from the fact that Be stars, by definition, show strong Balmer emission lines.
Their discs are thought to be gaseous with little or no dust, in stark contrast to 
primordial, pre-transitional or transitional discs (e.g. Silaj et al.~2010).
Near-IR spectra available for three of the Arches excess sources display strong 
CO bandhead emission (Stolte et al.~2010), such that these sources host both optically thick 
molecular gas as well as hot dust contributing the $L$-band excess emission.
However, all three excess sources show no evidence for hydrogen emission (or absorption)
at the wavelength of the near-infrared Brackett $\gamma$ line.
Given both the substantial presence of dust and the absence of hydrogen-line
emission in these three Arches sources, we conclude that classical Be stars -- 
or stars with similar circumstellar properties -- are unlikely candidates for 
the origin of the discs around these sources.

The second scenario for a secondary disc origin is related to binary evolution.
In this scenario, the mass ejection by the primary in a close binary system leads
to the formation of a circumbinary disc. Such a scenario is suggested for the young, 
evolved B[e] supergiants GG Carinae (Kraus, M. et al.~2013) and HD\,327083 (Wheelwright
et al.~2012). Both objects host high-mass $\sim 25\,M_\odot$ primaries, and disc models
indicate an inner rim of gaseous CO emission at 3 AU at a temperature of $T_{gas}\sim 1700$K.
These gaseous discs are located inside  the dust emitting region, which has an inner 
radius of 5 AU (Wheelwright et al.~2012), in reasonable agreement with the radii
estimated for the $L$-excess origin above. Both sources lack dust close to the central
star (inner holes), and the modelled inner radii are larger than the binary separation,
arguing for circumbinary discs. The study of GG Car reveals that standard Roche lobe
overflow from the primary to the secondary is unlikely, as the primary has not filled
its expected Roche lobe given its evolutionary state (Kraus, M. et al.~2013). 
The same authors argue that the primary was in a classical Be phase towards the 
end of its main-sequence evolution, such that matter from an equatorial decretion 
disc could have overflowed the Roche surface of the binary and streamed into 
circumbinary orbits. In the case of the Arches and Quintuplet main-sequence
excess sources, such a decretion disc had to evolve at an earlier stage
in the stellar lifetime. Such a scenario would argue for tight binary systems,
where the smaller common Roche surface could be filled 
by mass loss from a high-mass primary, possibly enforced by rotation. 

A secondary disc origin in a binary system appears even more likely as 
primordial discs in close binary 
systems disperse on timescales of less than 1-2 Myr, while the survival time for 
discs around low-mass single stars is found to be longer, with a mean of 3-5 Myr
(Kraus, A. et al.~2012). Given the strong UV radiation fields and the added tidal
torques, close, high-mass binaries can be expected to deplete their primordial discs 
even faster than their lower-mass counterparts. A binary origin of a secondary 
disc is intriguing in view of the recent suggestion that most high-mass stars
undergo binary interactions during their lifetime.
For Galactic O-type stars, Sana et al.~(2012) find a fraction of 70\% to be 
affected by binary evolution, with 50\% of all high-mass stars in their sample
undergoing envelope stripping, accretion, or common envelope evolution. Combining
their observations with binary evolution models, they suggest that 26\%
of O-star binaries have a high likelihood of being affected by mass transfer 
events already on the main sequence. While this ratio might be smaller for 
the B-type disc host stars investigated here, a significant fraction of B-type 
main-sequence cluster members is expected to be located in a binary system.
Reviewing recent studies of B-star binary properties, Duch\^ene \& Kraus (2013)
suggest 60\% as a lower limit for the binary fraction of B6 to B2 main-sequence stars.
However, they also find that the fraction decreases towards lower masses, 
with 30-45\% found for late B- to early A-type stars. 
If a few per cent of the $\sim 700$ main-sequence stars found in the
Quintuplet reference sample and of the $\sim 400$ main-sequence stars
observed in the Arches are affected by binary interaction at the present epoch,
the survival of secondary circumbinary discs with molecular gas and hot inner
rims at this level might be expected from binary mass transfer alone.
Such a scenario would naturally explain the similarity of the disc extent 
and properties with B[e] supergiants, although the mass transfer would happen at 
a much earlier stage during the main-sequence evolution of the primary component, 
and the material is provided by the wind mass loss of the high-mass star.

Whether the circumstellar discs in the Arches and Quintuplet clusters are of a 
primordial or a secondary origin cannot be distinguished with the current 
observations. High-resolution 
spectra delivering rotational velocities, mass loss rates, and abundance ratios
of the disc host stars are required to definitively answer the question of the disc
origin. For instance, for equatorial discs of rapidly rotating stars,
the N/C abundance ratios should be enhanced compared to non-rapid rotators of 
the same population, and a high rotational velocity is maintained in initially 
rapid rotators over the entire main-sequence lifetime (Granada et al.~2013).
For pre-transitional discs, the study of the mid-infrared SED would provide
the same distinction mechanism used to define this subclass. 
While MIR surveys such as Spitzer do not provide
the spatial resolution to construct NIR to MIR SEDs for the GC cluster disc sources,
the veiling in NIR spectra as suggested by Espaillat et al.~(2012) can confirm the
existence of thick inner discs as proposed here from the $L$-band excess emission.
The characteristics of secondary binary mass transfer discs are not well constrained,
as there are no high-mass transfer disc models so far. The first step towards
constraining the nature of the discs would therefore be to probe the sample for 
spectroscopic binaries, especially as only tight binaries show mass transfer already
during their main-sequence evolution. Confirming the binary nature of the disc candidates
and hence catching these objects in the act
of mass-transfer would imply that a very special phase of high-mass stellar evolution
can be observationally analysed in detail for the first time.
\\

In summary, we conclude that the discs observed in the Arches and 
Quintuplet clusters have either originated from massive primordial
discs that are in a warm, pre-transitional phase or, more likely, 
that they are composed of secondary discs caused by binary mass transfer.

\section{Summary}
\label{summary}

We investigate deep VLT/NACO $K_sL'$ in combination with HST/WFC3 $JH$ 
imaging of the Arches and Quintuplet clusters to derive disc fractions
from $L$-band excess emission. Multi-epoch NACO $K_s$ observations
are used to derive proper motion membership for $L$-band excess 
sources and the main-sequence reference samples. The results are 
summarised for each cluster below.
\\

Circumstellar disc candidates in the Quintuplet cluster:
\\

 1. A total of 26 $L$-band excess sources as candidates of
    circumstellar discs are discovered in the Quintuplet cluster.
\\

 2. The $L$-band excess fraction in the Quintuplet is observed to 
    be $4.0\pm 0.7$\%, where the uncertainty originates from the selection 
    of the main-sequence reference sample. This fraction is consistent
    in all three fields where excess sources are detected, and -- in contrast
    to earlier results in the Arches cluster (Stolte et al.~2010) -- no trend
    of a varying excess fraction with radius is found. While several excess 
    sources are found out to a radius of 1.2 pc, 
    by far most of the excess sources are found in the central $r < 0.8$pc.
\\

 3. The lack of excess sources in Quintuplet Fields 3 and 4 and at larger radii
    in Fields 2 and 5 indicates a rapid decline
    of the cluster profile at the largest radii, $r > 1.5$ pc, covered by 
    our member survey. The detection of only three non-proper motion members
    featuring an $L$-band excess suggests that the fraction of field 
    excess sources is very low. This conclusion is supported by the 
    sparsity of excess sources in Arches Fields 2 and 4.
\\

Circumstellar disc candidates in the Arches cluster:
\\

 4. We find a mean disc fraction of $9.2 \pm 1.2$\% out to the predicted tidal radius,
    $r < 1.5$pc, in the Arches cluster, which is slightly larger than the mean disc 
    fraction of $6\pm 2$\% detected previously in the cluster centre (Stolte et al.~2010).
\\

 5. In contrast to our earlier study of the Arches core, no trend of the 
    disc fraction with radius is found in either cluster at the larger radii
    investigated here, suggesting that
    the dominant disc destruction mechanism predominantly acts in the 
    densest part of the cluster core.
\\

 6. A total of 21 excess sources are detected over the entire cluster area
    at the NACO $L'$ sensitivity and spatial resolution. Of these, 8 excess
    sources in the cluster centre were previously known. 
    Combining the newly detected 13 disc sources with the previously known
    24 excess sources resolved with Keck/NIRC2 in the cluster centre increases
    the total number of discs in the Arches cluster to 37.
\\

Properties and disc origin:
\\

The stellar mass range of the disc host stars is approximated to be 
$2.2 - 17\,M_\odot$ in both clusters, with a mean mass of $6-7\,M_\odot$, 
corresponding to early A to early B main-sequence stars.
The inner radius of the dust emission region is estimated to be 
in the range 3-12 AU. For a B2V host star, we estimate the minimum 
dust mass in the discs to be on the order of 
$10^{-10}$ to $10^{-7}\,M_{Jup}$. Comparing 
the fractional $L$-band flux to observed evolved discs indicates
that our disc sources could be the higher-mass counterparts to 
the recently identified class of pre-transitional discs. 
As these arguments ignore the effects of UV evaporation expected 
to severely influence the disc lifetime of B-type stars, we discuss
the possibility of a secondary disc origin. We conclude that mass
transfer discs in interacting high-mass binary systems provide a likely 
origin of the $L$-band excess emission. As mass transfer from the 
primary donor to the secondary companion alters the chemical 
composition on the surface of the acceptor, these two different disc 
mechanisms need to be investigated further with high-resolution 
spectroscopy to obtain abundance ratios in order to reach a final conclusion
on the disc origin of B-type main-sequence stars in the Arches and 
Quintuplet clusters. 
\\

\begin{acknowledgements}
We sincerely thank the referee for the careful reading of the manuscript and 
especially for thoughtful comments on improvements to the discussion section.
AS, BH, and MH are grateful for generous support from the DFG Emmy Noether programme
under grant STO 496/3-1. We also wish to thank the Argelander Institute for 
its hospitality, and are indebted to Julien Girard and the NAOS-CONICA staff
for their persistent and unique support with the long-term observational programmes.
CO appreciates funding by the German Research Foundation (DFG) grant OL 350/1-1 and SFB~881. 
AMG and MRM gratefully acknowledge NSF support under grant AST-0909218.
Some of the data presented in this paper were obtained from the Mikulski Archive for 
Space Telescopes (MAST). STScI is operated by the Association of Universities for 
Research in Astronomy, Inc., under NASA contract NAS5-26555.
\end{acknowledgements}

\noindent
{\bf References}
\\
{\small
Adams, F. C., Hollenbach, D., Laughlin, G., Gorti, U. 2004, ApJ, 611, 360 \\
Alexander, R. D., Clarke, C. J., Pringle, J. E.2006, MNRAS, 369, 229 \\
Alonso-Albi, T., Fuente, A., Bachiller, R., et al. 2009, A\&A, 497, 117 \\
Anderson, K. R., Adams, F. C., Calvet, N. 2013, ApJ, 774, 9 \\
Benjamin, R.~A.,Churchwell, E., Babler, B.~L., et al. 2003, PASP, 115, 953 \\
Cesaroni, R., Galli, D., Lodato, G., Walmsley, C. M., Zhang, Q. 2007, Protostars \& Planets V, 
(B. Reipurth, D. Jewitt, and K. Keil (eds.)), University of Arizona Press (Tucson), 197-212 \\
Churchwell, E.,Babler, B.~L., Meade, M.~R., et al. 2009, PASP, 121, 213 \\
Cieza, L. A., Schreiber, M. R., Romero, G. A., Williams, J. P., Rebassa-Mansergas, A. 2012, ApJ, 750, 157 \\
Currie, T., Hern\'andez, J., Irwin, J., Kenyon, S. J., et al.~2010, ApJS, 186, 191 \\
Duch\^ene, G., Kraus, A. 2013, ARA\&A, 51, 269 \\
Espaillat, C., D'Alessio, P., Hern\'andez, J., Nagel, E., et al. 2010, ApJ, 717, 441 \\
Espaillat, C., Furlan, E., D'Alessio, P., et al. 2011, ApJ, 728, 49 \\
Espaillat, C., Ingleby, L, Hern\'andez, J., Furlan, E., et al. 2012, ApJ, 747, 103 \\ 
Figer, D. F., McLean, I. S., Morris, M. 1999, ApJ, 514, 202 \\
Fritz, T., Gillessen, S., Trippe, S., et al. 2010, MNRAS, 401, 1177 \\
Fruchter, A. S., Hook, R. N. 2002, PASP, 114, 144 \\
Fruchter, A., Sosey, M., Hack, W., Dressel, L., et al. 2009, {\sl The MultiDrizzle Handbook}, version 3.0, (Baltimore, STScI) \\
Furlan, E., Watson, Dan M., McClure, M. K., Manoj, P., et al. 2009, ApJ, 703, 1964 \\
Fujiwara, H., Ishihara, D., Onaka, T., Takita, S., et al. 2013, A\&A, 550, A45 \\
Gorti, U., Dullemond, C. P., Hollenbach, D. 2009, ApJ, 705, 1237 \\
Granada, A., Eckstr\"om, S., Georgy, C., Krti\v{c}ka, J., et al. 2013, A\&A, 553, 25 \\
Habibi, M., Stolte, A., Brandner, W., Hu{\ss}mann, B., Motohara, K. 2013, A\&A, 556, 26 \\
Habibi, H. 2014, PhD thesis, University of Bonn, Germany \\
Haisch, K. E., Jr., Lada, E. A., Lada, C. J. 2001, ApJL, 553, 153 \\
Hartmann, L. 2000, {\sl Accretion Processes in Star Formation}, 1st paperback edition,
Cambridge University Press (New York) \\
Hern\'andez, J., Calvet, N., Hartmann, L., Brice\~no, C., et al. 2005, ApJ, 652, 472 \\
Hern\'{a}ndez, J., Hartmann, L, Megeath, T., et al. 2007, ApJ, 662, 1067 \\
Hern\'andez, J., Hartmann, L., Calvet, N., Jeffries, R. D., et al. 2008, ApJ, 686, 1195 \\
Hodgkin, S. T., Irwin, M. J., Hewett, P. C., Warren, S. J. 2009, MNRAS, 394, 675 \\
Hoffmeister, V. H., Chini, R., Scheyda, C. M., et al. 2006, A\&AL, 457, 29 \\
Hollenbach, D., Johnstone, D., Lizano, S., Shu, F. 1994, ApJ, 428, 654 \\
Hu{\ss}mann, B., Stolte, A., Brandner, W., Gennaro, M., Liermann, A. 2012, A\&A, 540, 57 \\
Hu{\ss}mann, B. 2014, PhD Thesis, University of Bonn, http://hss.ulb.uni-bonn.de/2014/3485/3485.htm \\
Ilee, J. D., Wheelwright, H. E., Oudmaijer, R. D., et al. 2013, MNRAS, 429, 2960 \\
Johnstone, D., Hollenbach, D., Bally, J. 1998, ApJ, 499, 758 \\
Koekemoer, A., Fruchter, A., Hook, R., Hack, W. 2002, HST Calibration Workshop, p. 337 \\
Kraus, A. L., Ireland, M. J., Hillenbrand, L. A., Martinache, F. 2012, ApJ, 745, 19 \\
Kraus, M., Oksala, M. E., Nickeler, D. H., Muratore, M. F., et al. 2013, A\&A, 549, 28 \\
Lawrence, A., Warren, S. J., Almaini, O., Edge, A. C., et al. 2007, MNRAS, 379, 1599 \\
Lenzen, R., Hartung, M., Brandner, W., Finger, G., et al. 2003, SPIE, 4841, 944 \\
Liermann, A., Hamann, W.-R., Oskinova, L. M. 2012, A\&A, 540, 14 \\
Liermann, A., Hamann, W.-R., Oskinova, L. M., Todt, H., Butler, K. 2010, A\&A, 524, 82 \\
Liermann, A., Hamann, W.-R., Oskinova, L. M. 2009, A\&A, 494, 1137 \\
Lucas, P. W., Hoare, M. G., Longmore, A., Schröder, A. C., et al. 2008, MNRAS, 391, 136 \\
Maaskant, K. M., Honda, M., Waters, L. B. F. M., Tielens, A. G. G. M., et al. 2013, A\&A, 555, 64 \\
Maercker, M., Burton, M. G. 2005, A\&A, 438, 663 \\
Maercker, M., Burton, M. G., Wright, C. M. 2006, A\&A, 450, 253 \\
Martins, F., Hillier, D. J., Paumard, T., Eisenhauer, F., et al. 2008, A\&A, 478, 219 \\
Mo\'or, A., \'Abrah\'am, P, Juh\'asz, A., Kiss, Cs., et al. 2011, ApJ Letters, 740, 7 \\
Monnier, J. D., Millan-Gabet, R. 2002, ApJ, 579, 694 \\
Muzerolle, J., Allen, L. E., Megeath, S. T., Hernández, J., Gutermuth, R. A. 2010, ApJ, 708, 1107 \\
Nishiyama, S., Tamura, M., Hatano, H. et al. 2009, ApJ, 696, 1407 \\
Olczak, C., Kaczmarek, T., Harfst, S., Pfalzner, S., Portegies Zwart, S. 2012, ApJ, 756, 123 \\
Owen, J. E., Clarke, C. J., Ercolano, B. 2011, MNRAS, 422, 1880 \\
Richling, S., Yorke, H. W. 2000, ApJ, 539, 258 \\
Rousset, G., Lacombe, F., Puget, P., Hubin, N. N., et al. 2003, SPIE, 4839, 140 \\
Sana, H., de Mink, S. E., de Koter, A., Langer, N., et al. 2012, Science, 337, 444 \\
Scally, A., Clarke, C. J. 2001, MNRAS, 325, 449 \\
Schlegel, D. J., Finkbeiner, D. P., Davis, M. 1998, ApJ, 500, 525 \\
Schneider, F. R. N., Izzard, R. G., de Mink, S. E., Langer, N., et al.~ApJ, 780, 117 \\
Silaj, J., Jones, C. E., Tycner, C., Sigut, T. A. A., Smith, A. D. 2010, ApJS, 187, 228 \\
Teixeira, P. S., Lada, C. J., Marengo, M., Lada, E. A. 2012, A\&A, 540, 83 \\
Tokunaga, A., Vacca, W. 2005, PASP, 117, 421 \\ 
Trippe, S., Gillessen, S., Gerhard, O. E., et al. 2008, A\&A, 492, 419 \\
Wheelwright, H. E., de Wit, W. J., Weigelt, G., Oudmaijer, R. D., Ilee, J. D. 2012, A\&A, 543, 77 \\
Williams, J. P., Cieza, L. A. 2011, ARA\&A, 49, 67 \\
}



\clearpage


\clearpage

\begin{figure}
\includegraphics[width=8.4cm]{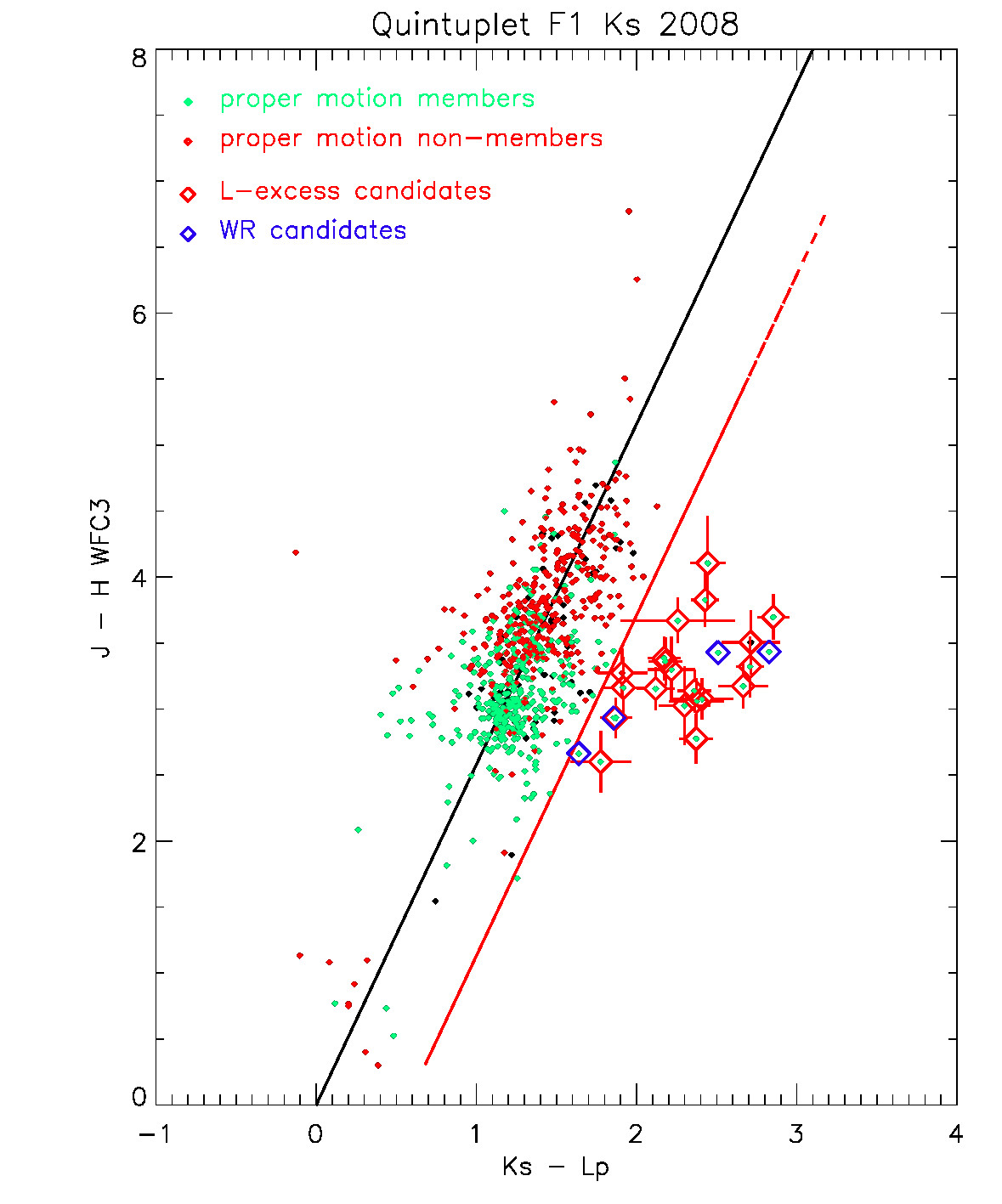}
\hspace*{4mm}
\begin{minipage}{8cm}
\caption{\label{ccds} 
$J-H, K_s-L'$ colour-colour diagram of the central field of the Quintuplet cluster.
The solid black line depicts the extinction vector (Nishiyama et al.~2009),
and the red line is offset by 3 $\times$ the standard deviation of main-sequence
cluster members without excess sources (green points). All $L'$ excess candidates 
are labelled (diamonds), with Wolf-Rayet candidates shown in blue ($K_s < 12$ mag),
while fainter excess sources marked in red are candidates for circumstellar disc emission.
Non-cluster members are shown in red, and black symbols depict sources with unknown membership
status. The $J-H, K_s-L'$ colour-colour diagrams of the outer fields can be found 
in  Appendix (\ref{ccds_app}).}
\end{minipage}
\end{figure}

\begin{figure}
\includegraphics[width=8.4cm]{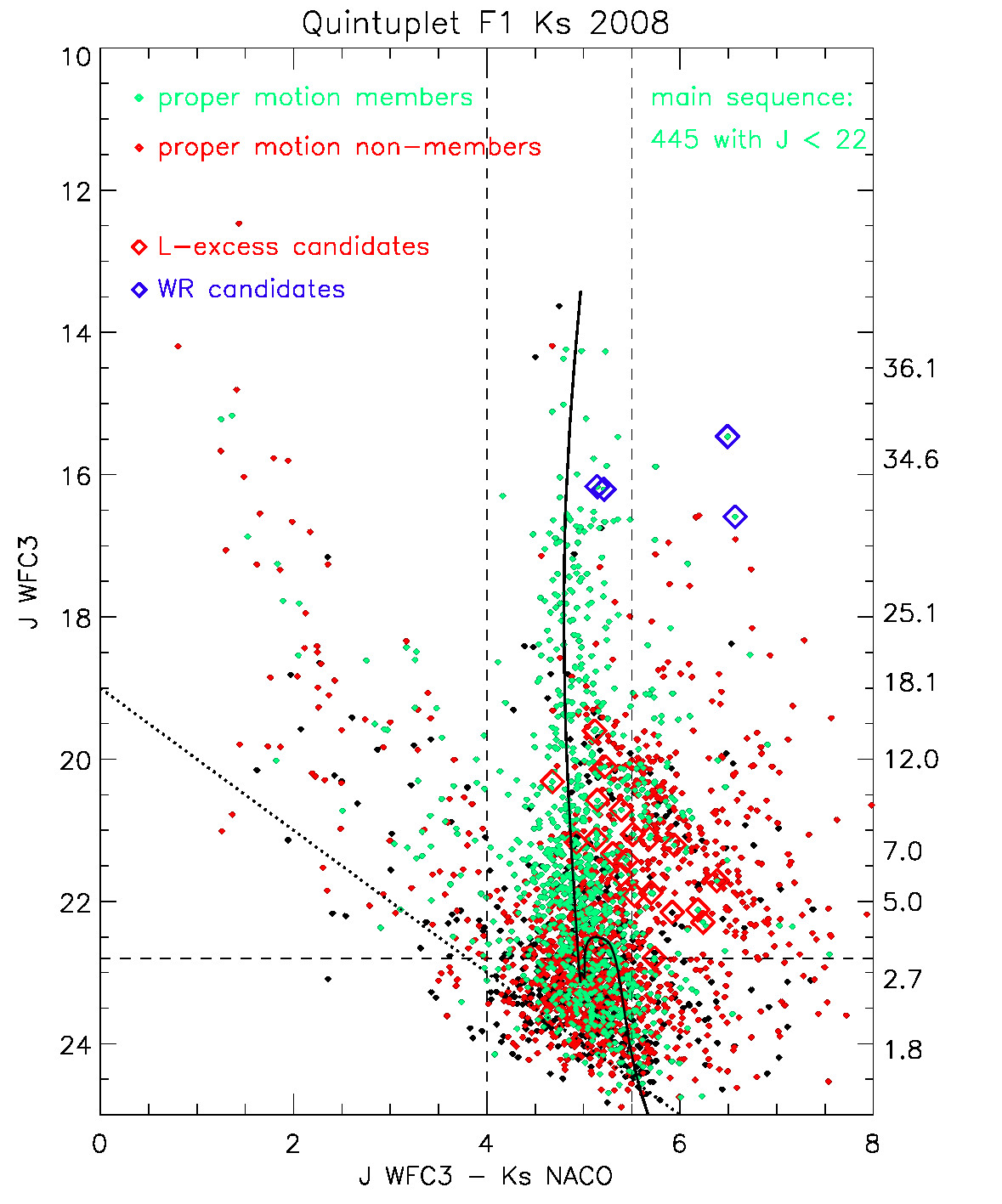}
\hspace*{4mm}
\begin{minipage}{8cm}
\caption{\label{jjk_cmds} $J$, $J-K_s$ colour-magnitude diagram of the central
field of the Quintuplet cluster.
$L'$-excess sources derived from Fig.~\ref{ccds} are marked as red diamonds, while
Wolf-Rayet candidates are marked in blue. Proper motion members are shown in green,
while non-members are shown in red and sources with unknown membership in black.
The imposed magnitude limit to allow for proper-motion member 
selection, $K_s < 17.5$ mag, is shown as a dotted line. The horizontal dashed line
marks the $J$-band completeness limit at $J=22$ mag above which the combined
cluster excess fraction was derived.
The vertical dashed lines indicate the main-sequence colour selection applied in addition to 
the proper motion membership criterion to select the main-sequence reference sample
to calculate the disc fraction. The $J$, $J-K_s$ colour-magnitude diagrams of
the outer fields can be found in  Appendix (\ref{jjk_cmds_app}).}
\end{minipage}
\end{figure}

\begin{figure}
\includegraphics[width=8.4cm]{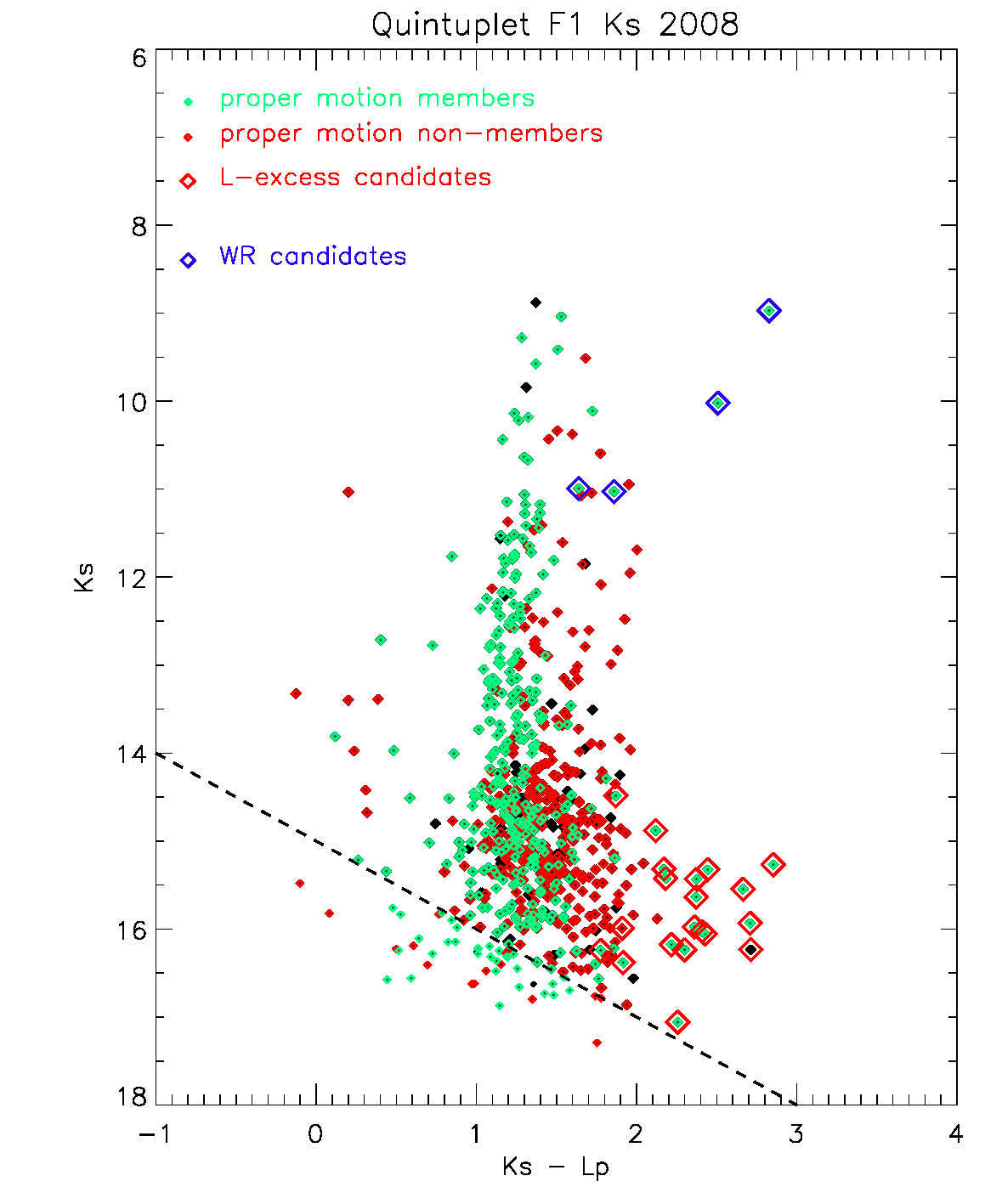}
\hspace*{4mm}
\begin{minipage}{8cm}
\caption{\label{kkl_cmds} 
$K_s$, $K_s-L'$ colour-magnitude diagram of the Quintuplet cluster centre.
All sources are labelled as in Fig.~\ref{jjk_cmds}. One faint $L'$ detection
has no membership information, but is consistent with the remainder of the 
excess sources regarding its $K_s-L'$ colour.
The $K_s$, $K_s-L'$ colour-magnitude diagrams of
the outer fields can be found in  Appendix (\ref{kkl_cmds_app})}.
\end{minipage}
\end{figure}

\begin{figure}
\includegraphics[width=8.4cm]{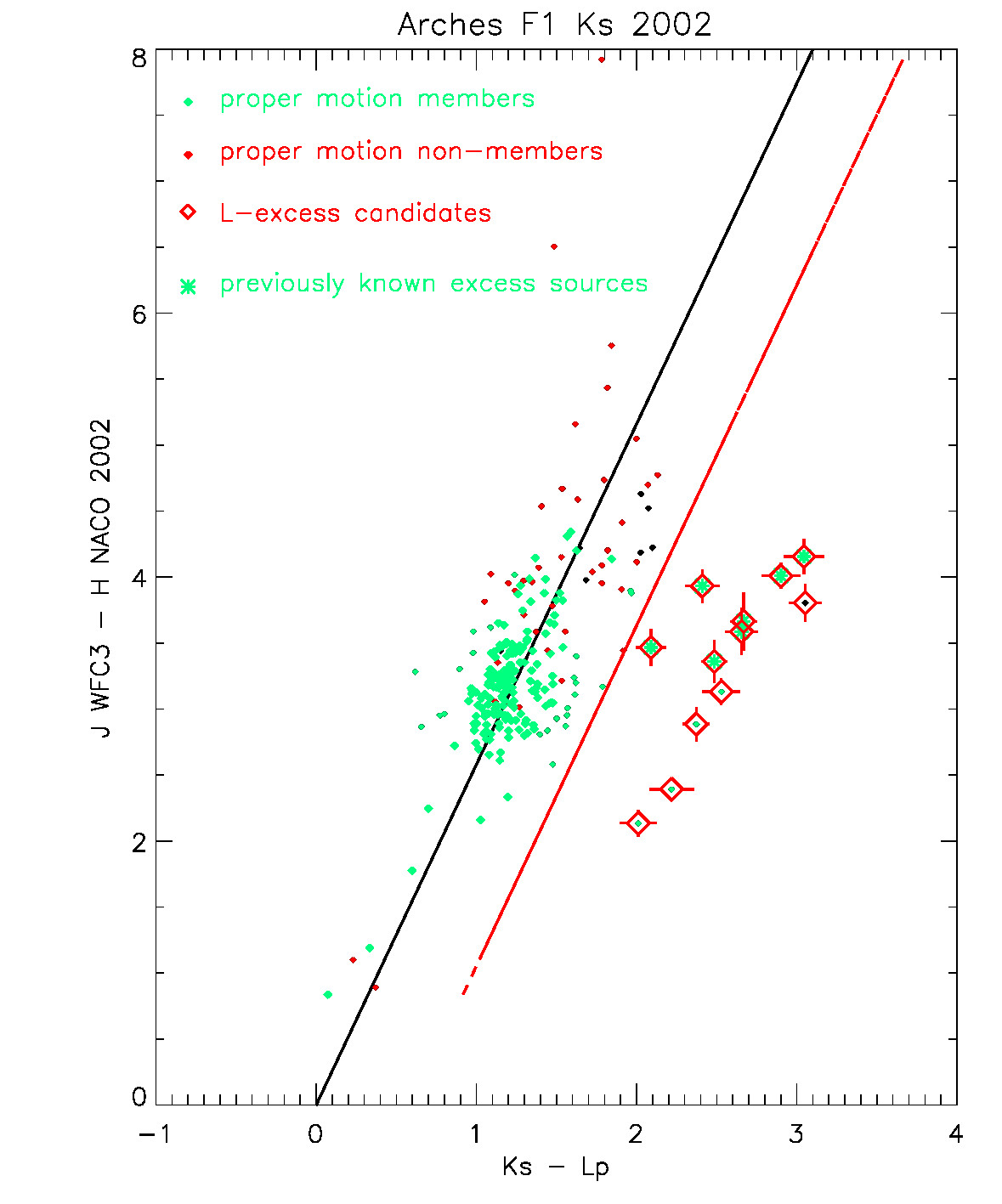}
\hspace*{4mm}
\begin{minipage}{8cm}
\caption{\label{ccds_arch} 
$J-H, K_s-L'$ colour-colour diagram of the central field of the Arches cluster.
The solid black line depicts the extinction vector (Nishiyama et al.~2009),
and the red line is offset by 3 $\times$ the standard deviation of main-sequence
cluster members (green points) without excess sources. 
Excess sources fainter than $K_s = 12$ mag marked as red diamonds
are candidates for circumstellar disc emission.
Non-cluster members are shown in red; black symbols depict sources with unknown membership
status. As many of the excess sources in the cluster centre are located close to brighter
stars because of the high stellar density in the cluster core, only 7 of the 23 $L$-band excess 
sources found in our Keck investigation (Stolte et al.~2010) are detected in the 
lower-resolution NACO images (green asterisks). At larger radii, the fraction of excess sources 
is expected to be more complete owing to the less severe crowding effects.
The $J-H, K_s-L'$ colour-colour diagrams of the outer fields can be found 
in  Appendix (\ref{ccds_arch_app}).}
\end{minipage}
\end{figure}

\begin{figure}
\includegraphics[width=8.4cm]{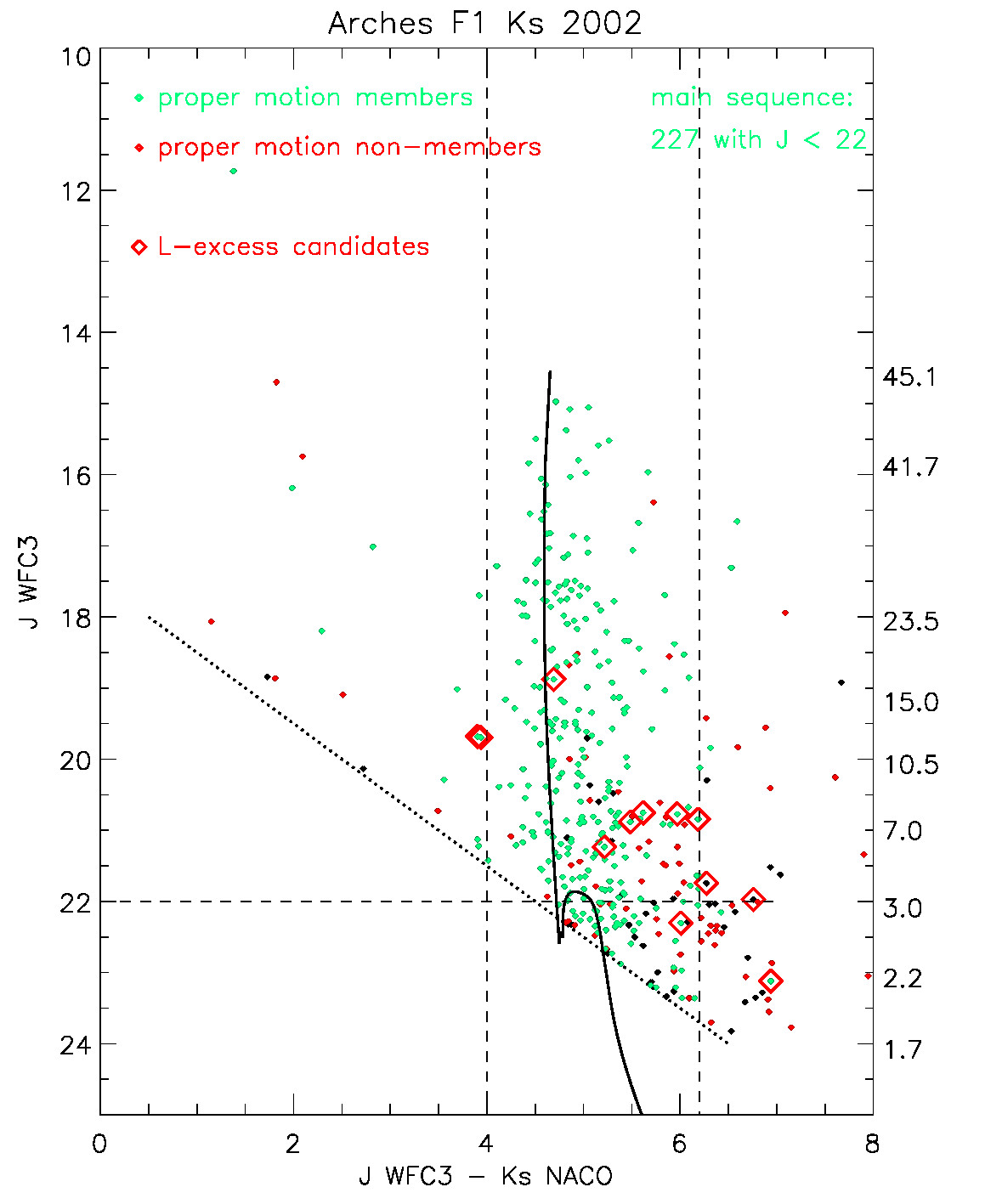}
\hspace*{4mm}
\begin{minipage}{8cm}
\caption{\label{jjk_cmds_arch} 
$J$, $J-K_s$ colour-magnitude diagram of the central field of the Arches cluster.
$L'$-excess sources derived from Fig.~\ref{ccds_arch} are marked as red diamonds.
Proper motion members are shown in green,
 non-members are shown in red, and sources with unknown membership in black.
The imposed magnitude limit to allow for proper-motion member 
selection, $K_s < 17.5$ mag, is shown as the dotted line. The horizontal dashed line
marks the $J$-band completeness limit at $J=22$ mag above which the combined
cluster excess fraction was derived.
The dashed vertical lines indicate the main-sequence colour selection applied in addition to 
the proper motion membership criterion to select the main-sequence reference sample
to calculate the excess fraction. 
Note that several $L'$-excess sources do not show excess emission at $K_s$.
The $J$, $J-K_s$ colour-magnitude diagrams of
the outer fields can be found in  Appendix (\ref{jjk_cmds_arch_app}).}
\end{minipage}
\end{figure}

\begin{figure}
\includegraphics[width=8.4cm]{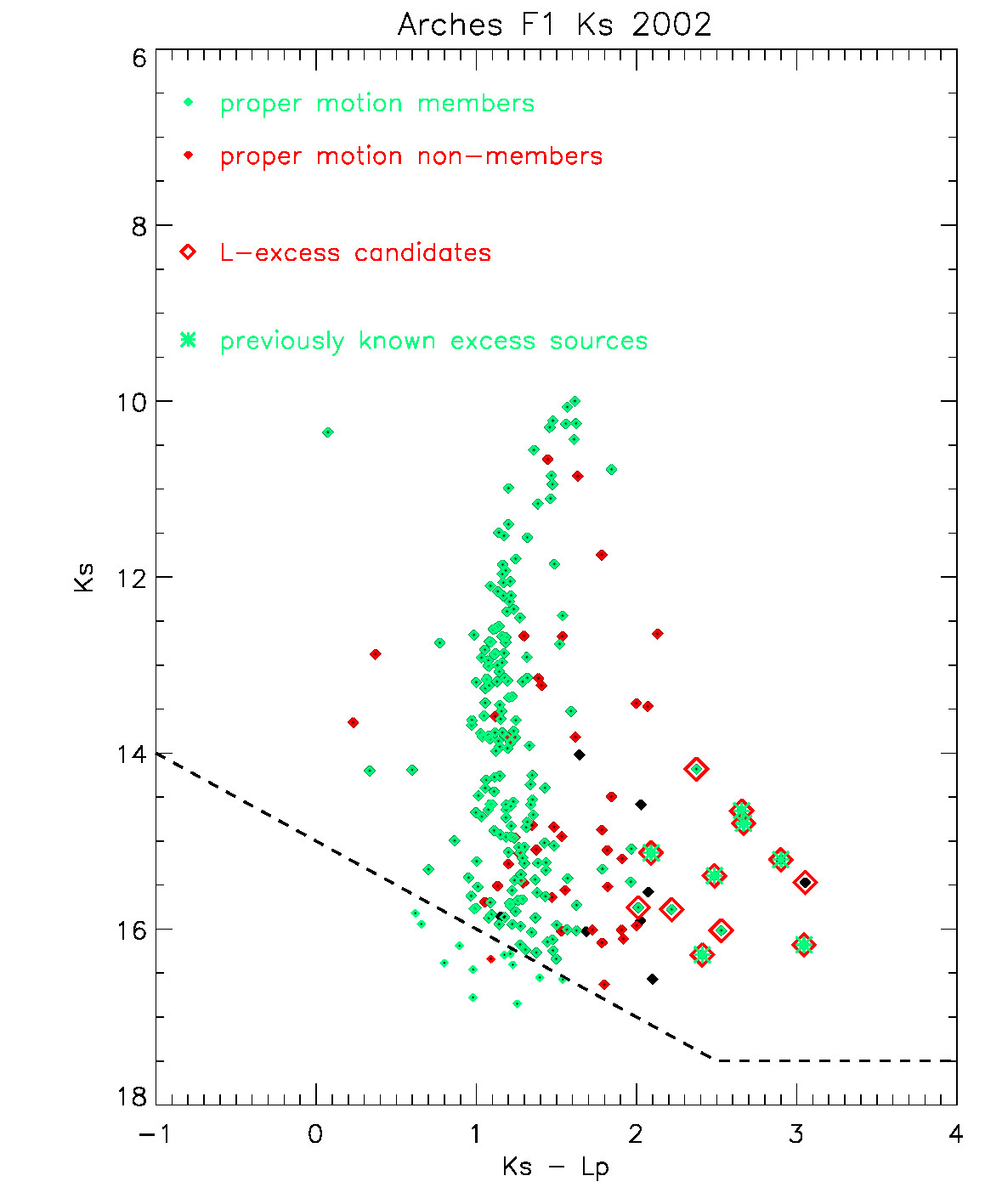}
\hspace*{4mm}
\begin{minipage}{8cm}
\caption{\label{kkl_cmds_arch} 
$K_s$, $K_s-L'$ colour-magnitude diagram of the central field of the Arches cluster.
All sources are labelled as in Fig.~\ref{jjk_cmds_arch}. A pronounced 
cluster main sequence composed of proper motion members (green) 
is seen in this central field. The $K_s$, $K_s-L'$ colour-magnitude diagrams of
the outer fields can be found in  Appendix (\ref{kkl_cmds_app}).}
\end{minipage}
\end{figure}

\clearpage

\begin{figure*}
\hspace*{24mm}
\includegraphics[width=12cm]{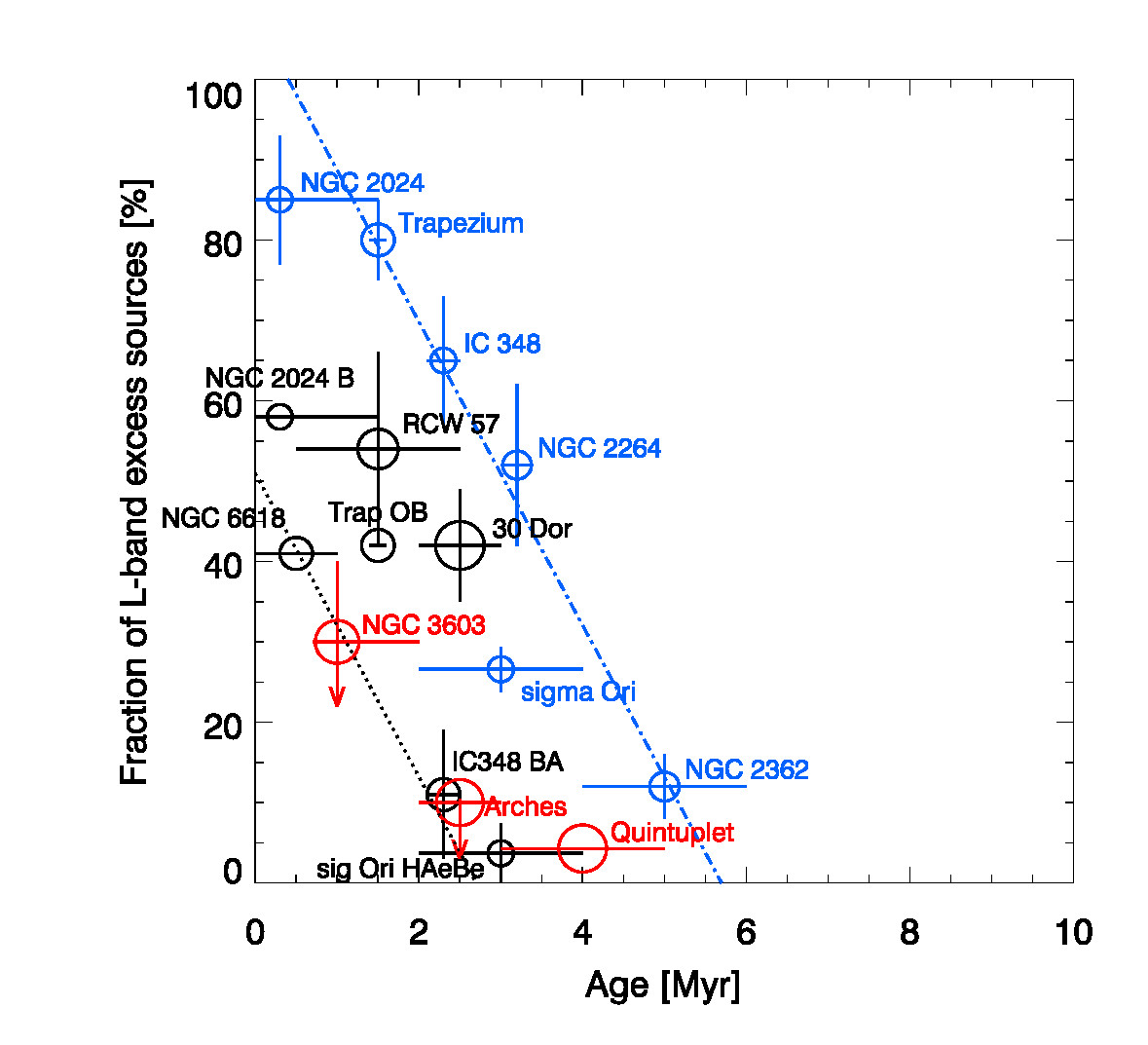}
\caption{\label{discage} 
Disc fraction as a function of cluster age, 
extracted from Haisch et al.~(2001; NGC 2024, Trapezium, IC 348,
NGC 2264, NGC 2362, Taurus, Chamaeleon I), Hern\'{a}ndez et al.~(2007) ($\sigma$ Ori), 
Hoffmeister et al.~(2006) (M17: NGC 6618), Maercker \& Burton (2005) (30 Dor region), 
and Maercker et al.~(2006) (NGC 3576: RCW 57). All disc fractions are derived
from $L$-band excesses with the exception of $\sigma$ Ori, where IRAC SED slopes 
were used to select optically thick discs.
Symbols are scaled to the logarithm of the cluster mass, from the least massive
with $\sim 200\,M_\odot$ in stars (NGC\,2024, IC\,348, $\sigma$ Ori), 
to the most massive with $> 30000\,M_\odot$ (30 Dor, see Stolte et al.~2010 for details).
Black and red circles mark disc fractions derived from high-mass stars of types OBA
only, while blue (light grey) circles mark populations dominated by low-mass stars.
The three starburst clusters Arches, Quintuplet, and NGC 3603 are highlighted in red.
The 30 Dor disc fraction covers the extended HII region, including
star-forming ridges harbouring YSO candidates, but does not resolve the central 
cluster, and is therefore an upper limit to the disc fraction in this environment.
The dash-dotted line corresponds to the linear decrease in disc fraction vs.
cluster age suggested by Haisch et al. (lighter circles only).
The dotted line represents the same relation, 
shifted to lower ages and disc fractions, indicating that disc
depletion progresses more rapidly in the most massive clusters.
The error bars in the disc fraction represent radial variations
in those clusters where a radial dependence is observed (see also
Stolte et al.~2010).
In the case of the Arches and NGC\,3603, the downward arrow indicates 
the radial decrease in the fraction of discs from larger radii toward the cluster
core (NGC 3603 outer cluster region: Stolte et al.~2004, resolved core: 
Harayama et al.~2008).}
\end{figure*}

\begin{figure*}
\centering
\includegraphics[width=9cm]{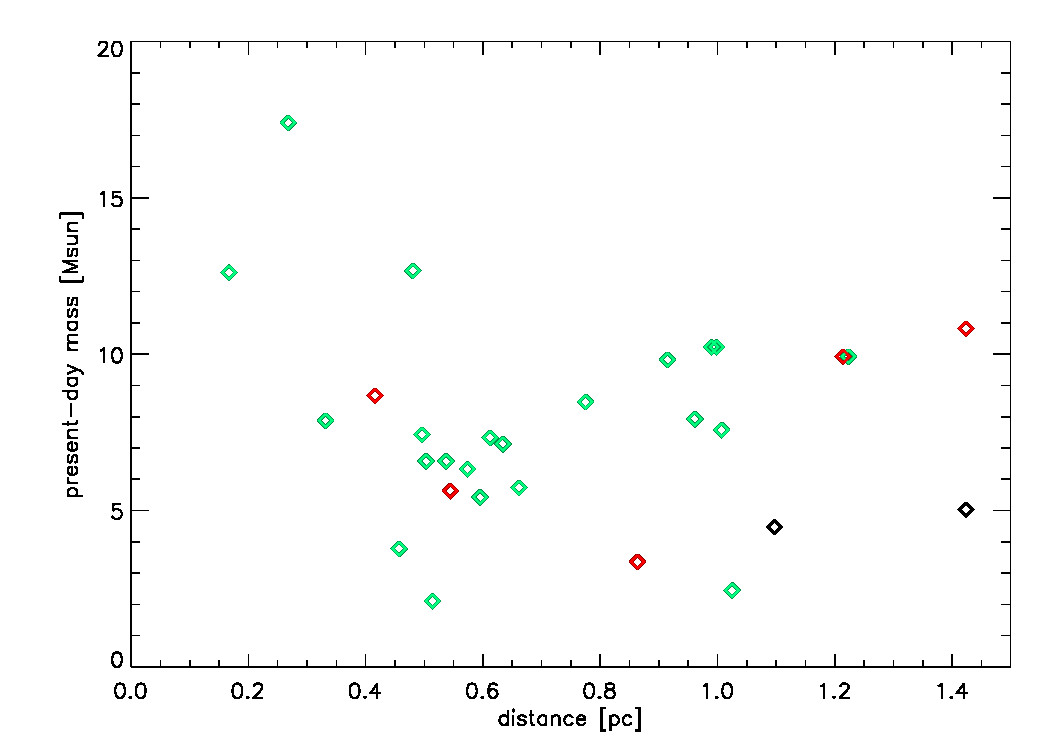}
\includegraphics[width=9cm]{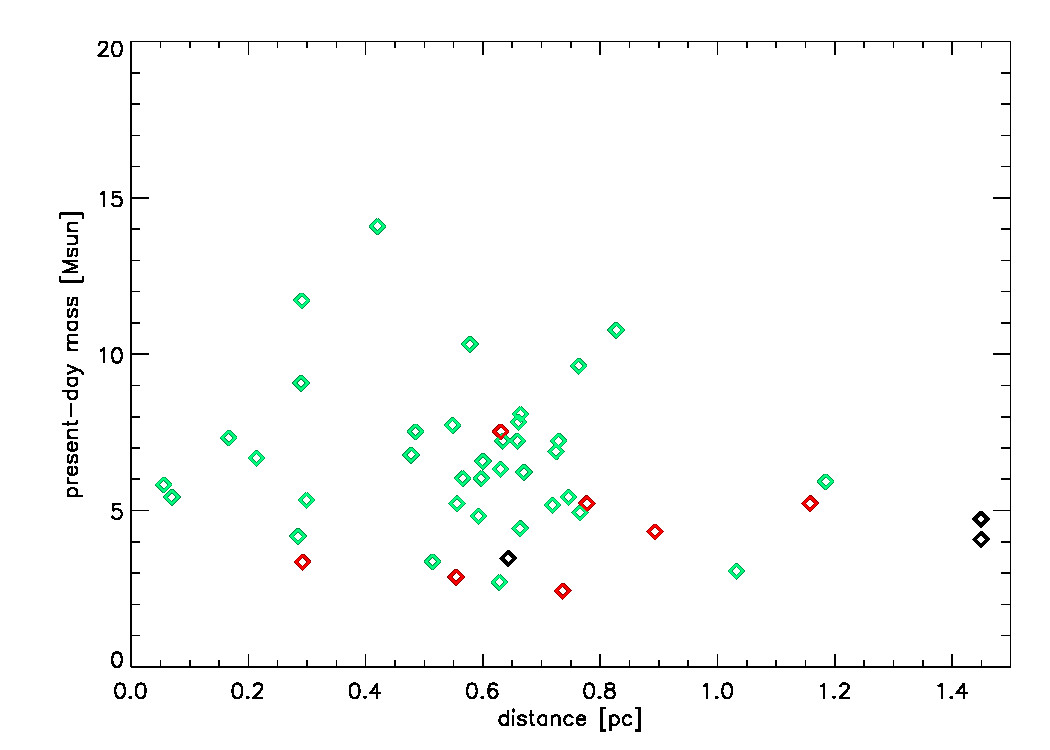}
\caption{\label{massrad} Radial distribution of disc host star
masses, as derived from the $J$-band luminosity and a 2.5 Myr
(Arches, top panel) and 4 Myr (Quintuplet, bottom panel) Geneva isochrone. 
Green symbols mark proper motion members, red symbols non-members, and black 
symbols represent stars without membership information.}
\end{figure*}

\clearpage



\begin{table*}
\centering
\caption{\label{obstab} VLT/NACO observations}
\begin{tabular}{lcccccccccc}
\hline
Date-Obs & Filter & DIT & NDIT & EXPTIME & $N_{frames}$ & $N_{used}$ & $t_{driz}$ & Seeing & FWHM & NGS  \\
         &        & [s] &      &   [s]   &            &           &  [min]    & [``]    & [mas] & [mag] \\
\hline
\multicolumn{11}{c}{Arches Field 1} \\
\hline
2002-Mar-31 & $K_s$ & 0.5 & 8 & 4.0 & 10 & 10 & 0.7 & 0.8 & 84 & V=16.3  \\
2002-Mar-31 & $K_s$ & 15.0 & 4 & 60.0 & 20 & 7 & 7.0 & 0.8 & 84 & V=16.3  \\
2010-Aug-09 & $K_s$ & 10.0 & 2 & 20.0 & 40(80) & 11(22) & 3.7 & 0.7 & 130 & Ks=10.4  \\
2011-Mar-31 & $K_s$ & 3.0 & 23 & 69.0 & 31(713) & 27(621) & 31.0 & 0.8 & 108 & Ks=10.4  \\
2012-Jun-15 & $L_p$ & 0.2 & 57 & 11.4 &   91  &  67   & 12.73 & 0.6 & 162 & Ks=10.4  \\
\hline
\multicolumn{11}{c}{Arches Field 2} \\
\hline
2008-Jun-06 & $K_s$ & 15.0 & 2 & 30.0 & 40 & 38 & 19.0 & 0.7 & 87 & Ks=9.2  \\
2010-Aug-09 & $K_s$ & 10.0 & 2 & 20.0 & 40(80) & 36(72) & 12.0 & 0.7 & 122 & Ks=9.2  \\
2011-Sep-09 & $L_p$ & 0.175 & 170 & 29.75 & 22 & 22 & 10.9 & 0.9 & 122 & Ks=9.2  \\
\hline
\multicolumn{11}{c}{Arches Field 3} \\
\hline
2008-Jun-06 & $K_s$ & 15.0 & 2 & 30.0 & 40 & 32 & 16.0 & 0.8 & 108 & Ks=9.8  \\
2011-Sep-17 & $K_s$ & 3.0 & 23 & 69.0 & 20(460) & 10(224) & 11.2 & 0.9 & 157 & V=16.0  \\
2012-Jun-12 & $K_s$ & 3.0 & 20 & 60.0 & 20(400) & 19(381) & 19.0 & 0.8 & 87 & Ks=9.8  \\
2012-Aug-05 & $L_p$ & 0.175 & 170 & 29.75 & 22 & 22 & 10.9 & 1.2 & 145 & Ks=9.8  \\
\hline
\multicolumn{11}{c}{Arches Field 4} \\
\hline
2008-Jun-06 & $K_s$ & 15.0 & 2 & 30.0 & 40 & 33 & 16.5 & 0.9 & 87 & Ks=10.4  \\
2011-Sep-18 & $K_s$ & 6.0 & 11 & 66.0 & 20(220) & 20(220) & 22.0 & 0.7 & 81 & Ks=10.4  \\
2012-Jun-12 & $K_s$ & 3.0 & 20 & 60.0 & 20(400) & 17(339) & 17.0 & 0.9 & 108 & Ks=10.4  \\ 
2012-Jun-15 & $L_p$ & 0.2 & 58 & 11.6 & 40 & 31 & 6.0 & 0.9 & 122 & Ks=10.4  \\
\hline
\multicolumn{11}{c}{Arches Field 5} \\
\hline
2008-Jun-10 & $K_s$ & 15.0 & 2 & 30.0 & 40 & 29 & 14.5 & 0.8 & 135 & Ks=10.3  \\
2011-Sep-19 & $K_s$ & 3.0 & 23 & 69.0 & 20(460) & 15(352) & 17.6 & 0.6 & 135 & V=16.1  \\
2012-Jun-12 & $K_s$ & 3.0 & 20 & 60.0 & 20(400) & 16(319) & 16.0 & 0.8 & 92 & Ks=10.3  \\
2012-Jun-15 & $L_p$ & 0.2 & 58 & 11.6 & 40 & 40 & 7.7 & 1.0 & 122 & Ks=10.3  \\
\hline
\multicolumn{11}{c}{Quintuplet Field 1} \\
\hline
2003-Jul-22 & $K_s$ & 2.0  & 30 & 60.0 & 16 & 16 & 16.0 & 0.4 & 87 & Ks=6.6 \\
2003-Jul-22 & $K_s$ & 20.0 & 2 & 40.0 & 16 & 16 & 10.7 & 0.4 & 76 & Ks=6.6 \\ 
2008-Jul-24 & $K_s$ & 2.0 & 15 & 30.0 & 44 & 33 & 16.5 & 0.5 & 87 & Ks=6.6 \\
2009-Jun-17 & $L_p$ & 0.175 & 170 & 29.75 & 36 & 36 & 17.9 & 0.6 & 106 & Ks=6.6 \\
\hline
\multicolumn{11}{c}{Quintuplet Field 2} \\
\hline
2008-Jul-24 & $K_s$ & 2.0 & 15 & 30.0 & 47 & 34 & 17.0 & 0.5 & 70 & Ks=7.3 \\
2011-Sep-19 & $K_s$ & 2.0 & 30 & 60.0 & 22 & 13(397) & 13.2 & 0.9 & 108 & V=16.2 \\
2012-Jun-14 & $K_s$ & 2.0 & 30 & 60.0 & 38 & 30(901) & 30.0 & 0.9 & 119 & Ks=7.3 \\
2011-Sep-10 & $L_p$ & 0.175 & 170 & 29.75 & 21 & 21 & 10.4 & 1.2 & 108 & Ks=7.3 \\
\hline
\multicolumn{11}{c}{Quintuplet Field 3} \\
\hline
2008-Aug-25 & $K_s$ & 2.0 & 15 & 30.0 & 63 & 17 & 8.5  & 0.7 & 108 & Ks=7.8 \\
2009-Apr-07 & $K_s$ & 2.0 & 15 & 30.0 & 87 & 32 & 16.0 & 0.5 & 108 & Ks=7.8 \\
2012-Jun-14 & $K_s$ & 2.0 & 30 & 60.0 & 34 & 27(827) & 27.6 & 0.8 & 108 & Ks=7.8 \\
2011-Sep-11 & $L_p$ & 0.175 & 170 & 29.75 & 22 & 22 & 10.9 & 0.9 & 108 & Ks=7.8 \\
\\
\hline
\multicolumn{11}{c}{Quintuplet Field 4} \\
\hline
2009-Apr-10 & $K_s$ & 2.0 & 15 & 30.0 & 44 & 32 & 16.0 & 0.5 & 81 & Ks=7.1  \\ 
2012-Jun-14 & $K_s$ & 2.0 & 30 & 60.0 & 20 & 17(515) & 17.2 & 0.9 & 108 & Ks=7.1  \\
2012-Jun-13 & $L_p$ & 0.2 & 66 & 13.2 & 63 & 20 & 4.4 & 1.5 & 154 & Ks=7.1  \\
\hline
\multicolumn{11}{c}{Quintuplet Field 5} \\
\hline
2009-Apr-10 & $K_s$ & 2.0 & 15 & 30.0 & 44 & 31 & 15.5 & 0.5 & 84 & Ks=7.5  \\
2012-Aug-02 & $K_s$ & 2.0 & 30 & 60.0 & 12 & 9(269) & 9.0 & 0.7 & 108 & Ks=7.5  \\
2012-Jun-13 & $L_p$ & 0.2 & 66 & 13.2 & 44 & 38 & 8.4 & 1.0 & 114 & Ks=7.5  \\
\end{tabular}
\tablefoot{Seeing refers to the optical seeing measured with the DIMM 
facility at the time of observations. DIT: Detector integration time, 
NDIT: Number of DITs obtained per frame, EXPTIME: resulting exposure
time per position, $N_{frames}$: Number of images obtained, $N_{used}$:
Number of images with good quality, $t_{driz}$: total integration time
in the final, drizzled image, $FWHM$: Full width at half maximum of 
the PSF, NGS: brightness of the natural guide star, 
$LF_{peak}$: peak of the luminosity function as indicator for the 
detection limit.
\\
The Arches 2002 data are described in detail in Stolte et al.~2002.}
\end{table*}


%

\begin{table*}
\caption{\label{phottab} 
Photometric sensitivity limits (in calibrated magnitudes), estimated from the peak of the luminosity functions, and 
number of sources in matched $JHK_s$ and $JHK_sL'$ catalogues.}
\begin{tabular}{lcccccrr}
\hline
Field & $J_{lim}$ & $H_{lim}$ & $K_{s,lim}$ Ep1 & $K_{s,lim}$ Ep2 & $L'_{lim}$ & $N_{stars}$ $JHK_s$ & $N_{stars}$ $JHK_sL'$ \\
\hline
\multicolumn{8}{c}{Arches} \\
\hline
1    &  24  &  21  & 18 & 18 & 15 & 467 & 253 \\
2    &  24  &  21  & 19 & 18 & 15 & 251 & 142 \\
3    &  24  &  21  & 18 & 19 & 15 & 446 & 168 \\
4    &  24  &  21  & 19 & 18 & 15 & 358 & 131 \\
5    &  24  &  21  & 18 & 19 & 15 & 352 & 121 \\
\hline
\multicolumn{8}{c}{Quintuplet} \\
\hline
1    &  24  &  21  & 19 & 19 & 16 & 3123 & 784 \\
2    &  24  &  21  & 20 & 19 & 15 & 1764 & 341 \\
3    &  24  &  21  & 19 & 19 & 15 & 1680 & 248 \\  
4    &  24  &  21  & 20 & 19 & 14 & 1341 &  81 \\ 
5    &  24  &  21  & 20 & 19 & 15 & 2402 & 231 \\ 
\hline
\end{tabular}
\end{table*}


\begin{table*}
\caption{\label{hsttab} HST/WFC3 observations}
\begin{tabular}{lcccccc}
\hline
Date-Obs & Filter & EXPTIME & $N_{frames}$ & $N_{used}$ & $t_{driz}$ & FWHM \\
         &        &   [s]   &            &           &  [min]    & [mas] \\
\hline
\multicolumn{7}{c}{WFC3 Arches} \\
\hline
2010-Aug-13 & F127M & 600 &  4 & 4 &  40 & 220 \\
2010-Aug-13 & F127M & 600 &  4 & 4 &  40 & 220 \\
2010-Aug-13 & F127M & 600 &  4 & 4 &  40 & 220 \\
2010-Aug-09 & F153M & 350 & 21 & 21 & 120 & 200 \\
2011-Sep-07 & F153M & 350 & 21 & 21 & 120 & 200 \\
\hline
\multicolumn{7}{c}{WFC3 Quintuplet} \\
\hline
2010-Aug-10 & F127M & 600 & 12 & 12 & 120 & 220 \\
2010-Aug-16 & F153M & 350 & 21 & 21 & 120 & 200 \\
\hline
\end{tabular}
\tablefoot{See Table \ref{obstab} for table notes.}
\end{table*}



\begin{table*}
\caption{\label{disctab} L-band excess sources in the Quintuplet cluster}
\begin{tabular}{lcccccccccccc}
\hline
  Seq & Field &  dRA   &   dDEC  &    J    & $\sigma_J$ & H & $\sigma_H$ &  $K_s$  & $\sigma_{Ks}$ & $L'$ & $\sigma_{L}$ & mem \\
      &       & [asec] &  [asec]  &  [mag]  &   [mag] &   [mag] &   [mag] &   [mag] &   [mag] &   [mag] &   [mag] &   \\
\hline
   1 &   1 &   10.68 &   -1.74 &   19.60 &    0.04 &   16.66 &    0.07 &   14.48 &    0.00 &   12.61 &    0.03 &   1 \\
   2 &   1 &    4.72 &    5.81 &   20.10 &    0.04 &   16.95 &    0.08 &   14.88 &    0.00 &   12.76 &    0.05 &   1 \\
   3 &   1 &  -18.62 &  -10.33 &   20.31 &    0.09 &   17.54 &    0.07 &   15.64 &    0.03 &   13.26 &    0.02 &   1 \\
   4 &   1 &  -18.84 &   -5.63 &   20.58 &    0.07 &   17.52 &    0.07 &   15.43 &    0.01 &   13.06 &    0.10 &   1 \\
   5 &   1 &   -5.86 &    4.61 &   20.71 &    0.05 &   17.33 &    0.08 &   15.32 &    0.01 &   13.15 &    0.03 &   1 \\
   6 &   1 &   -3.25 &  -13.71 &   21.06 &    0.06 &   17.88 &    0.07 &   15.55 &    0.02 &   12.88 &    0.08 &   1 \\
   7 &   1 &    1.73 &   12.35 &   21.11 &    0.06 &   17.75 &    0.09 &   15.42 &    0.00 &   13.24 &    0.03 &   1 \\
   8 &   1 &   -1.38 &  -18.67 &   21.21 &    0.05 &   17.51 &    0.09 &   15.27 &    0.02 &   12.41 &    0.02 &   1 \\
   9 &   1 &    0.40 &    4.28 &   21.17 &    0.14 &   18.57 &    0.09 &   16.24 &    0.12 &   14.46 &    0.02 &   1 \\
  10 &   1 &   18.59 &   -1.34 &   21.33 &    0.05 &   18.25 &    0.07 &   16.02 &    0.01 &   13.61 &    0.12 &   1 \\
  11 &   1 &  -13.45 &   -7.59 &   21.42 &    0.06 &   18.28 &    0.07 &   15.98 &    0.01 &   13.61 &    0.03 &   1 \\
  12 &   1 &  -12.22 &    7.91 &   21.63 &    0.19 &   18.60 &    0.11 &   16.24 &    0.02 &   13.94 &    0.09 &   1 \\
  13 &   1 &    0.62 &    1.25 &   21.71 &    0.23 &   17.60 &    0.17 &   15.32 &    0.04 &   12.88 &    0.01 &   1 \\
  14 &   1 &   10.29 &  -16.15 &   21.88 &    0.13 &   18.58 &    0.12 &   16.18 &    0.02 &   13.96 &    0.07 &   1 \\
  15 &   1 &   -4.33 &    6.38 &   21.91 &    0.09 &   18.75 &    0.13 &   16.38 &    0.00 &   14.46 &    0.06 &   1 \\
  16 &   1 &  -15.24 &   -0.78 &   22.12 &    0.10 &   18.79 &    0.12 &   15.93 &    0.01 &   13.22 &    0.01 &   1 \\
  17 &   1 &  -13.66 &   10.27 &   22.29 &    0.09 &   18.46 &    0.10 &   16.05 &    0.01 &   13.63 &    0.01 &   1 \\
  18 &   1 &   -8.33 &  -10.27 &   22.79 &    0.05 &   19.12 &    0.09 &   17.06 &    0.00 &   14.80 &    0.29 &   1 \\
  19 &   1 &   17.79 &   -8.60 &   22.15 &    0.13 &   18.65 &    0.10 &   16.23 &    0.04 &   13.52 &    0.10 &  -1 \\
  20$^a$&2 &   -1.38 &  -18.67 &   21.21 &    0.05 &   17.51 &    0.09 &   15.89 &    0.01 &   13.13 &    0.00 &   1 \\
  21 &   2 &   -7.02 &  -26.19 &   21.21 &    0.05 &   17.97 &    0.10 &   15.69 &    0.00 &   13.28 &    0.01 &   1 \\
  22 &   2 &  -10.77 &  -29.58 &   21.97 &    0.19 &   18.85 &    0.14 &   16.96 &    0.01 &   14.74 &    0.04 &   1 \\
  23 &   4 &   28.73 &   18.35 &   17.67 &    0.09 &   15.22 &    0.06 &   13.87 &    0.00 &   12.34 &    0.04 &   1 \\
  24 &   5 &   26.71 &   -0.95 &   20.97 &    0.07 &   18.02 &    0.09 &   16.11 &    0.00 &   13.90 &    0.10 &   1 \\
  25 &   5 &   29.98 &   -6.28 &   21.07 &    0.05 &   17.92 &    0.07 &   16.15 &    0.01 &   13.71 &    0.01 &   1 \\
  26 &   5 &   18.59 &   -1.33 &   21.33 &    0.05 &   18.25 &    0.07 &   16.65 &    0.01 &   14.26 &    0.03 &   1 \\
  27$^b$&5 &   17.79 &   -8.60 &   22.15 &    0.13 &   18.65 &    0.10 &   17.04 &    0.00 &   14.23 &    0.07 &   1 \\
  28 &   5 &   13.65 &   -9.39 &   20.67 &    0.09 &   17.62 &    0.11 &   16.64 &    0.01 &   14.39 &    0.13 &  -1 \\
\hline
\end{tabular}
\tablefoot{Positional offsets in right ascension and declination are given in arcseconds, 
relative to the Wolf-Rayet member Q12 in the cluster core, RA 17:46:15.13, DEC -28:49:34.7.
Sources identified as proper motion members have a membership index of 1, while sources without
reliable membership information are labelled as -1. Non-members are likely field stars and are 
not shown here. \newline
$^a$ Source 20 in Field 2 is identical to source 8 in Field 1. \newline
$^b$ Source 27 in Field 5 is identical to source 19 in Field 1.}
\end{table*}


\begin{table*}
\caption{\label{extab} $L'$ excess fractions in the Quintuplet cluster} 
\begin{tabular}{lcccccccccl}
Field & main seq $J-K_s$ & $n_{ms}$ & $n_{ms,corr}$ & $n_{ex}$ & $n_{ex,corr}$ & $f_{ex}$ & $\sigma_{f_{ex},low}$ & $\sigma_{f_{ex},high}$ & $f_{exc,corr}$ & comment \\
\hline
1  & 4.0-5.5 & 599 +40/-57 & 684 & 18 (+1) &  20.5 (22)&  3.0 (3.2) & -0.2 & +0.3 & 3.0 (3.2) & $J < 22.8$ mag    \\
   & 4.0-5.5 & 445 +32/-39 & 484 & 15 (+1) &  21.2 (+2) &  3.4 (3.6) & -0.3 & +0.3 & 4.4 (4.8) & $J < 22$ mag    \\
2  & 4.0-5.0 &  80  +4/-16 &  96 &  3      &  4.2    &  3.8       & -0.2 & +0.2 & 4.4 & $J < 22$ main seq F2-F5   \\
3  & 3.7-4.5 &  45  +9/-8  &  47 &  0      &  --     &  0.0       &  -   &   -  &     & no excess members \\
4  & 3.7-5.0 &  54  +9/-12 &  69 &  1      &  1.1    &  1.8       & -0.2 & +0.6  & 1.6: & low-sensitivity $L'$ \\
5  & 3.7-4.7 &  99 +13/-9  & 115 &  4 (+1) &  5 (6) &  4.0 (5.0) & -0.4 & +0.4 & 4.2 (5.2) & excess at J=22.1 included \\
\hline
1,2,5$^a$  &   -   &  591 +49/-64 & 658 & 24 (+1) & 27 (+1) & 4.1 (4.2) & -0.3 & +0.5 & 4.1 (4.3) & main seq: F1/2/5 $J < 22$ \\
all$^a$    &   -   &  681 +67/-84 & 766 & 25 (+1) & 28 (+1) & 3.7 (3.8) & -0.4 & +0.5 & 3.7 (3.8) & main seq: F1/2/3/4/5 $J < 22$ \\
\hline
\end{tabular}
\tablefoot{Column 2 contains the main-sequence colour selection for each field; columns 3-4 and 5-6 
are the observed and completeness corrected number of main-sequence stars and excess sources, respectively;
Col. 7 provides the observed excess fraction in per cent, while Col. 10 contains
the completeness corrected excess fractions in per cent; Cols. 8 and 9 
are the uncertainties in the derived excess fraction when shifting the red main-sequence boundary 
by $\pm 0.2$ mag: adding these values to the excess fraction in Col. 5 yields the lower and upper 
limits of the observed excess fractions in each field and across the cluster area.\newline
$^a$ The total number counts used to calculate the combined excess fraction of the cluster
in the final two rows are reduced by 2 duplicate excess sources in the overlap regions between 
Fields 2 and 5 with Field 1, and main-sequence number counts are reduced by 33 duplicates
in F2 and F5 (plus 6 duplicates in F4), and incompleteness number counts are adjusted accordingly.}
\end{table*}


\begin{table*}
\caption{\label{disctab_arch} L-band excess sources in the Arches cluster}
\begin{tabular}{lcccccccccccc}
\hline
  Seq & Field &  dRA   &   dDEC  &    J    & $\sigma_J$ & H$^a$ & $\sigma_H$ &  $K_s$  & $\sigma_{Ks}$ & $L'$ & $\sigma_{L}$ & mem \\
      &       & [asec] &  [asec]  &  [mag]  &   [mag] &   [mag] &   [mag] &   [mag] &   [mag] &   [mag] &   [mag] & \\
\hline
   1 &   1 &    3.10 &    5.54 &   18.87 &    0.06 &   15.99 &    0.01 &   14.18 &    0.01 &   11.81 &    0.01 &   1 \\
   2 &   1 &    7.67 &    3.46 &   19.68 &    0.01 &   17.29 &    0.01 &   15.78 &    0.01 &   13.56 &    0.07 &   1 \\
   3 &   1 &    3.21 &    1.17 &   19.69 &    0.03 &   17.56 &    0.01 &   15.75 &    0.02 &   13.74 &    0.04 &   1 \\
   4 &   1 &   -8.60 &    9.55 &   20.75 &    0.07 &   17.28 &    0.00 &   15.13 &    0.00 &   13.04 &    0.03 &   1 \\
   5 &   1 &    5.81 &    2.25 &   20.77 &    0.15 &   17.10 &    0.01 &   14.80 &    0.01 &   12.13 &    0.01 &   1 \\
   6 &   1 &    3.44 &   15.29 &   20.84 &    0.11 &   17.26 &    0.01 &   14.65 &    0.01 &   12.00 &    0.03 &   1 \\
   7 &   1 &    7.04 &   -0.79 &   20.88 &    0.09 &   17.52 &    0.01 &   15.39 &    0.01 &   12.91 &    0.01 &   1 \\
   8 &   1 &   -2.42 &    8.65 &   21.23 &    0.03 &   18.10 &    0.01 &   16.01 &    0.02 &   13.49 &    0.04 &   1 \\
   9 &   1 &    7.44 &    2.19 &   22.30 &    0.05 &   18.37 &    0.01 &   16.29 &    0.01 &   13.88 &    0.04 &   1 \\
  10 &   1 &    9.08 &   13.35 &   23.12 &    0.06 &   18.96 &    0.01 &   16.18 &    0.01 &   13.13 &    0.05 &   1 \\
  11 &   1 &  -14.59 &   11.22 &   21.74 &    0.07 &   17.94 &    0.01 &   15.47 &    0.01 &   12.42 &    0.03 &  -1 \\
  12 &   1 &  -12.53 &    1.40 &   21.97 &    0.03 &   17.96 &    0.01 &   15.21 &    0.00 &   12.31 &    0.05 &  -1 \\
  13 &   3 &   24.73 &   13.40 &   19.90 &    0.04 &   17.01 &    0.01 &   14.64 &    0.00 &   11.79 &    0.05 &   1 \\
  14 &   3 &   31.11 &   13.68 &   19.95 &    0.18 &   16.83 &    0.03 &   14.66 &    0.00 &   11.95 &    0.03 &   1 \\
  15 &   3 &   17.14 &    7.73 &   20.67 &    0.05 &   17.54 &    0.02 &   15.44 &    0.00 &   13.34 &    0.05 &   1 \\
  16 &   3 &   14.85 &   -1.37 &   20.83 &    0.08 &   17.50 &    0.00 &   15.50 &    0.00 &   13.30 &    0.11 &   1 \\
  17 &   3 &   15.43 &   -3.98 &   21.27 &    0.05 &   17.80 &    0.00 &   15.54 &    0.00 &   12.94 &    0.04 &   1 \\
  18 &   4 &   -6.63 &   24.36 &   20.89 &    0.05 &   17.14 &    0.01 &   14.44 &    0.00 &   11.83 &    0.00 &   1 \\
  19$^b$&5 &   24.73 &   13.40 &   19.90 &    0.04 &   17.01 &    0.01 &   14.67 &    0.00 &   12.01 &    0.01 &   1 \\
  20 &   5 &   10.36 &   20.33 &   20.31 &    0.02 &   17.40 &    0.01 &   15.30 &    0.00 &   12.74 &    0.01 &   1 \\
  21 &   5 &    8.75 &   16.44 &   20.60 &    0.07 &   17.37 &    0.01 &   15.58 &    0.00 &   13.31 &    0.04 &   1 \\
  22$^c$&5 &    3.44 &   15.29 &   20.84 &    0.11 &   17.26 &    0.00 &   14.67 &    0.00 &   12.07 &    0.01 &   1 \\
  23 &   5 &    3.15 &   19.43 &   21.17 &    0.02 &   17.76 &    0.01 &   15.62 &    0.00 &   13.40 &    0.06 &   1 \\
  24$^d$&5 &    9.08 &   13.35 &   23.12 &    0.06 &   19.30 &    0.01 &   16.52 &    0.01 &   13.24 &    0.09 &   1 \\
\hline
\end{tabular}
\tablefoot{Positional offsets in right ascension and declination are given in arcseconds, 
relative to the brightest $K$-band source in the cluster core (see Fig.~\ref{ksmosaic}), 
RA 17:45:50.42, DEC -28:49:22.3.
The cluster centre is located at RA 17:45:50.54, DEC -28:49:19.8.
Sources identified as proper motion members have a membership index of 1, while sources without
reliable membership information are labelled as -1. Non-members are likely field stars and are 
not shown here.\newline
$^a$ H-band magnitudes are from NACO 2002 photometry in Field 1, and from WFC3 F153M photometry 
in all other fields. \newline
$^b$ Source 19 in Field 5 is identical to source 13 in Field 3, with $K_s$ and $L'$ measured on Field 5. 
\newline
$^c$ Source 22 in Field 5 is identical to source 6 in Field 1.
\newline
$^d$ Source 24 in Field 5 is identical to source 10 in Field 1.}
\end{table*}


\begin{table*}
\caption{\label{extab_arch} $L'$ excess fractions in the Arches cluster} 
\begin{tabular}{lcccccccccl}
Field & main seq $J-K_s$ & $n_{ms}$ & $n_{ms,corr}$ & $n_{ex}$ & $n_{ex,corr}$ & $f_{ex}$ & $\sigma_{f_{ex},low}$ & $\sigma_{f_{ex},high}$ & $f_{ex,corr}$ & comment \\
\hline
1  & 4.0-6.2 & 231  +2/-7   & 546 & 10 (+2) & 66 (+6) &  4.3 (5.2) & -0.1  & +0.2 & 12.1 (13.2) &   \\
2  & 5.0-6.2 &  21  +5/-4   &  46 &  0      & --      &  0.0       &  --   &  --  & -- &  no excess members \\
3  & 3.7-5.5 &  83  +2/-6   & 199 &  5      & 10      &  6.0       & -0.1  & +0.5 & 5.0 & 1 excess also in F5 \\
4  & 4.7-6.0 &  38  +1/-4   &  89 &  1      & 1.5     &  2.6       & -0.1  & +0.3 & 1.7: & just one excess \\
5  & 4.0-5.3 &  62  +2/-7   & 126 &  6      & 14      &  9.7       & -0.3  & +1.2 & 11.1 & 2 excess also in F1 \\
\hline
1,3,4,5$^a$ & -  &  400  +7/-24 & 906 & 19 (+2) & 85 (+6) & 4.8 (5.3) & -0.1 & +0.3 & 9.4 (10.0) & main seq: F1/3/4/5 \\
all$^a$     & -  &  421 +12/-28 & 952 & 19 (+2) & 85 (+6) & 4.5 (5.0) & -0.1 & +0.3 & 8.9 (9.6)  & main seq: F1/2/3/4/5 \\
%
\hline
\end{tabular}
\tablefoot{Column 2 contains the main-sequence colour selection for each field; columns 3-4 and 5-6 
are the observed and completeness corrected number of main-sequence stars and excess sources, respectively;
Col. 7 provides the observed excess fraction in per cent, while Col. 10 contains
the completeness corrected excess fractions in per cent; Cols. 8 and 9 
are the uncertainties in the derived excess fraction when shifting the red main-sequence boundary 
by $\pm 0.2$ mag: adding these values to the excess fraction in Col. 5 yields the lower and upper 
limits of the excess fractions in each field and across the cluster area. \newline
$^a$ The total number counts used to calculate the combined excess fraction of the cluster
in the final two rows are reduced by the number of 14 duplicate main sequence and 3 duplicate excess sources 
in the overlap regions between Fields 1, 3, and 5, and incompleteness number counts are
adjusted accordingly.}
\end{table*}

\clearpage


\onecolumn
\appendix

\section{Proper motion diagrams of the outer cluster fields}

As a consequence of the variation in stellar density across the 
wide-field coverage of the cluster images, the proper motion  
diagrams are populated with varying levels of source density.
In order to avoid confusion, especially given the dense environment
in the cluster centres, the proper motion diagrams of the individual
data sets are shown here. 

\begin{figure*}[h]
\begin{minipage}{18cm}
\includegraphics[width=9cm]{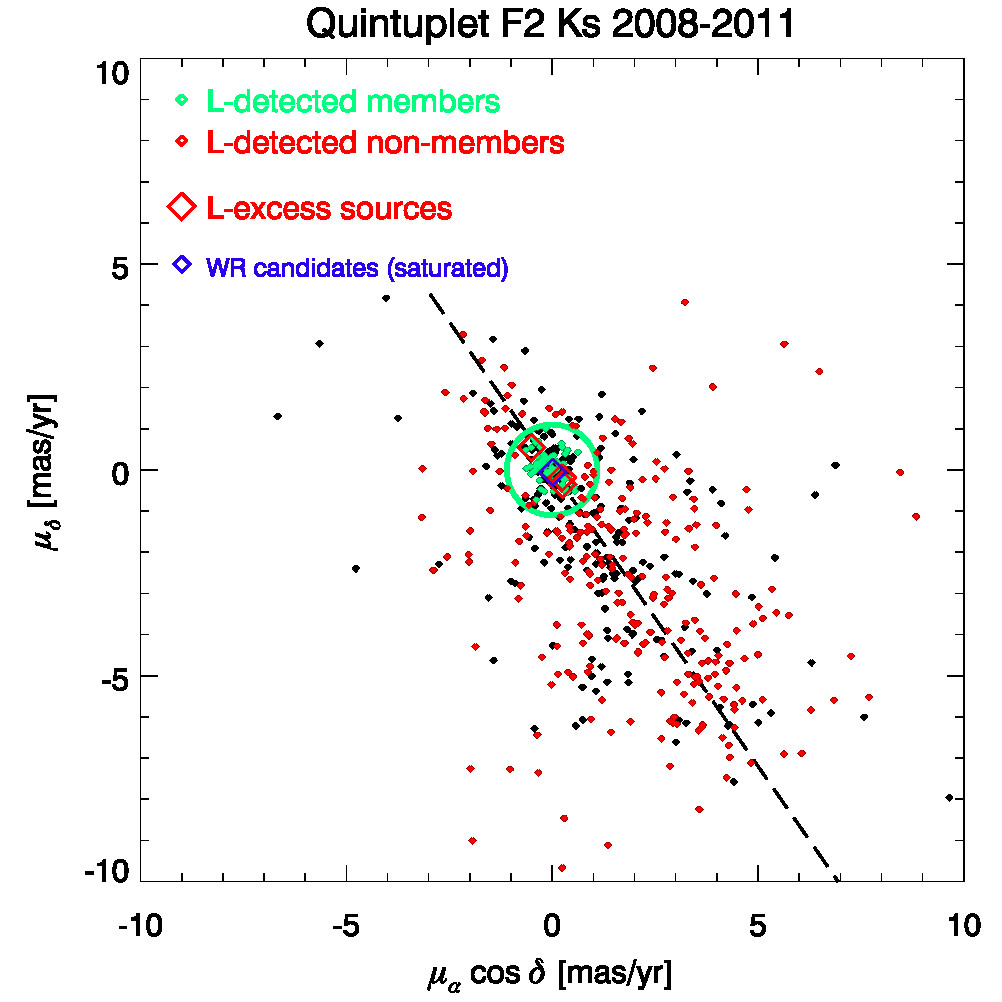}
\includegraphics[width=9cm]{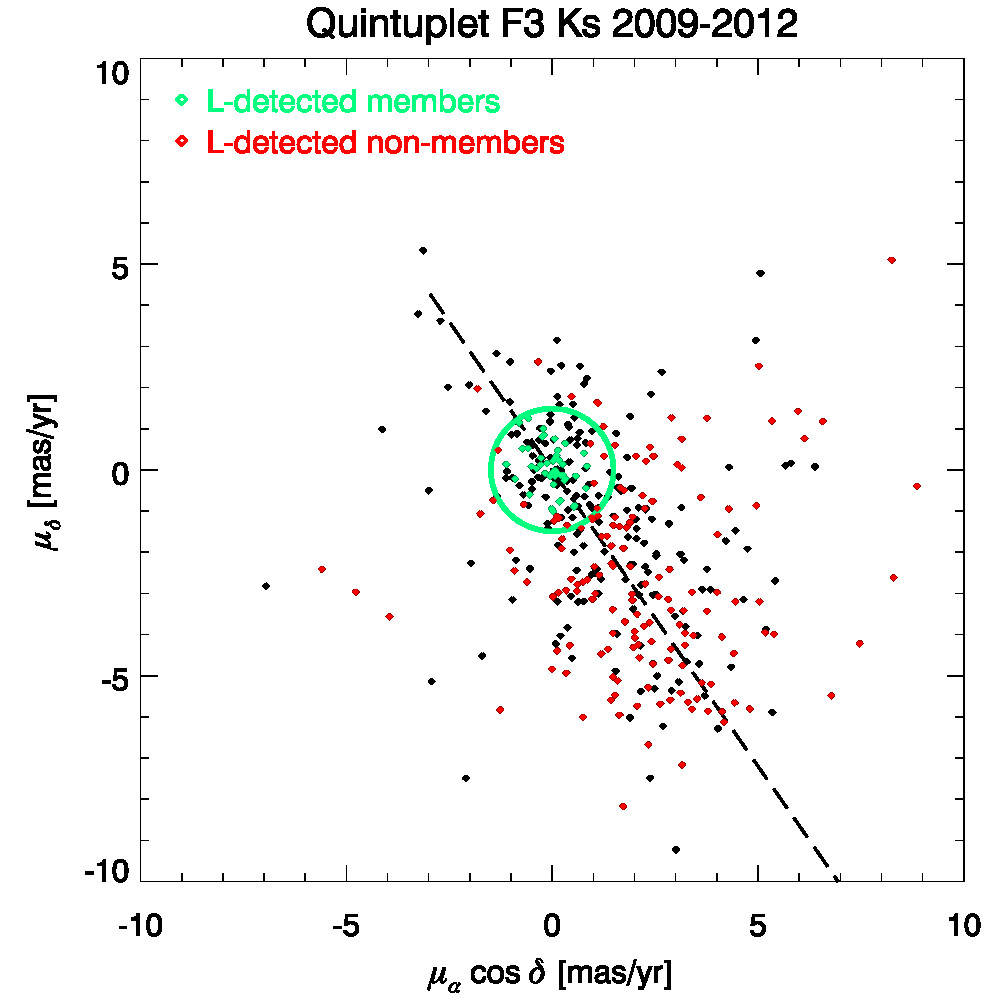}
\end{minipage}
\includegraphics[width=9cm]{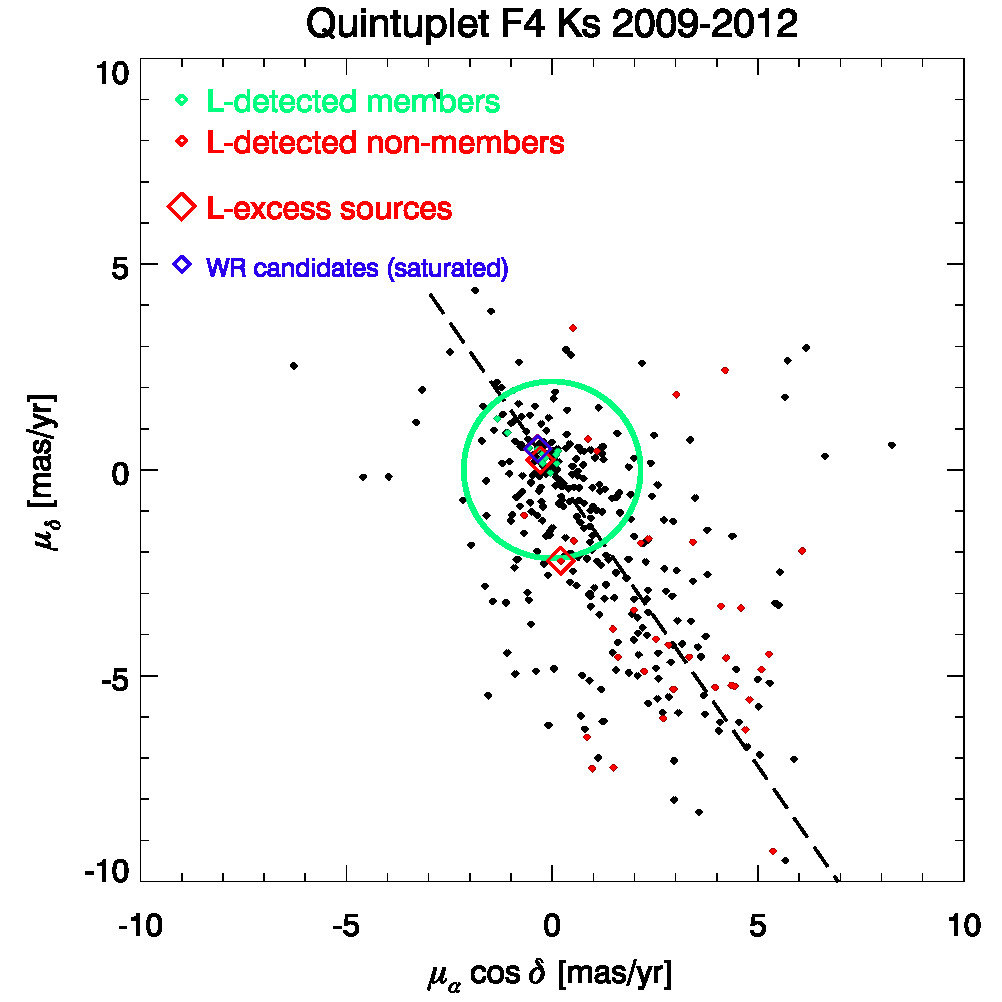}
\includegraphics[width=9cm]{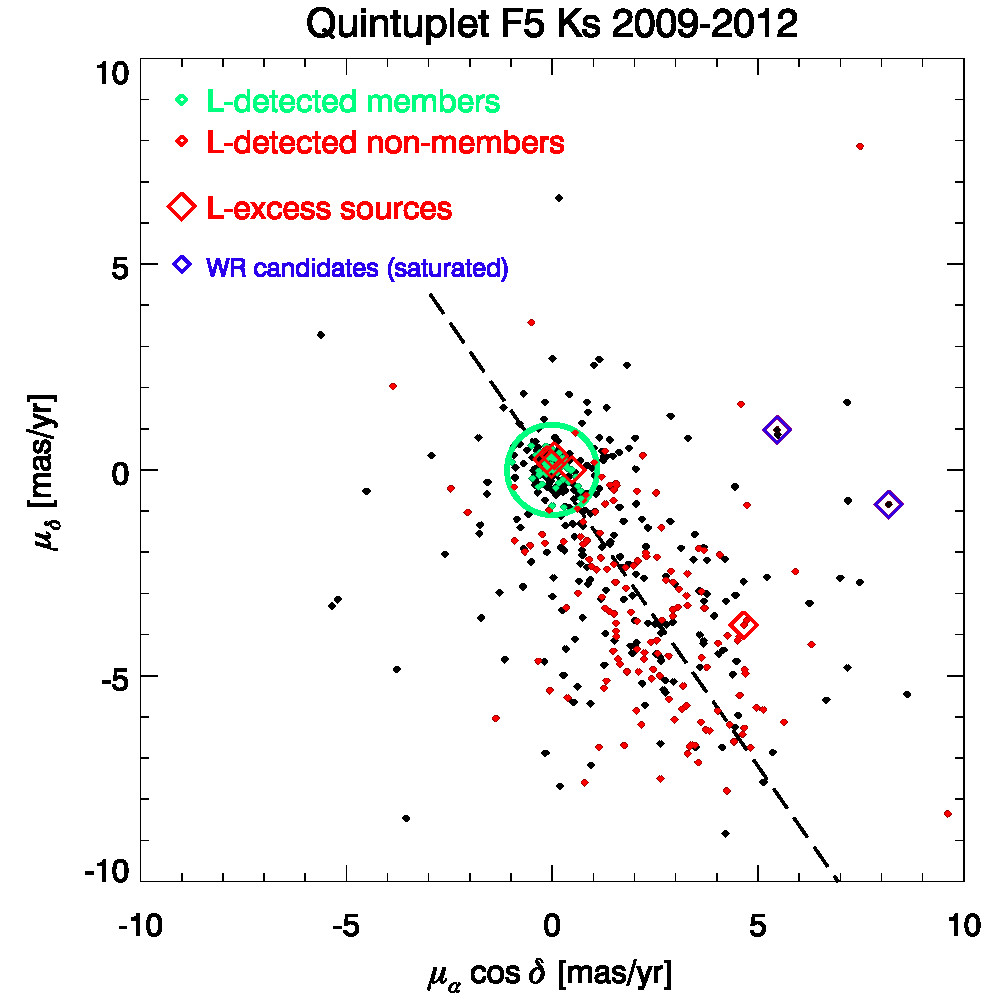}
\caption{\label{quinpm_app} 
Proper motion diagrams of the Quintuplet fields beyond the 
cluster centre. $L$-band excess sources are marked as red diamonds, 
while Wolf-Rayet candidates are shown in blue. The criterion 
for cluster member selection (circle) depends on the astrometric 
quality of each data set and on the corresponding dispersion in 
the cluster population around the origin. For the outer 
Quintuplet fields exclusively, membership probabilities were 
used to distinguish cluster members from field stars. The indicated
membership criterion is therefore approximate, as the probabilities
account for individual astrometric uncertainties and the location
of each star with respect to the cluster centre. See Fig.~\ref{quinpm}
for further details.}
\end{figure*}

\begin{figure*}
\begin{minipage}{18cm}
\includegraphics[width=9cm]{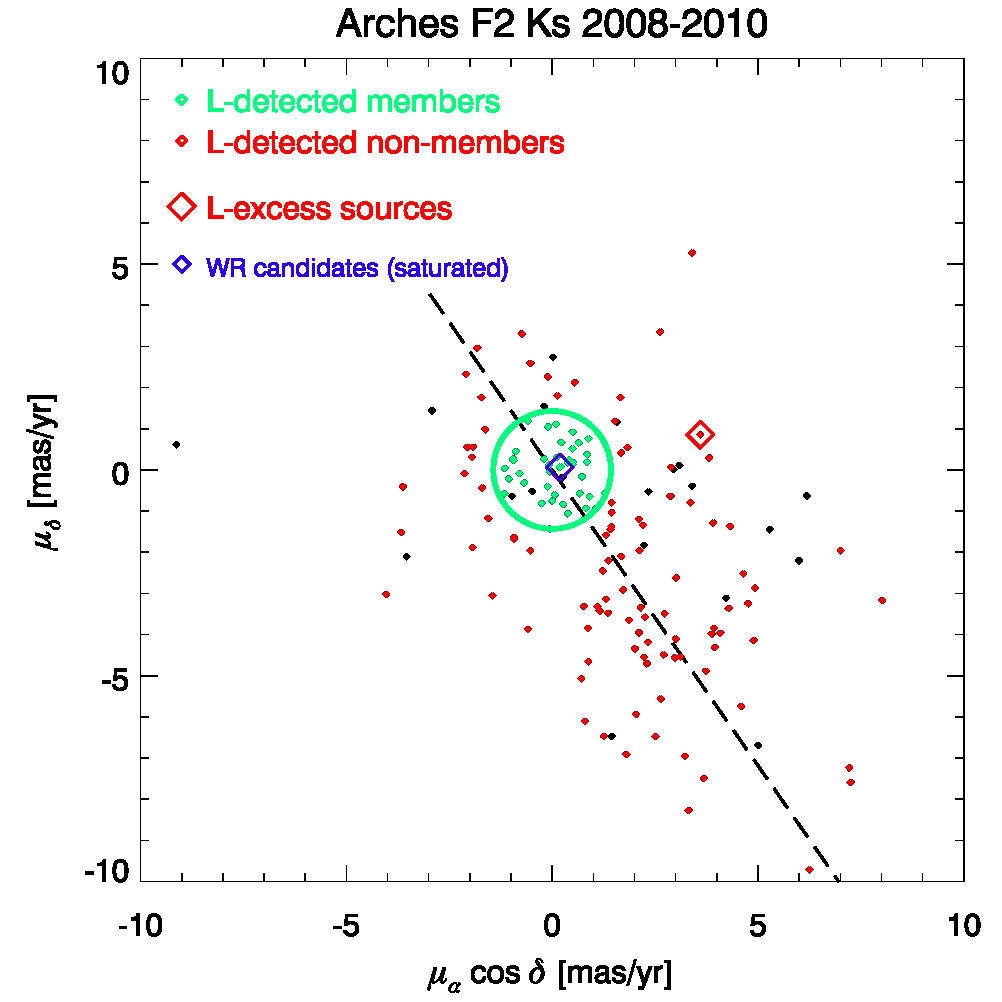}
\includegraphics[width=9cm]{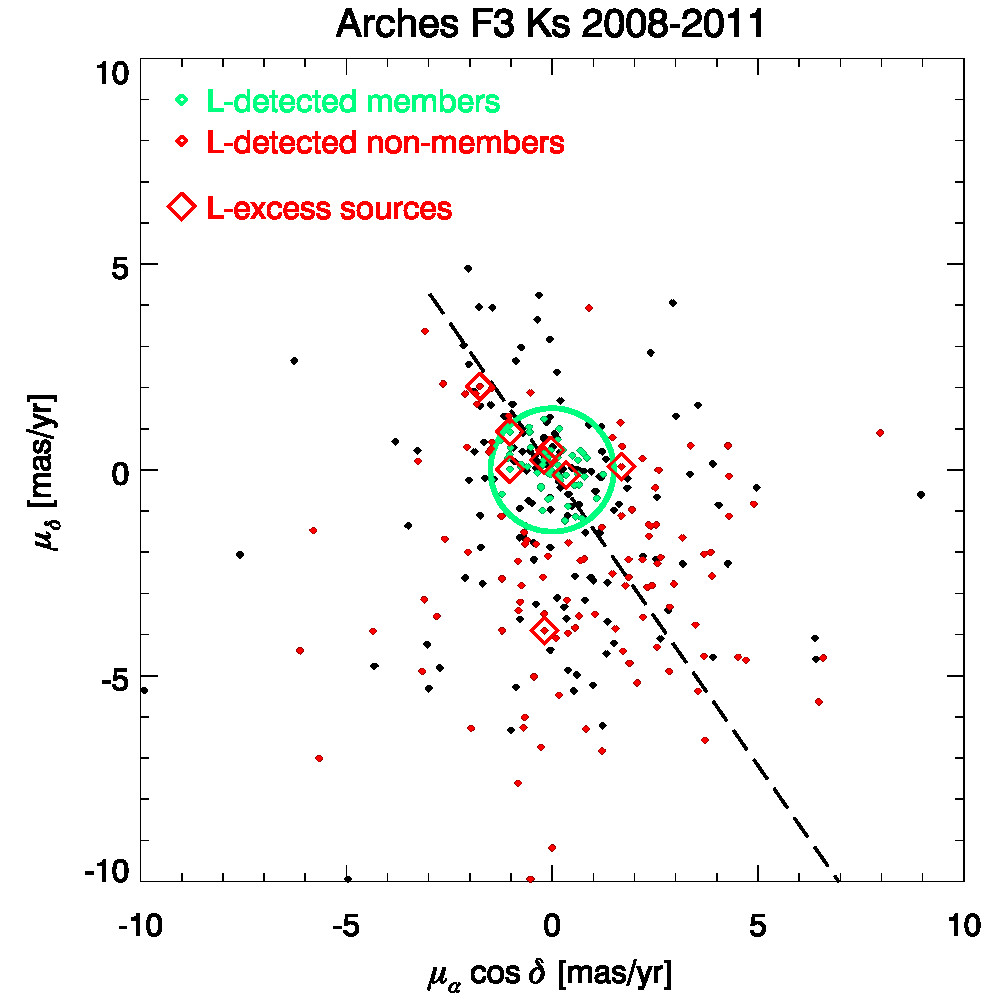}
\end{minipage}
\includegraphics[width=9cm]{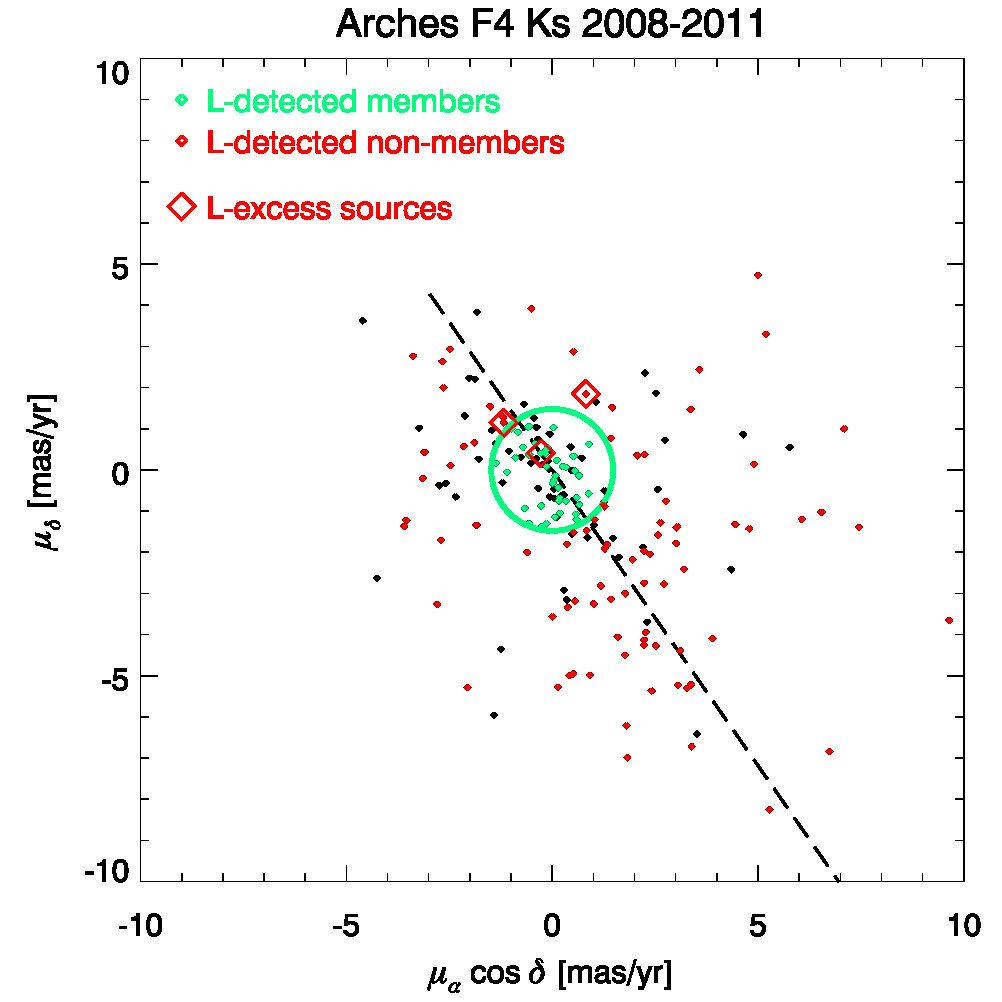}
\includegraphics[width=9cm]{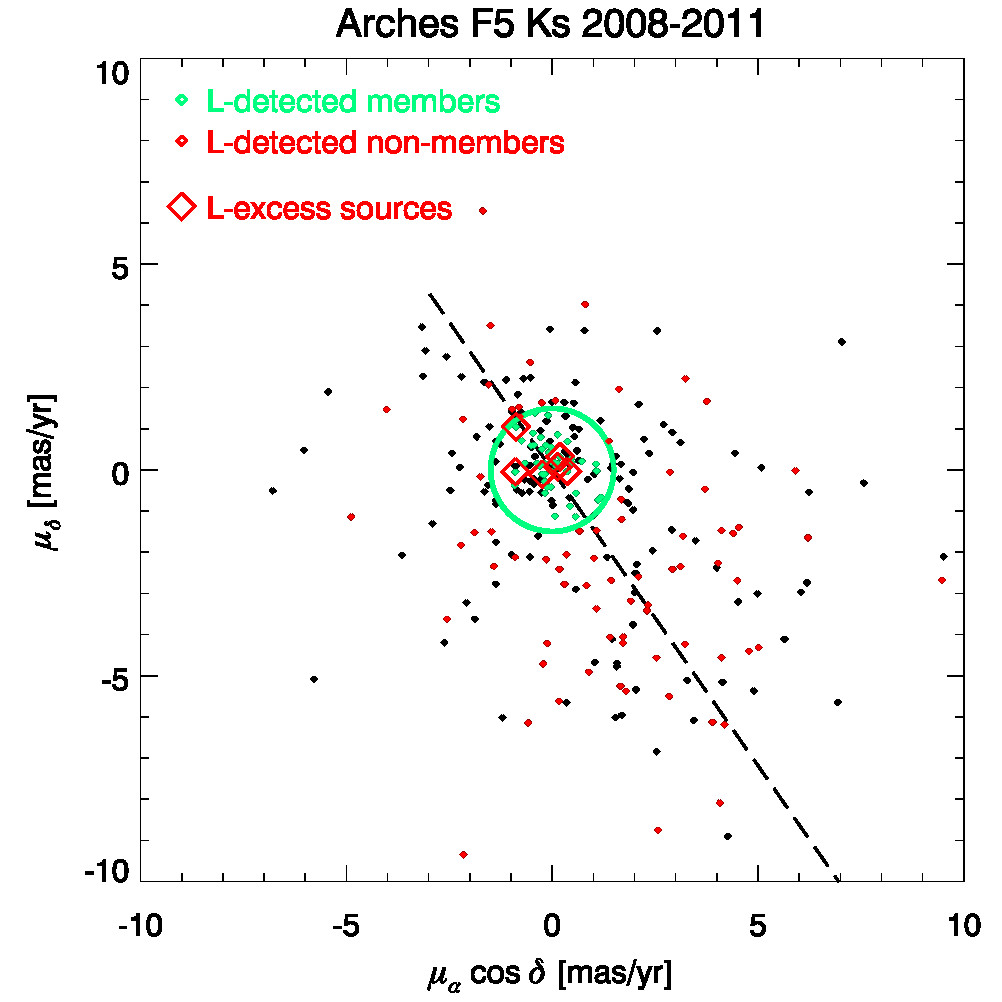}
\caption{\label{archpm_app} 
Proper motion diagrams of the outer Arches fields.
Labels are described in Fig.~\ref{quinpm_app}, and for further details
see Fig.~\ref{archpm}.
The larger scatter observed in the Arches proper motion diagrams  compared
to the Quintuplet proper motions is a consequence of the shallower Arches NACO 
photometry and mitigated adaptive optics performance. The circles indicate the membership 
selection criterion as derived from the central field in Fig.~\ref{pmhists}.}
\end{figure*}

\clearpage

\section{Colour-magnitude and colour-colour diagrams of the outer cluster fields}

The foreground extinction towards the Galactic centre is patchy and varies 
widely with position. Each of the Arches and Quintuplet fields feature
varying levels of foreground extinction, such that a single extinction 
value could not be used for the selection of the main-sequence reference
samples. The colour selection was adapted visually in each field to include
the proper motion members along the observed cluster main sequence
and exclude the red clump population starting at $J\sim 20$ mag.

\begin{figure*}[h]
\begin{minipage}{16.8cm}
\includegraphics[width=8.2cm]{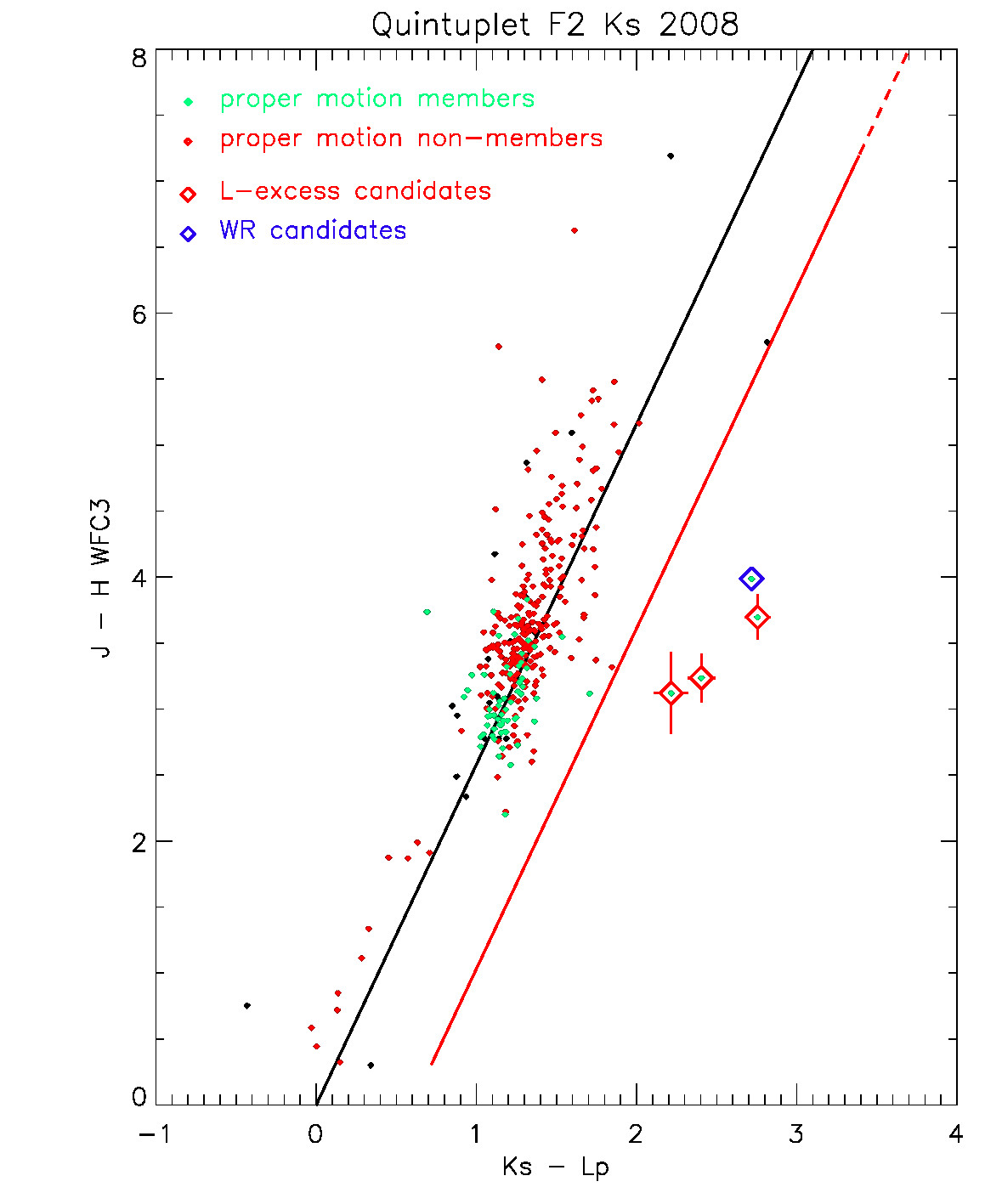}
\includegraphics[width=8.2cm]{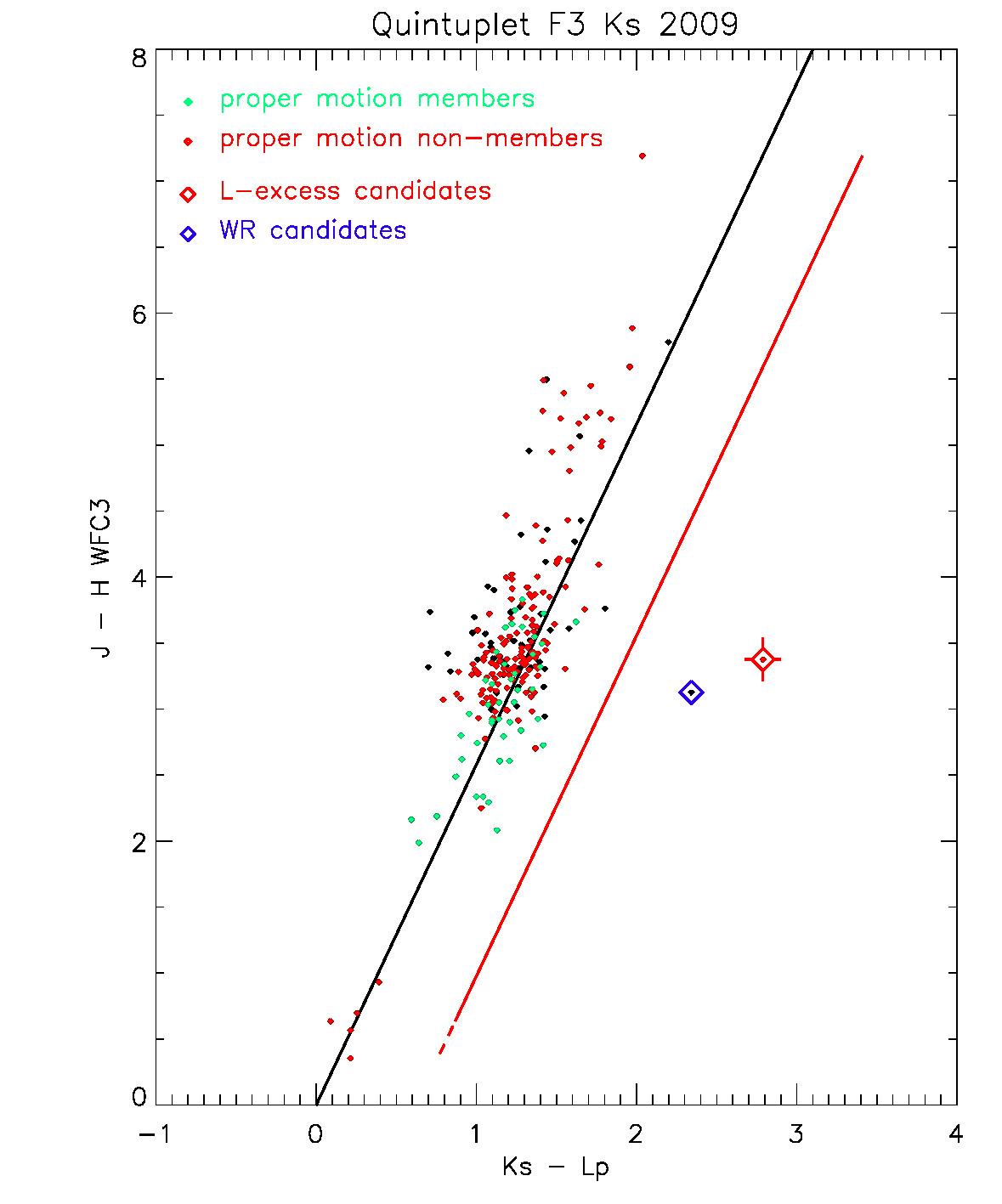}
\end{minipage}
\includegraphics[width=8.2cm]{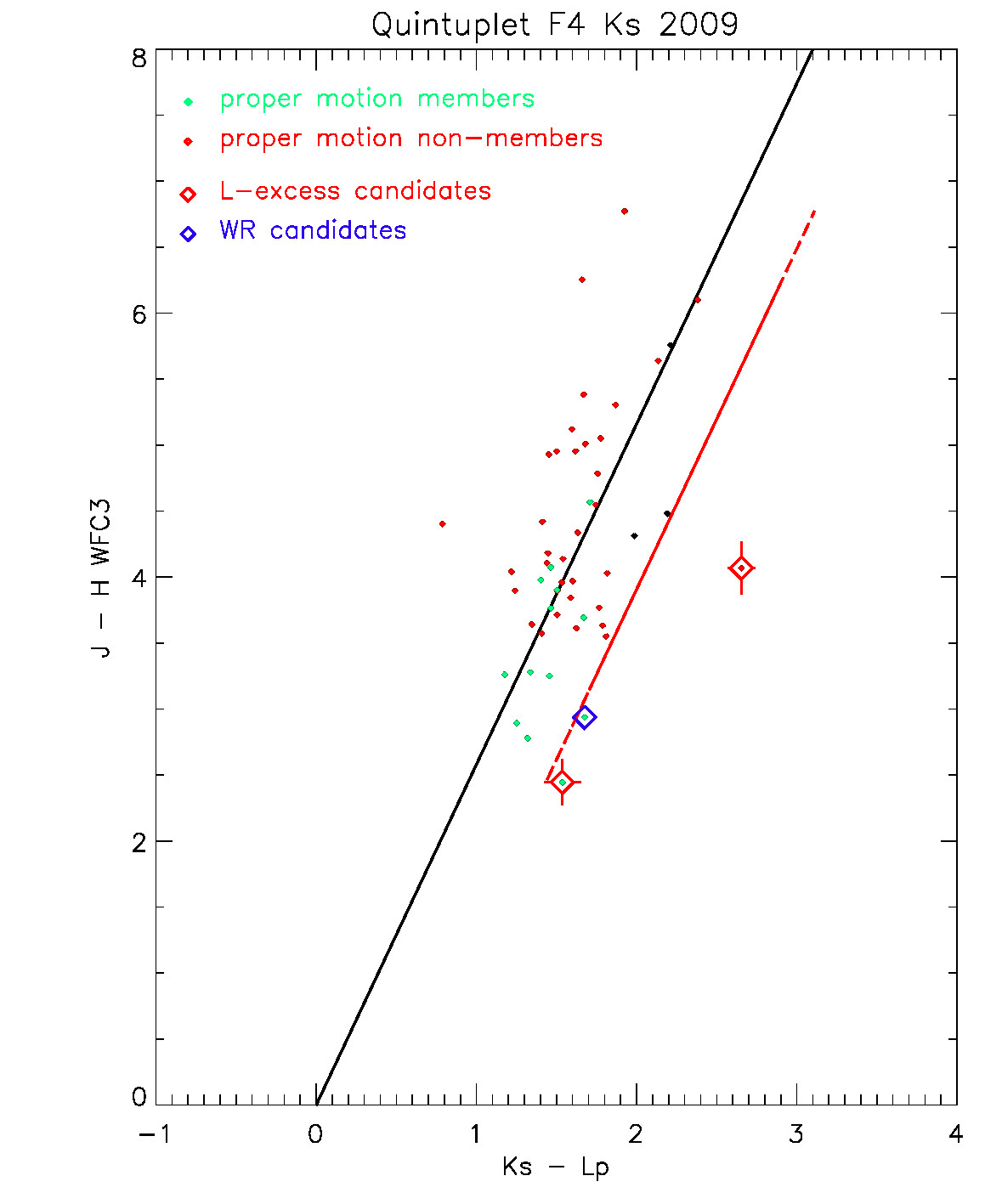}
\includegraphics[width=8.2cm]{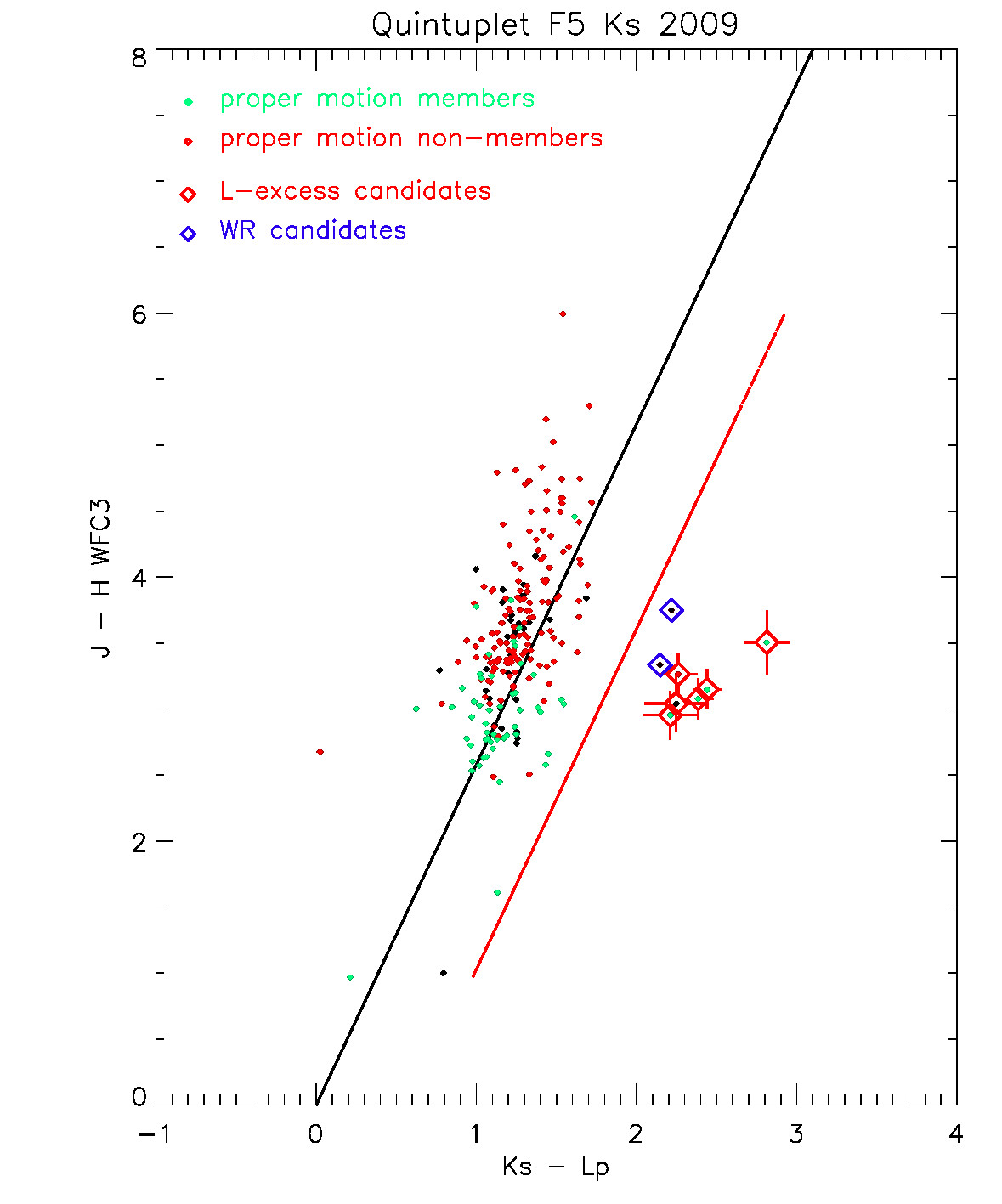}
\caption{\label{ccds_app} 
$J-H, K_s-L'$ colour-colour diagrams of the Quintuplet outer fields.
Proper motion members are shown in green, while non-members are 
shown in red. The black line denotes the extinction vector 
(Nishiyama et al.~2009). The red solid line shifted parallel to the 
reddening vector marks the 3$\sigma$ selection criterion for $L$-band 
excess sources (red diamonds). Wolf-Rayet candidates ($K_s<12$ mag) 
are marked in blue, while fainter excess sources marked as red diamonds 
are candidates for circumstellar disc emission.  
The $J-H, K_s-L'$ diagram of the Quintuplet central field 
can be found in Fig.~\ref{ccds}.}
\end{figure*}

\begin{figure*}
\begin{minipage}{16.8cm}
\includegraphics[width=8.2cm]{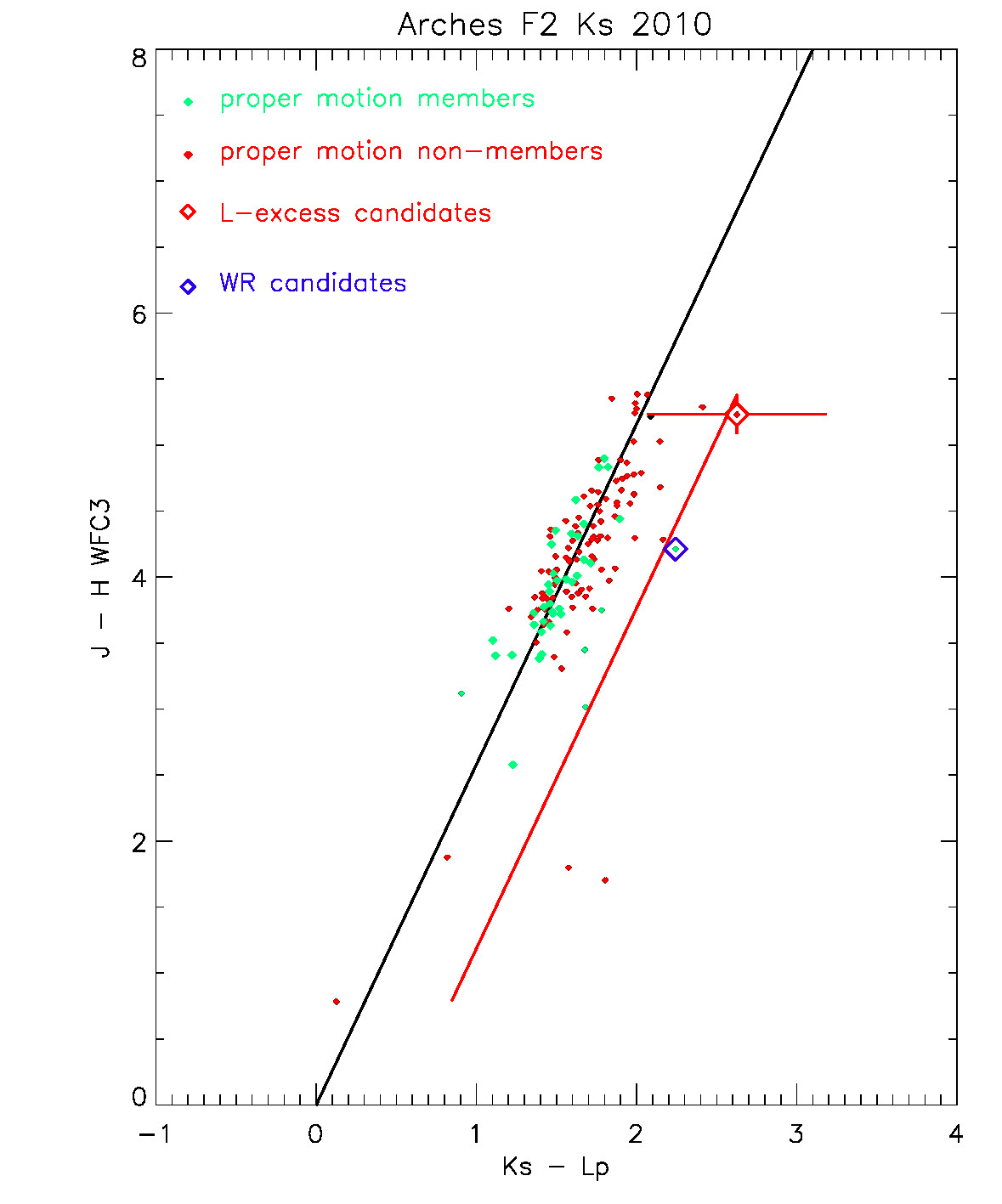}
\includegraphics[width=8.2cm]{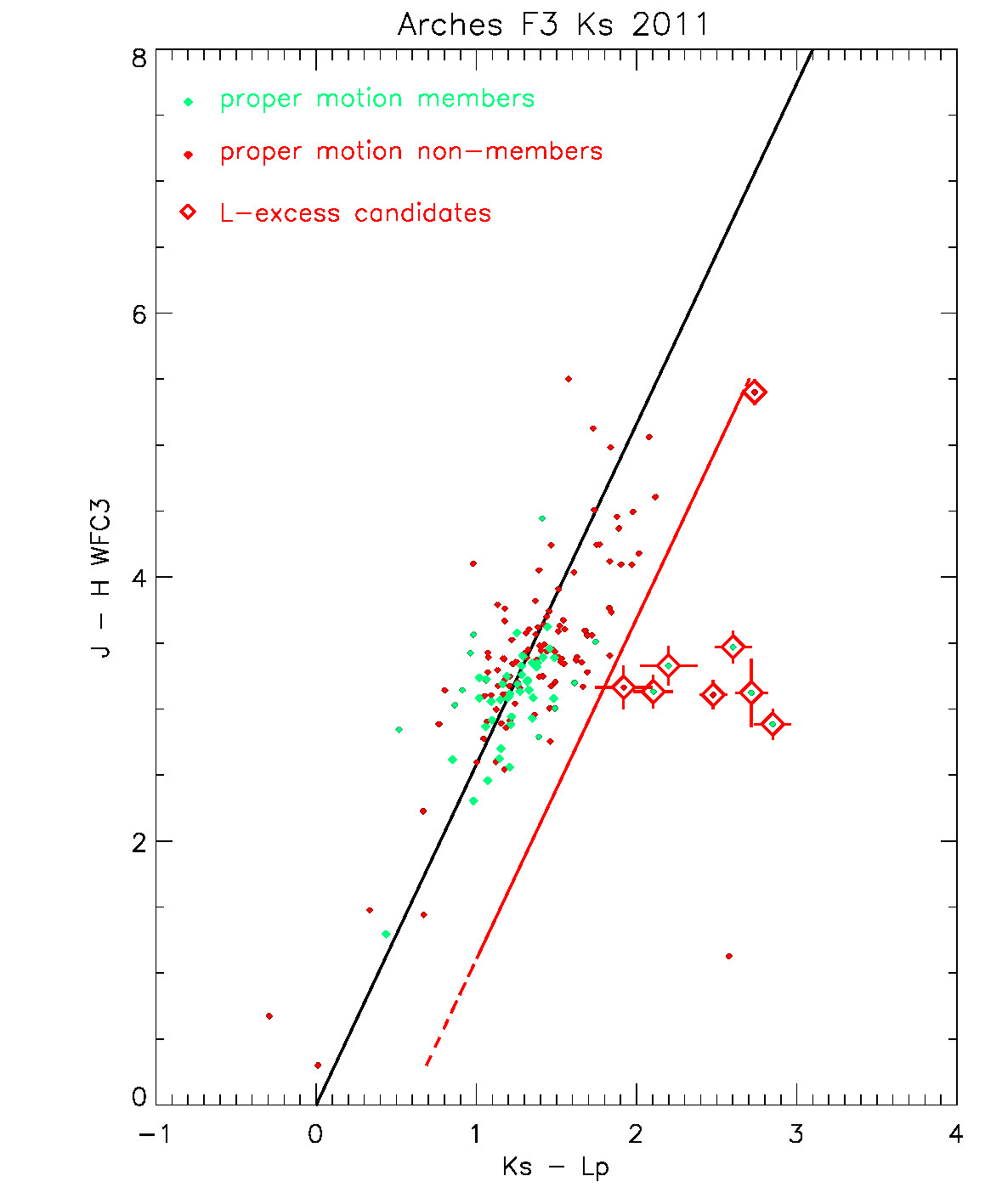}
\end{minipage}
\includegraphics[width=8.2cm]{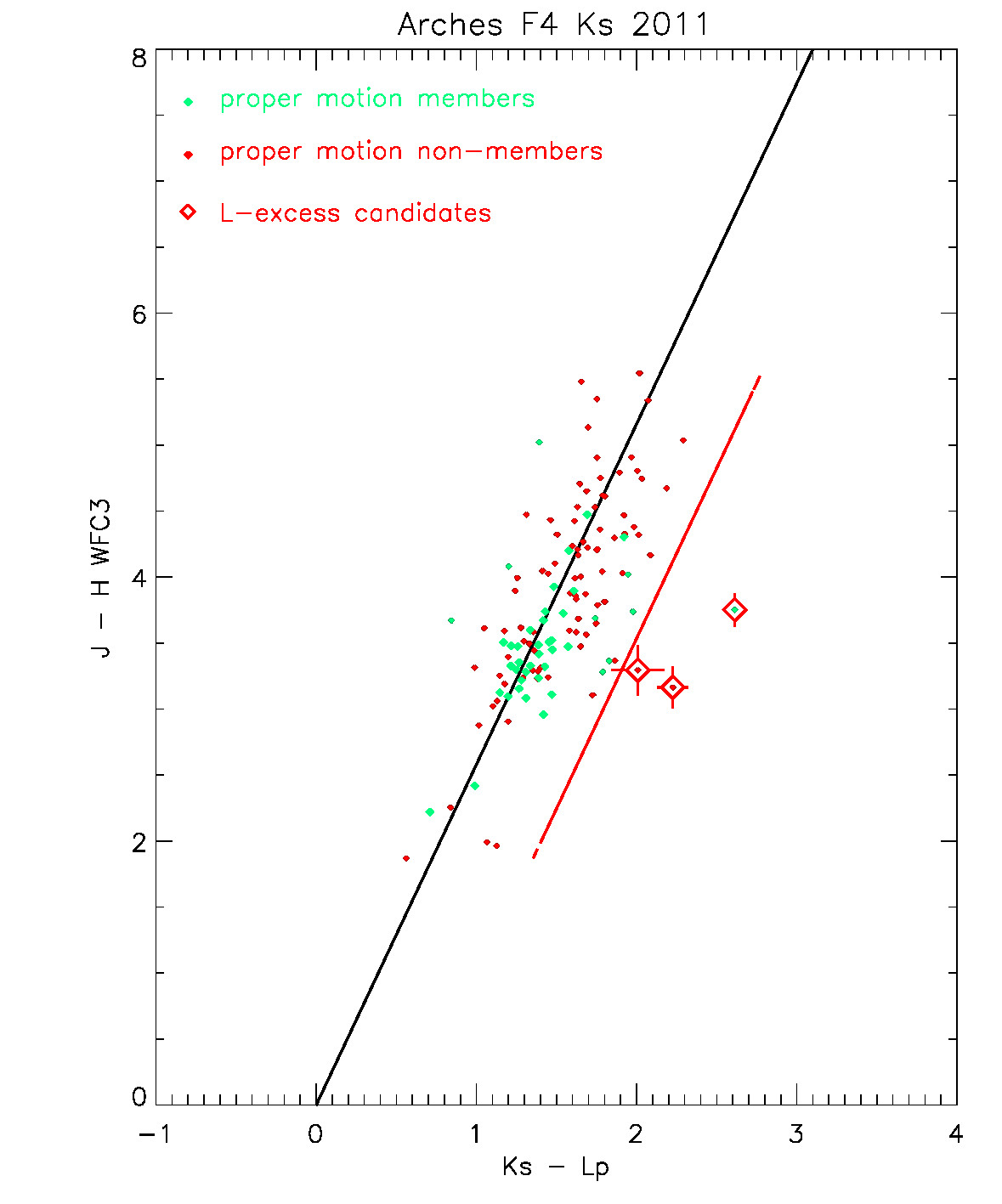}
\includegraphics[width=8.2cm]{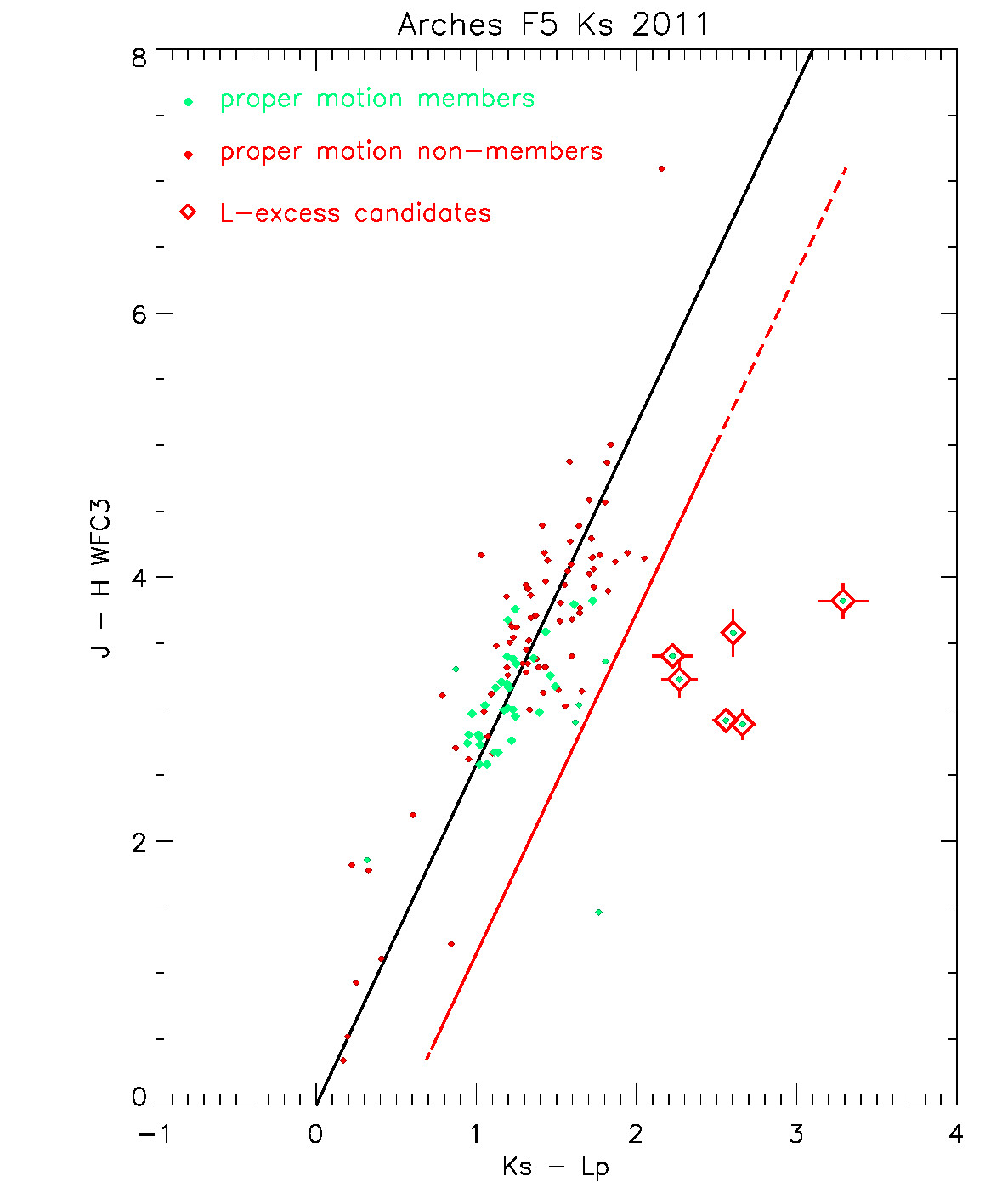}
\caption{\label{ccds_arch_app} 
$J-H, K_s-L'$ colour-colour diagrams of the Arches outer fields.
Labels are as described in Fig.~\ref{ccds_app}, and the $J-H, K_s-L'$ 
diagram of the Arches central field is shown in Fig.~\ref{ccds_arch}.}
\end{figure*}

\begin{figure*}
\begin{minipage}{18cm}
\includegraphics[width=9cm]{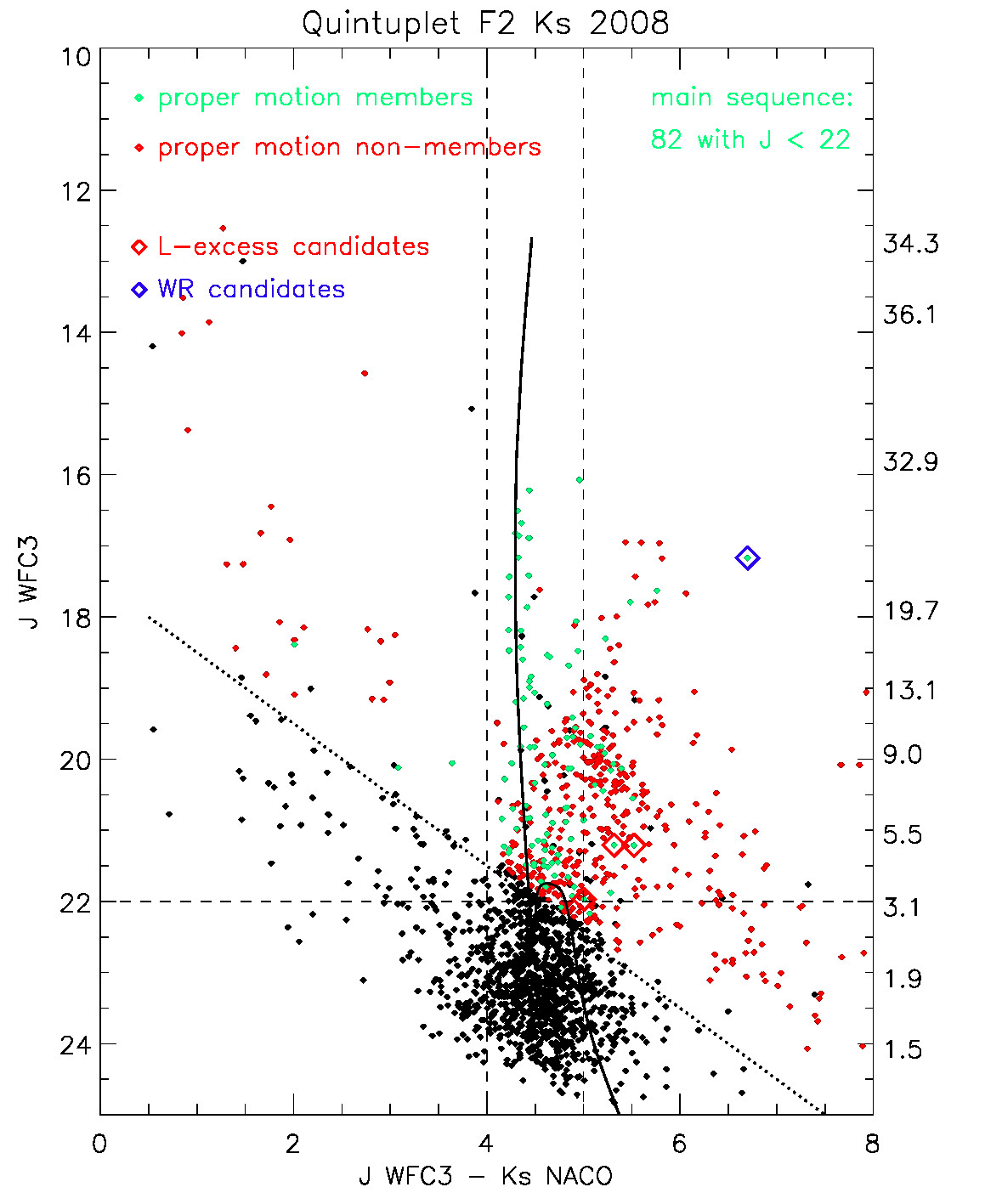}
\includegraphics[width=9cm]{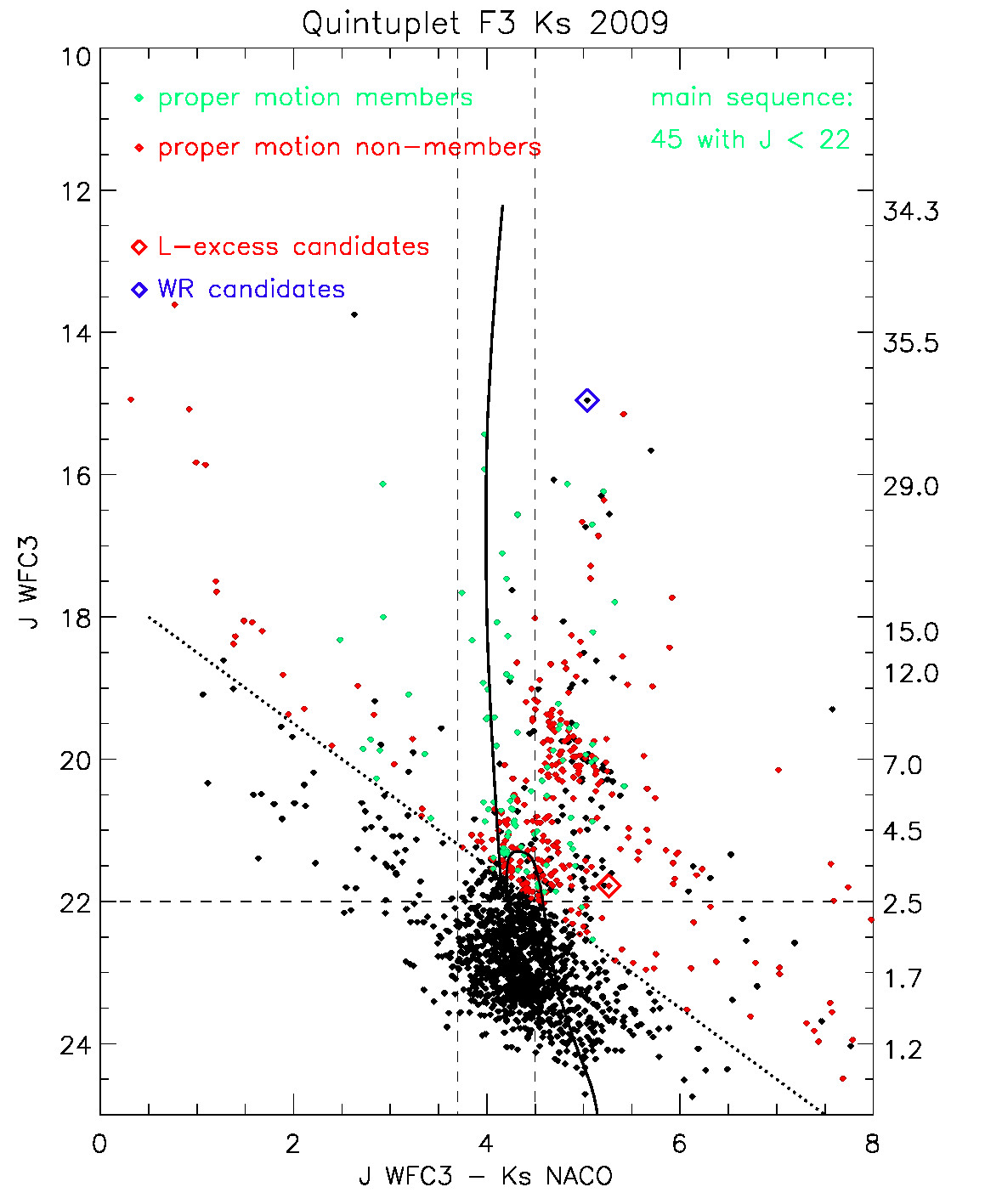}
\end{minipage}
\includegraphics[width=9cm]{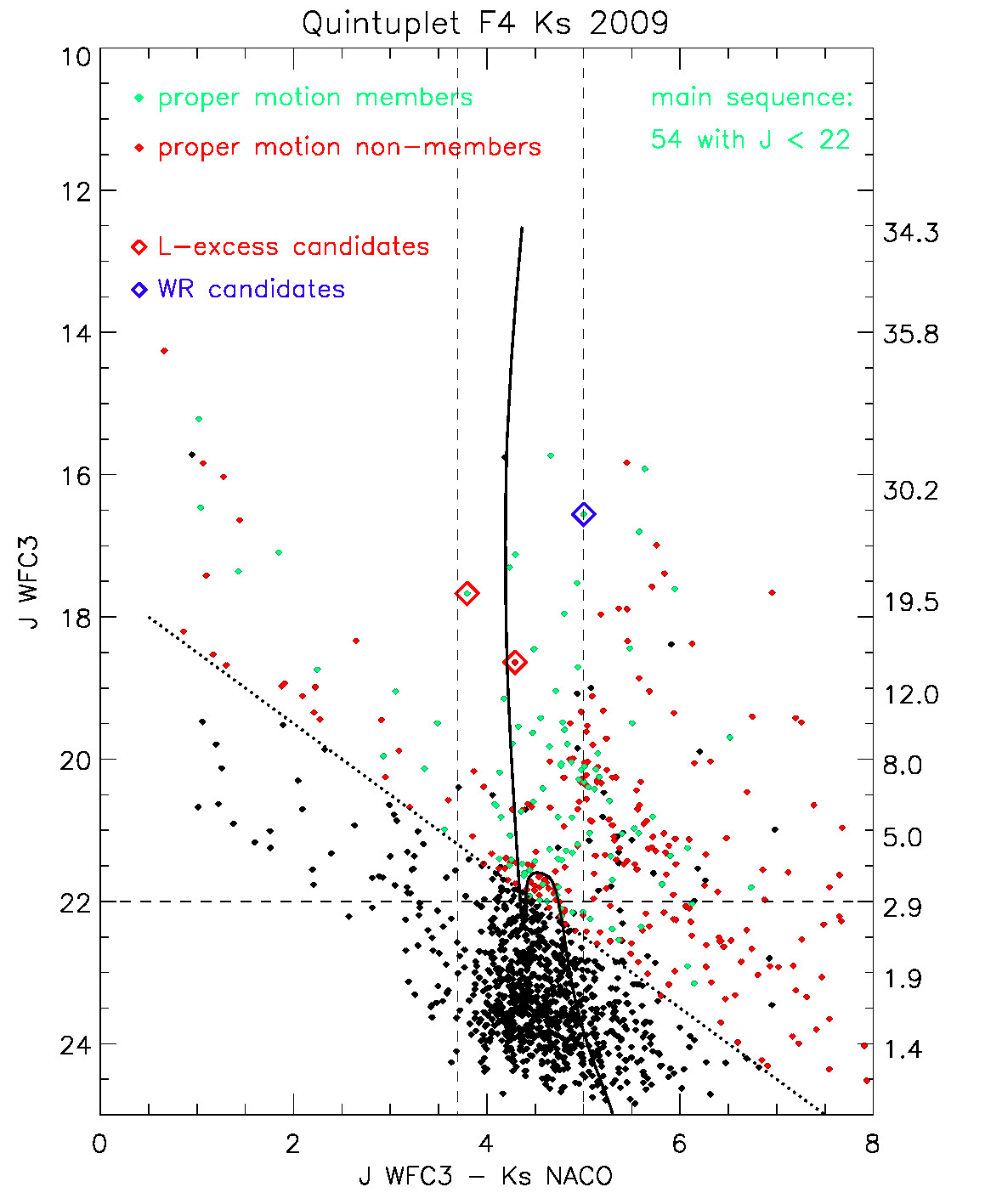}
\includegraphics[width=9cm]{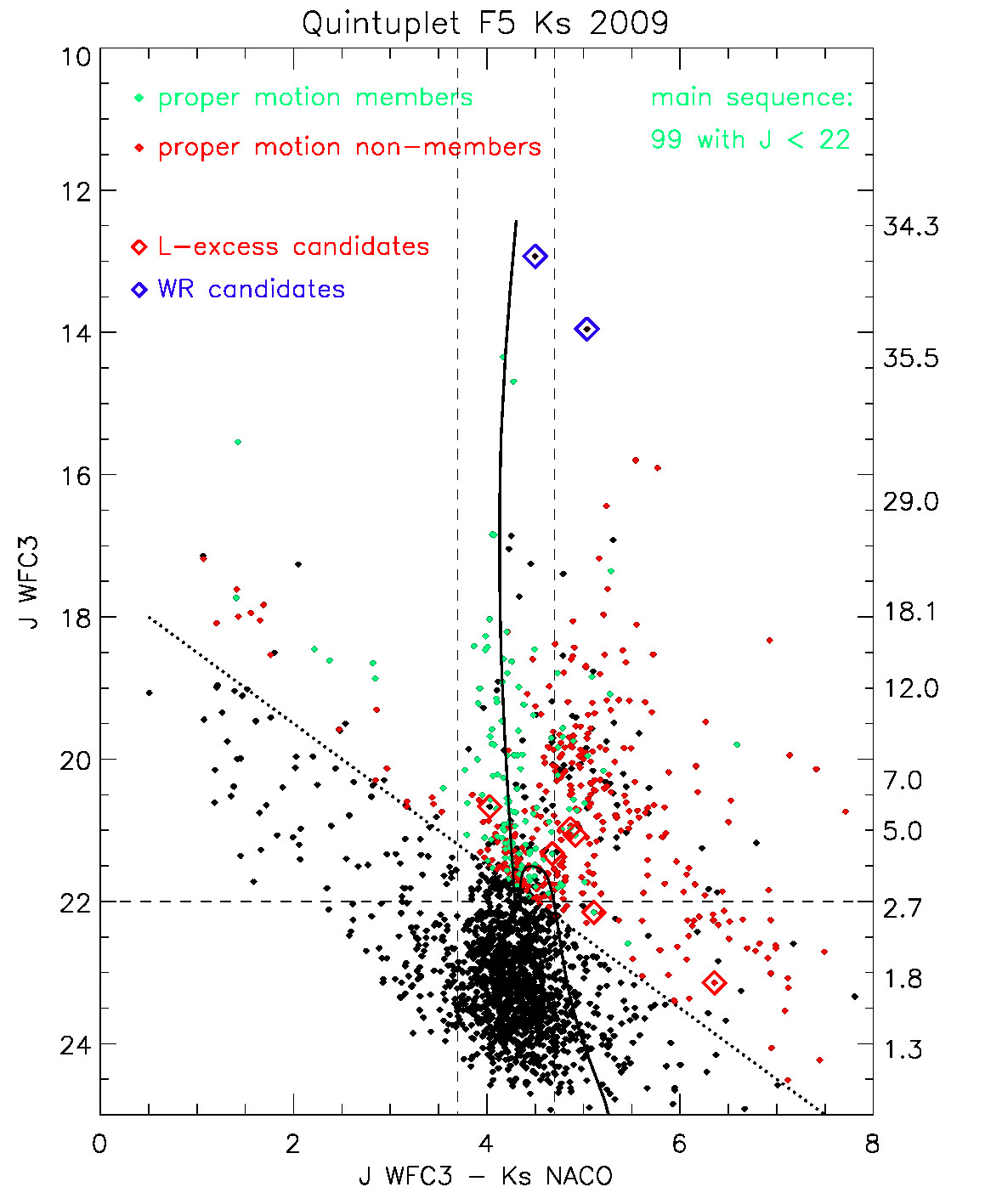}
\caption{\label{jjk_cmds_app} 
$J$, $J-K_s$ colour-magnitude diagrams of the Quintuplet outer fields.
Proper motion members are shown in green, while non-members are
shown in red. Disc and Wolf-Rayet candidates are also marked.
The diagonal overdensity at $J\sim 20$ mag marks the onset of the red clump.
The dotted lines mark the $K_s=17.5$ mag boundary for proper motion
member selection. 
The horizontal dashed line indicates the $J=22$ mag limit, above which 
the combined cluster disc fractions are derived. 
The main-sequence selection is indicated by vertical dashed lines.
The variation in main-sequence colour between fields
is caused by the locally varying extinction across the cluster area.
The $J$, $J-K_s$ diagram of the central Quintuplet field is
shown in Fig.~\ref{jjk_cmds}.}
\end{figure*}

\begin{figure*}
\begin{minipage}{18cm}
\includegraphics[width=9cm]{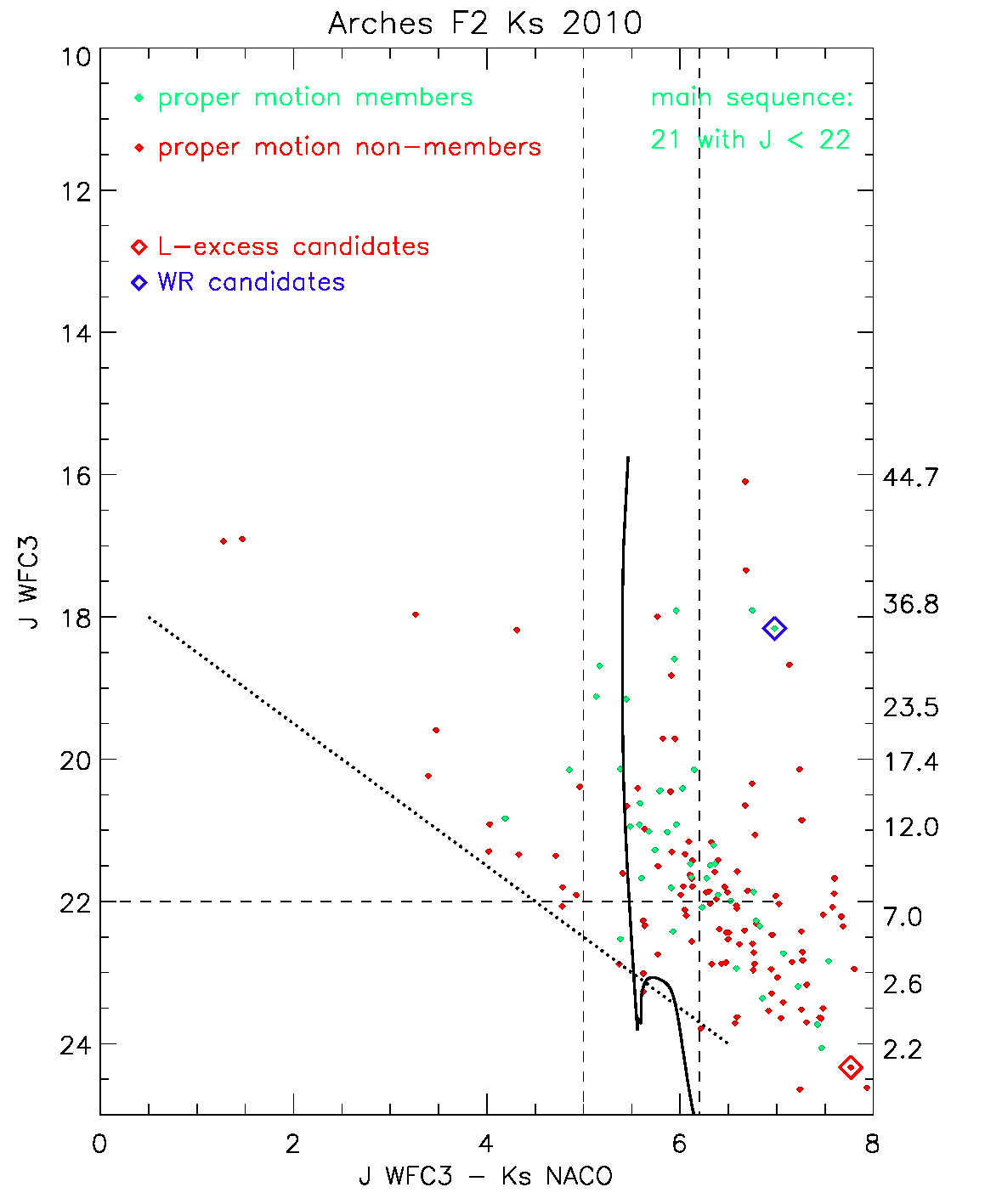}
\includegraphics[width=9cm]{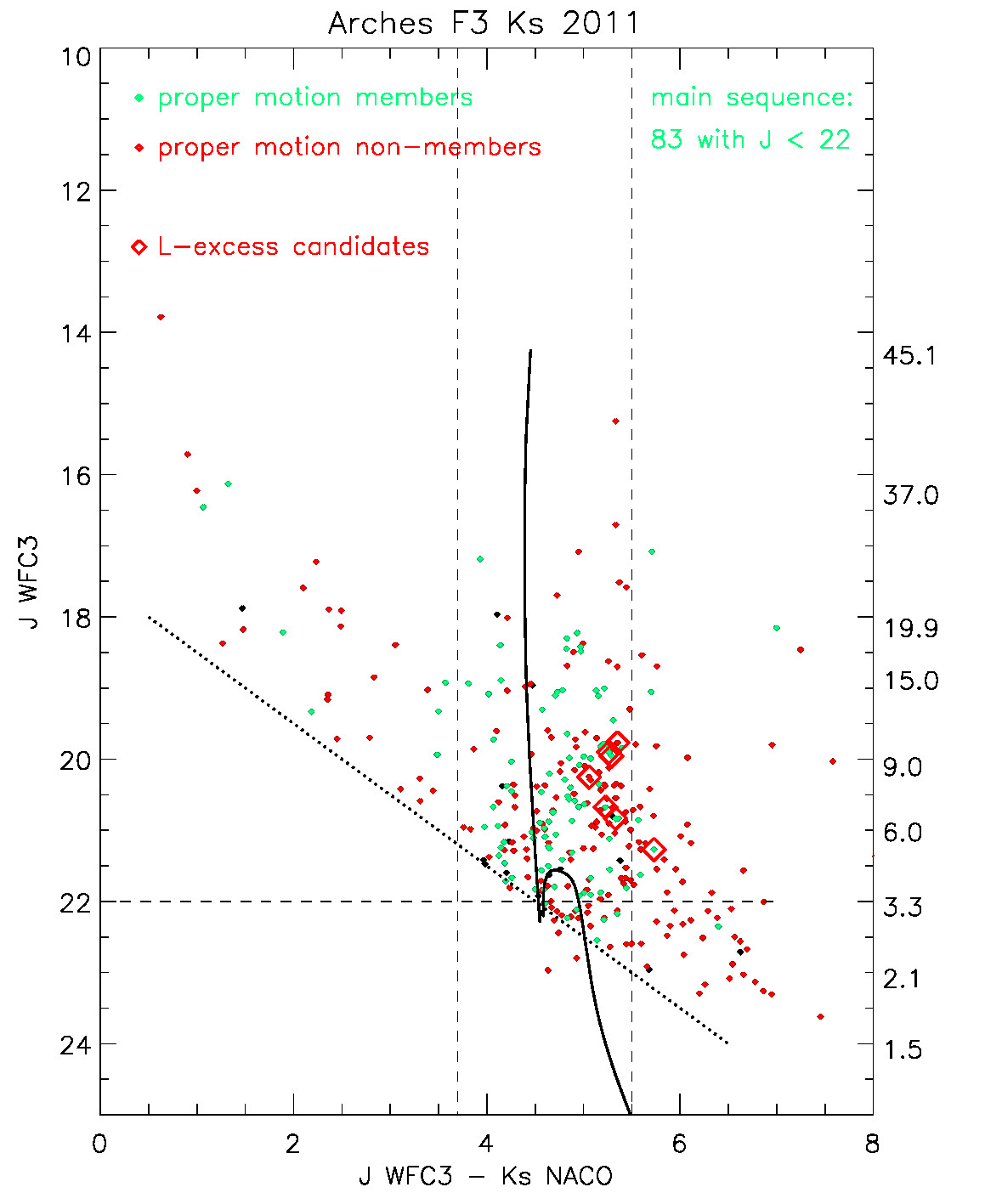}
\end{minipage}
\includegraphics[width=9cm]{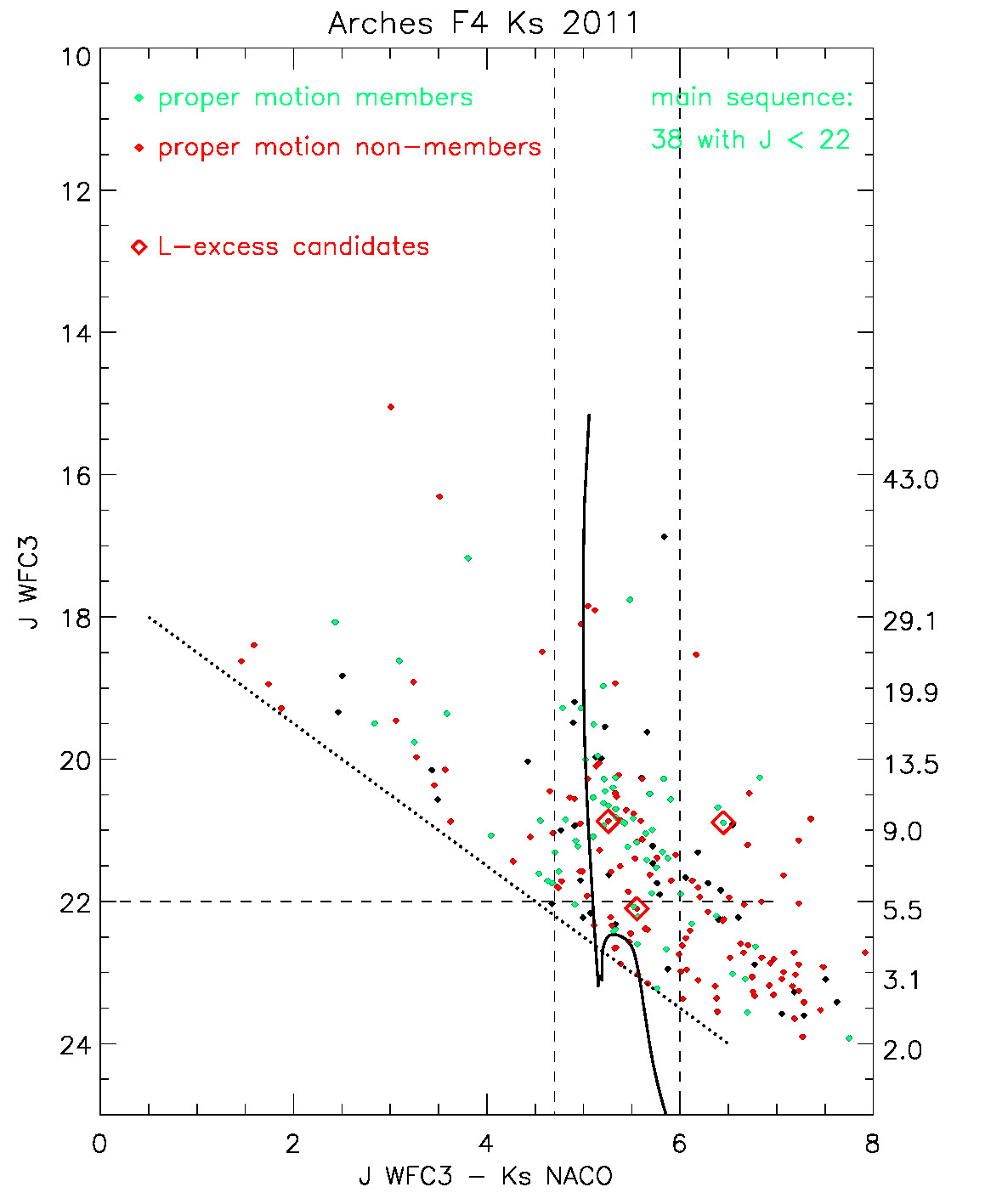}
\includegraphics[width=9cm]{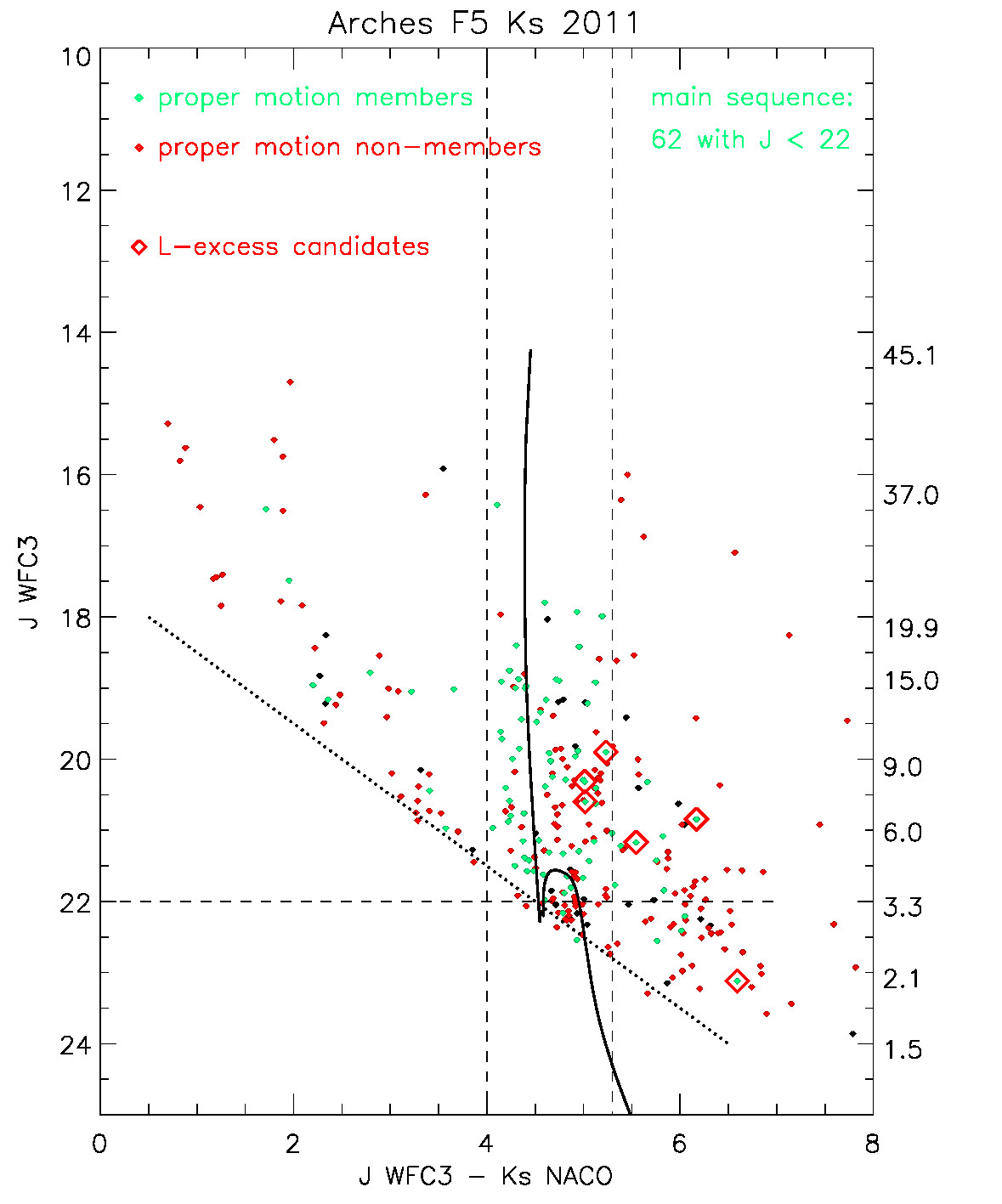}
\caption{\label{jjk_cmds_arch_app} 
$J$, $J-K_s$ colour-magnitude diagrams of the Arches outer fields.
Labels are as described in Fig.~\ref{jjk_cmds_app}, and the $J$, $J-K_s$ 
diagram of the Arches central field is shown in Fig.~\ref{jjk_cmds_arch}.
The variation in main-sequence colour between fields caused by the locally 
varying extinction is more pronounced than in the Quintuplet fields.
Notably, several $L'$-excess sources do not show excess emission at $K_s$.}
\end{figure*}

\begin{figure*}
\begin{minipage}{18cm}
\includegraphics[width=9cm]{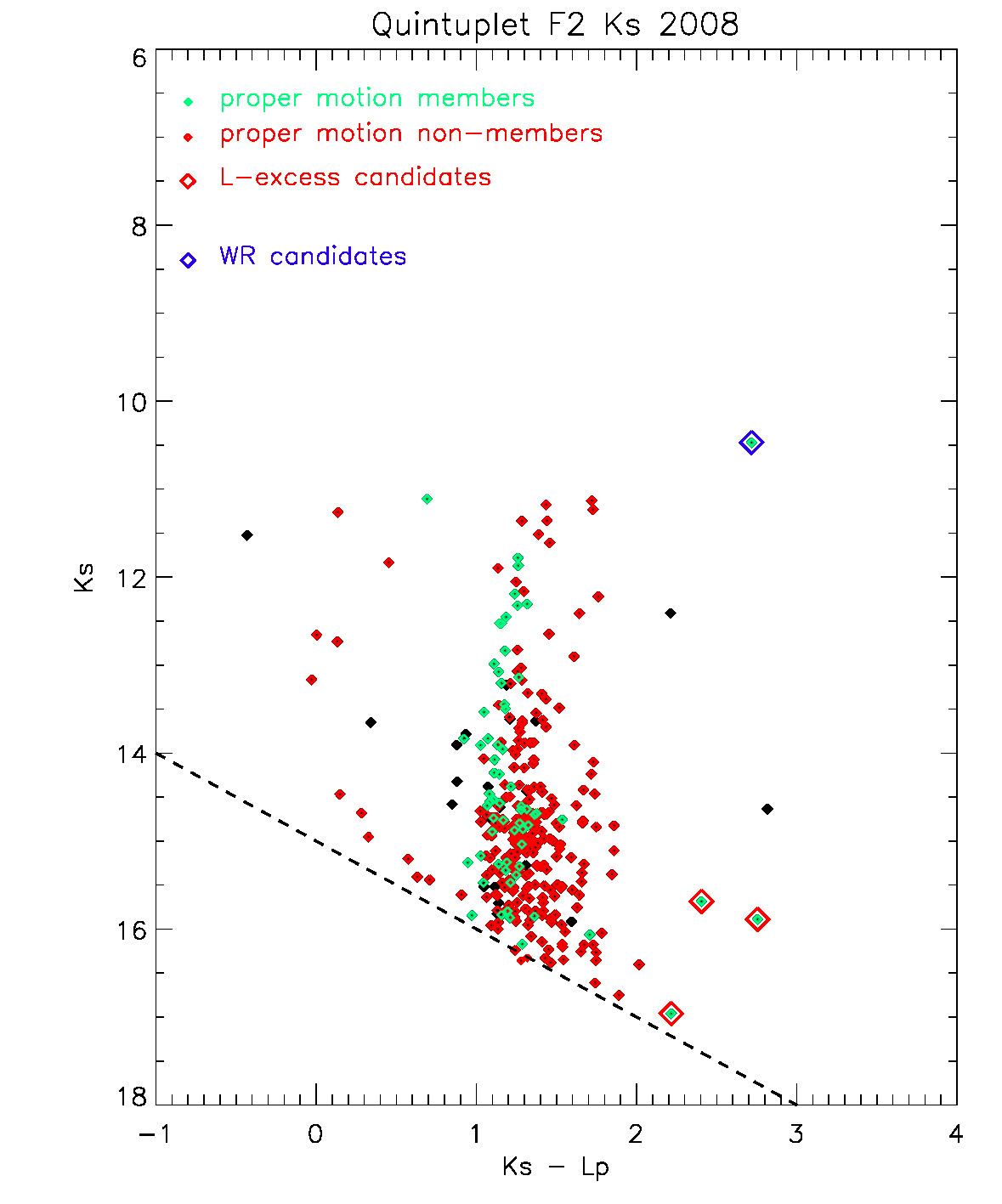}
\includegraphics[width=9cm]{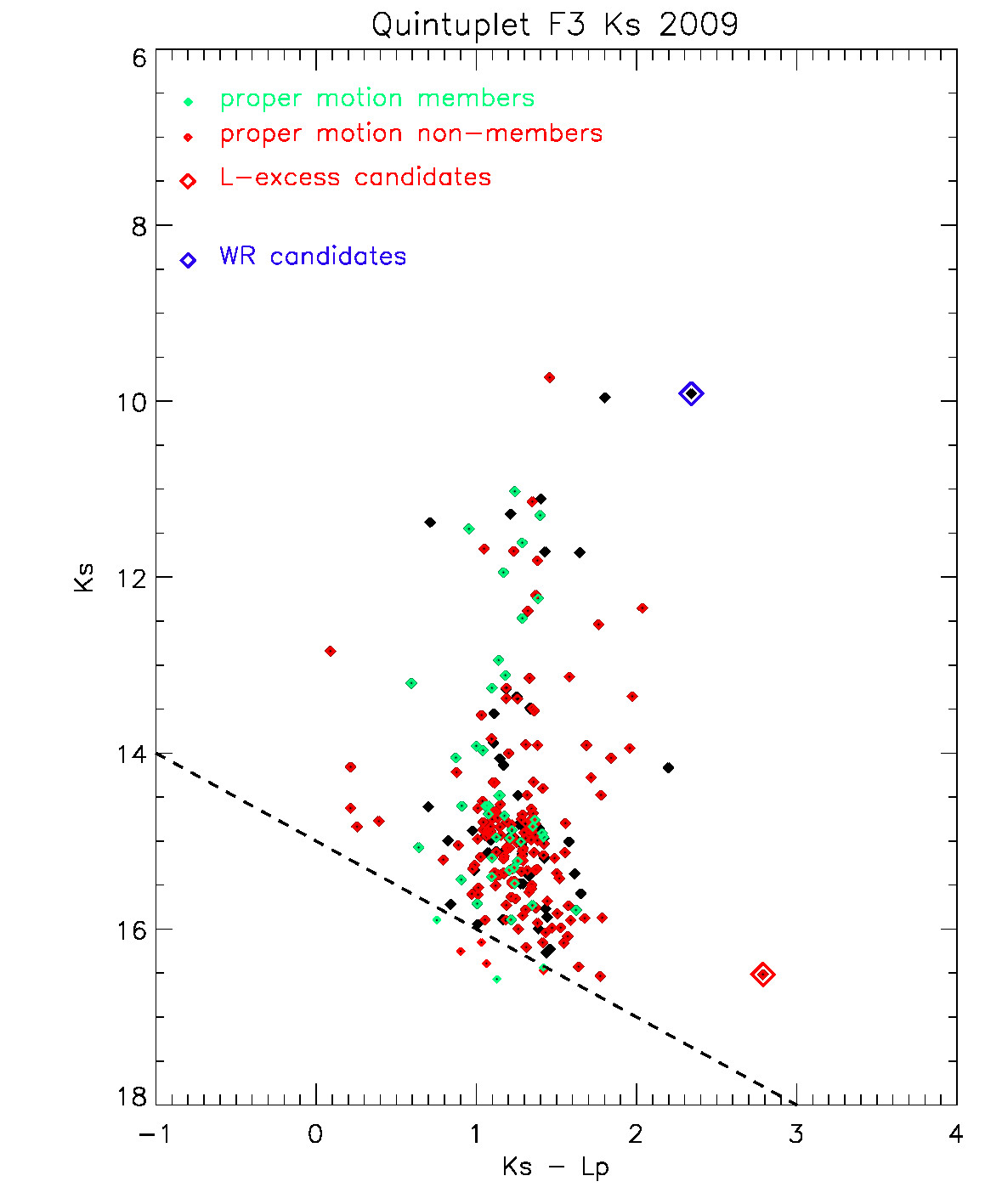}
\end{minipage}
\includegraphics[width=9cm]{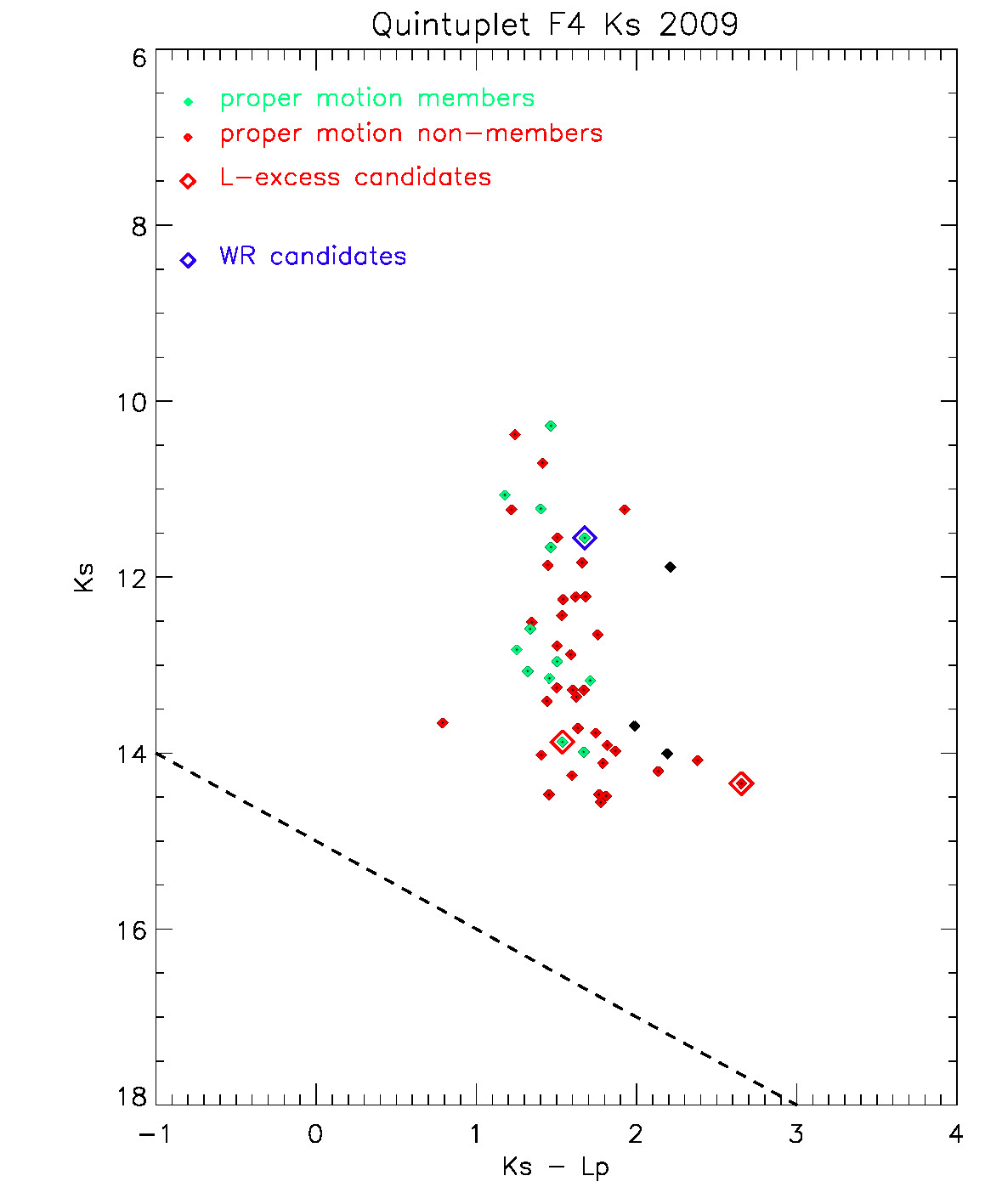}
\includegraphics[width=9cm]{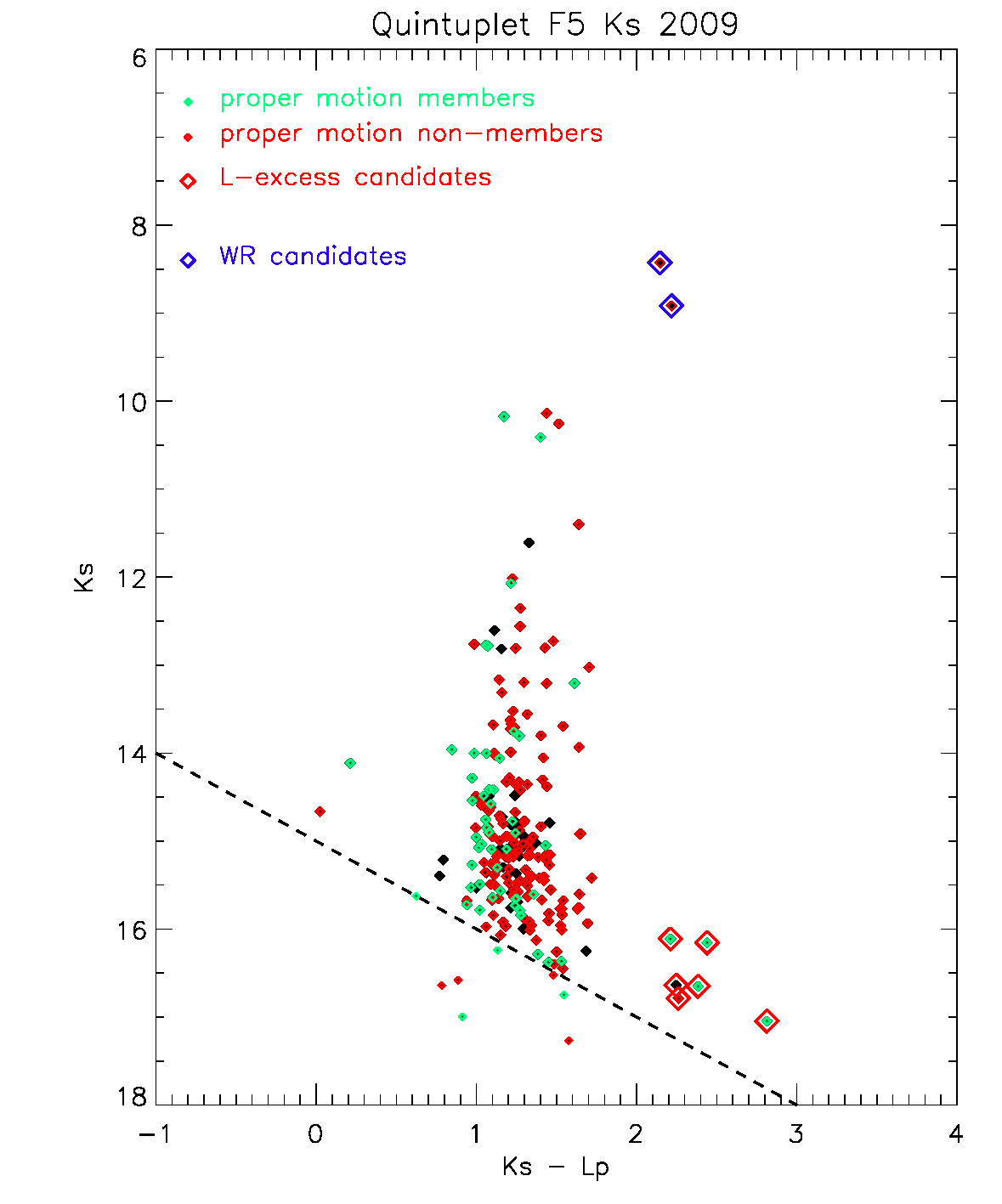}
\caption{\label{kkl_cmds_app} 
$K_s$, $K_s-L'$ colour-magnitude diagrams of the Quintuplet outer fields.
Labels are as in the previous figures, and the $K_s$, $K_s-L'$ diagram
of the central Quintuplet field can be found in Fig.~\ref{kkl_cmds}.
A distinct main sequence of cluster members can be seen in Field 2 (top left panel), 
while in all other fields the proper motion information is required to exhibit
the rare cluster stars at these increasing radii.}
\end{figure*}

\begin{figure*}
\begin{minipage}{18cm}
\includegraphics[width=9cm]{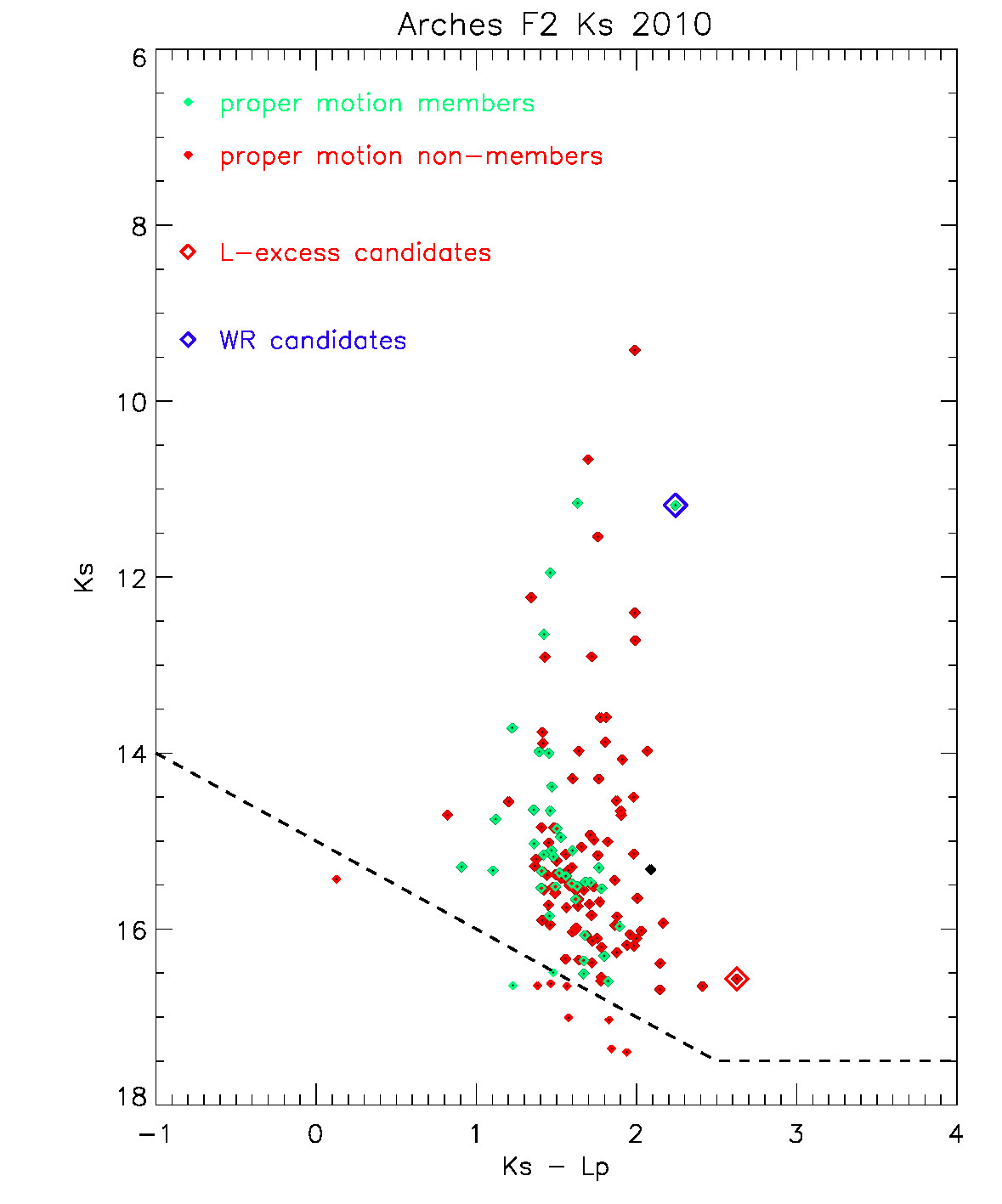}
\includegraphics[width=9cm]{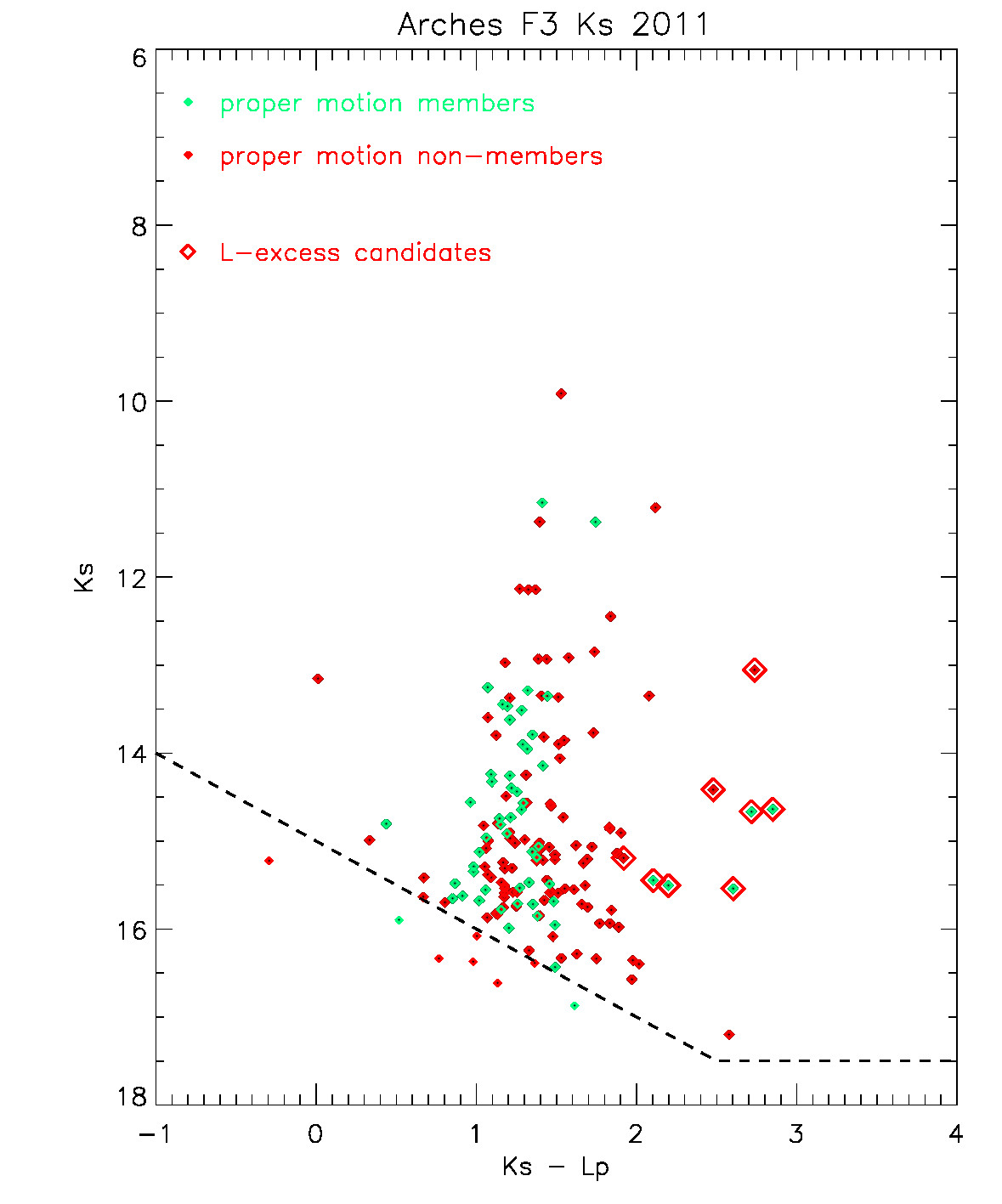}
\end{minipage}
\includegraphics[width=9cm]{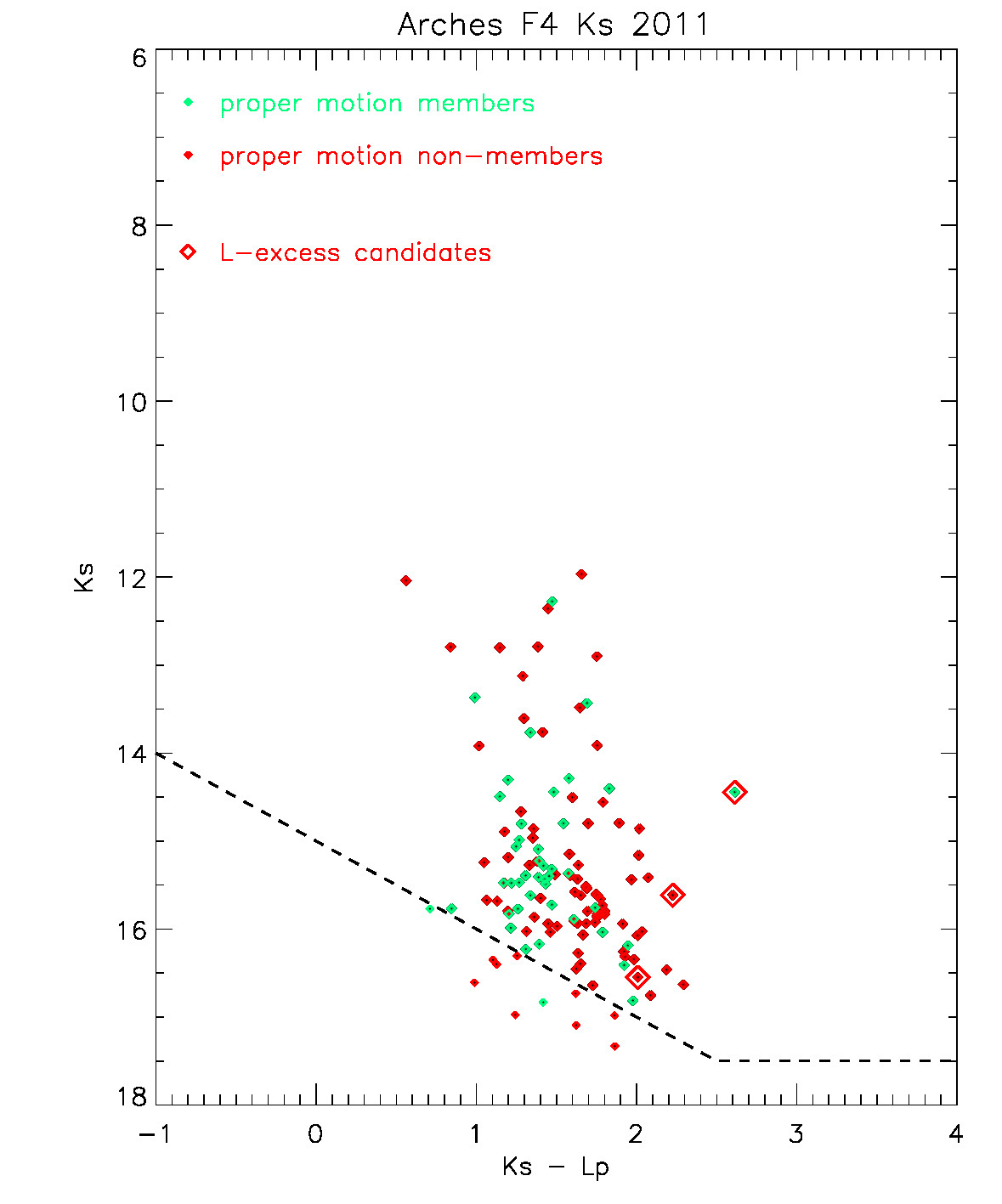}
\includegraphics[width=9cm]{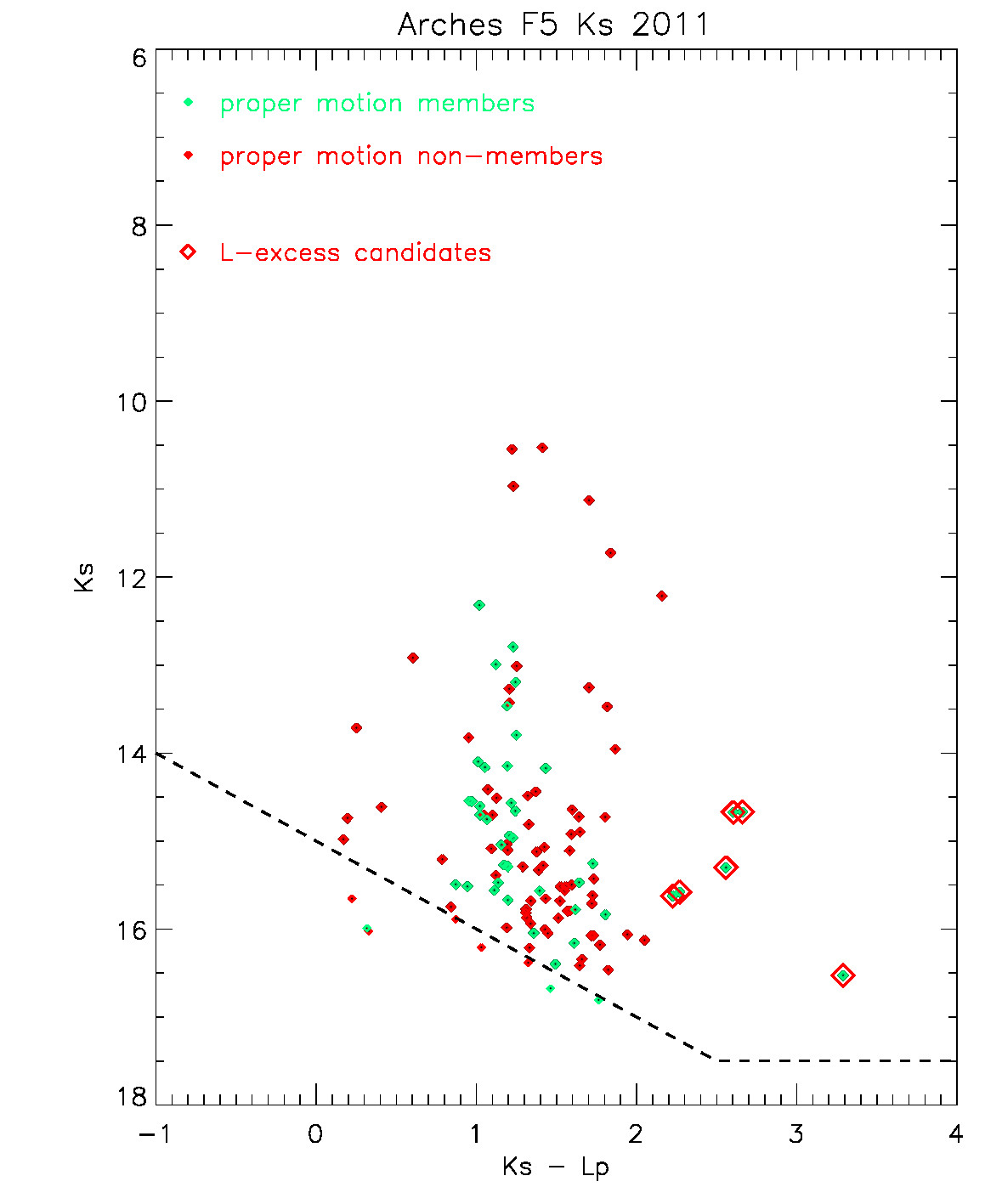}
\caption{\label{kkl_cmds_arch_app} 
$K_s$, $K_s-L'$ colour-magnitude diagrams of the Arches outer fields.
Labels are as in the previous figures, and the $K_s$, $K_s-L'$ diagram
of the central Arches field can be found in Fig.~\ref{kkl_cmds_arch}.
While a pronounced main sequence is observed in the central field,
in all outer fields the proper motion membership provides the best evidence 
for the distinction between cluster and field samples.}
\end{figure*}

\clearpage

\section{Variable stars}

From our multi-epoch $K_s$ proper motion campaign, we obtain a list of 
candidates for variable sources. Stars are defined as variable if the
standard deviation of their magnitudes across three epochs deviates 
by more than 3$\sigma$ from the median of the standard deviations of all
stars in their respective magnitude bin. The expected (median)
standard deviation is calculated in bins of $\Delta K_s = 1$ mag 
to account for the increase in photometric error towards fainter stars.
In the case that only two epochs are available, the magnitude difference 
between epoch 1 and epoch 2 has to exceed 3 times the standard 
deviation of the magnitude differences of all stars in the respective 
magnitude bin of a variable candidate. With the exception of one prominent
source in Arches Field 5, only stars brighter than $K_s < 16$ mag are searched
for variability, as photometric uncertainties in the crowded cluster fields
trigger many false detections beyond this limit. Two additional constraints are 
imposed to ensure that variable candidates are not affected by 
positional and/or PSF fitting uncertainties. The absolute magnitude 
difference or standard deviation for a star to be a variable candidate
has to exceed $\Delta K_s > 0.1$ mag, and the difference in the source
position in both the x and y coordinates has to be less than 0.5 pixel
after the geometric transformation to the reference epoch. By imposing 
this astrometric consistency, incorrect matches with very close neighbouring
stars are avoided. Finally, all variable candidates were inspected visually
in all $K_s$ epoch images to ensure that sources are not affected by 
background noise fluctuations, corrupt PSF cores, edge effects, and 
especially that their photometry is not compromised by the haloes of bright 
neighbours. 
As only two epochs were available in all Quintuplet fields except Field 2, no variable 
candidates could be identified unambiguously in the Quintuplet cluster fields.
The photometry of all variable stars identified in the Arches cluster
is summarised in Table \ref{vartab}. 


\begin{table*}[h]
\centering
\tiny
\caption{\label{vartab} Variable candidates in the Arches cluster}
\begin{tabular}{lccccccccccccccccc}
\hline
Field & dRA & dDEC & $J$ & $\sigma_J$ & $H$ & $\sigma_H$ & epoch 1 & $K_{s,1}$ & $\sigma_{K_s,1}$  & epoch 2 & $K_{s,2}$ & $\sigma_{K_s,2}$ & epoch 3 & $K_{s,3}$ & $\sigma_{K_s,3}$ & $L'$ & $\sigma_{L'}$ \\
\hline
1 & -8.95 & 14.97 & 18.53 & 0.05 & 15.00 & 0.01 & 2002 & 12.49 & 0.01 & 2010 & 12.22 & 0.01 & 2011 & 12.11 & 0.01 & -- & -- \\
1 &  0.72 &  2.34 &  --  &  -- &  -- & -- & 2002 & 12.51 & 0.01 & 2010 & 12.81 & 0.01 & 2011 & 12.83 & 0.00 & -- & -- \\
1 & -10.74 &  5.64 & 19.84 & 0.11 & 15.56 & 0.01 & 2008 & 13.52 & 0.01 & 2010 & 13.40 & 0.01 & 2011 & 13.23 & 0.01 & 11.93 & 0.01 \\ 
1 &  4.48 & -2.81 &  --  &  -- &  -- & -- & 2002 & 14.67 & 0.01 & 2010 & 15.02 & 0.05 & 2011 & 15.03 & 0.01 & -- & -- \\
1 &  7.71 &  3.38 & 19.68 & 0.01 & 17.36 & 0.01 & 2002 & 15.78 & 0.01 & 2010 & 15.06 & 0.08 & 2011 & 15.05 & 0.01 & 13.56 & 0.07 \\
\hline
2 & -20.23 & -21.86 &  --  &  -- &  -- & -- & 2008 & 15.01 & 0.01 & 2010 & 15.39 & 0.01 & -- & -- & -- & -- & -- \\
\hline
3 & 13.52 & -0.39 & 19.05 & 0.01 & 16.14 & 0.01 & 2008 & 14.22 & 0.05 & 2011 & 14.22 & 0.01 & 2012 & 14.04 & 0.01 & -- & -- \\
3 & 41.10 &  1.93 &   --  &  -- &  -- & -- & 2008 & 15.12 & 0.02 & 2011 & 15.19 & 0.01 & 2012 & 15.40 & 0.06 & -- & -- \\
\hline
4 & -21.05 & 31.06 & 21.39 & 0.06 & 17.95 & 0.08 & 2008 & 15.62 & 0.01 & 2011 & 15.86 & 0.01 & 2012 & 15.94 & 0.02 & 14.50 & 0.07 \\
\hline
5 &  5.36 & 29.45 & -- &  --  &  --  &  --  & 2008 & 14.22 & 0.01 & 2011 & 14.16 & 0.01 & 2012 & 14.68 & 0.02 & -- & -- \\
5 & 10.50 & 20.34 & 20.31 & 0.02 & 17.40 & 0.01 & 2008 & 14.82 & 0.01 & 2011 & 15.30 & 0.01 & 2012 & 15.12 & 0.02 & 12.44 & 0.01 \\
5 &  2.00 & 24.86 &  -- &  --  &  --  &  --  & 2008 & 15.95 & 0.09 & 2011 & 15.57 & 0.01 & 2012 & 15.48 & 0.07 & -- & -- \\
5 & 25.43 & 18.42 &  -- &  --  &  --  &  --  & 2008 & 17.74 & 0.08 & 2011 & 16.52 & 0.01 & 2012 & 16.60 & 0.07 & -- & -- \\
\hline
\end{tabular}
\end{table*}

\newpage

\section{$L$-band luminosity of the discs}
\label{disclum}

The $L$-band luminosity of the excess sources can be 
estimated for a central B2V star from the difference between 
the stellar brightness and the excess emission.

The 2.5 Myr Geneva isochrone yields an $L$-band brightness of 
$L(B2V) = -1.35$ mag for a $10\,M_\odot$ star. For our approximation 
here, we assume this $L$-band brightness to be close to the Vega
system, such that $L_{Vega} = 0$ mag, and
\\

$F_{B2V}/F_{Vega} = 10^{0.4\cdot 1.35} = 3.47$. \\
\\
Inserting the Vega flux of Tokunaga \& Vacca (2005) \\ 

$F(3.754\mu{\rm m})_{Vega} = 5.31\, 10^{-11} \frac{\rm W}{\rm m^2 \mu m}$
\\
\\
leads to
\\

$\Rightarrow\ F_{L'}(B2V) = 18.4\, 10^{-11} \frac{\rm W}{\rm m^2 \mu m}$ .
\\
\\
The absolute flux within the passband is given by the filter width 
$\Delta\lambda_{L'} = 0.62 \mu$m times the expected flux at the
central wavelength:
\\

\vspace*{2mm}
$\Rightarrow\ \lambda F_{L',abs}(B2V) = 11.4\, 10^{-11} \frac{\rm W}{\rm m^2}$ .
\\
\\
If the disc is about 1 mag brighter in $L'$ than the star,
then $\Delta L = 1$ mag implies a factor of 2.5 higher flux,
\\

$\lambda F_{L',disc} = 2.5 F_{L',abs}(B2V) 
\sim 3 10^{-10} \frac{\rm W}{\rm m^2} $,
\\
\\
which is already the flux we expect to measure within the passband
of $L'$ with filter width 0.62$\mu$m. In this estimate, no assumptions 
about the distance have been made, as all fluxes are derived for absolute
magnitudes only, hence at a standard distance of 10 pc.
This flux is used to derive an order of magnitude limit of 
the disc masses in Sect.~\ref{discmass}.

\section{Complete source lists}

The $JHK_s$ combined source lists for the Arches and Quintuplet clusters are made available
in the online version of the Journal, and in the CDS. Proper motion memberships are 
meant as indicator values for candidacy, and might change as further proper motion epochs 
become available. 

\clearpage


\begin{table*}[p]
\centering
 \rotatebox{90}{
 \begin{minipage}{\textheight}
\caption{\label{jhktab_quin} $JHK_s$ sources and $L'$ detections in the Quintuplet cluster}
 \end{minipage}}
\tiny
 \rotatebox{90}{
 \begin{minipage}{\textheight}
\tabcolsep1mm
\begin{tabular}{rcccrrrrrrrrrrrrcccccccccc}
\hline
  Seq & Field &  dRA   &   dDEC  &    J    & $\sigma_J$ & H & $\sigma_H$ & epoch1 & $K_s1$  & $\sigma_{Ks1}$ & epoch2 & $K_s2$ & $\sigma_{Ks2}$ & epoch3 & $K_s3$ & $\sigma_{Ks3}$ & $L'$ & $\sigma_{L}$ & $\mu_{\alpha cos\delta}$ & $\sigma_{\mu_{\alpha cos\delta}}$ & $\mu_\delta$ & $\sigma_{\mu_\delta}$ & mem & exc & $p_{clus}$ \\
      &       &  [``]  &   [``]  &  [mag]  &   [mag] &   [mag] &   [mag] & [year]  & [mag] &   [mag] & [year] &  [mag] & [mag] & [year] &[mag] & [mag] & [mag] & [mag] & [mas/yr] & [mas/yr] & [mas/yr] & [mas/yr] & & &  \\
\hline
\multicolumn{25}{c}{Field 1} \\
\hline
   1 &   1 &   3.255 &  -6.710 &  12.466 &   0.066 &  11.698 &   0.039 &2003.556 &  11.234 &   0.178 &2008.562 &  11.031 &   0.004 & -99.000 & -99.000 & -99.000 &  10.829 &   0.005 &  -1.321 &   0.557 &   6.373 &   0.720 &   0 &   0 & -99.000 \\
   2 &   1 &  -0.324 &   5.428 &  14.187 &   0.050 &  11.118 &   0.067 &2003.556 &   9.385 &   0.064 &2008.562 &   9.511 &   0.007 & -99.000 & -99.000 & -99.000 &   7.830 &   0.002 &  -3.090 &   0.283 &  -0.645 &   0.754 &   0 &   0 & -99.000 \\
... &&&&&&&&&&&&&&&&&&&&&&&&& \\
 723 &   1 &   0.000 &  -0.000 &  14.904 &   0.067 &  11.314 &   0.083 &2003.556 &   9.726 &   0.016 &2008.562 & -99.000 & -99.000 & -99.000 & -99.000 & -99.000 & -99.000 & -99.000 & -99.000 & -99.000 & -99.000 & -99.000 &  -1 & -99 & -99.000 \\
 724 &   1 &  -4.330 &   0.523 &  15.495 &   0.128 &  11.182 &   0.110 &2003.556 &   7.741 &   0.016 &2008.562 & -99.000 & -99.000 & -99.000 & -99.000 & -99.000 & -99.000 & -99.000 & -99.000 & -99.000 & -99.000 & -99.000 &  -1 & -99 & -99.000 \\
... &&&&&&&&&&&&&&&&&&&&&&&&& \\
\hline
\multicolumn{25}{c}{Field 2} \\
\hline
3133 &   2 &   6.599 & -26.482 &  16.218 &   0.069 &  13.084 &   0.074 &2008.562 &  11.780 &   0.001 &2011.718 &  11.873 &   0.006 &2012.452 &  11.773 &   0.022 &  10.521 &   0.005 &   0.229 &   0.359 &   0.641 &   0.403 &   1 &   0 &   0.623 \\
3134 &   2 &   8.071 & -26.940 &  16.506 &   0.058 &  13.580 &   0.064 &2008.562 &  12.189 &   0.002 &2011.718 &  12.270 &   0.003 &2012.452 &  12.167 &   0.001 &  10.949 &   0.004 &  -0.208 &   0.363 &   0.371 &   0.408 &   1 &   0 &   0.785 \\
... &&&&&&&&&&&&&&&&&&&&&&&&& \\
\hline
\multicolumn{25}{c}{Field 3} \\
\hline
4893 &   3 &  37.289 & -42.957 &  15.922 &   0.075 &  13.128 &   0.068 &2008.649 &  11.855 &   0.018 &2009.266 &  11.946 &   0.010 &2012.452 &  11.980 &   0.001 &  10.777 &   0.000 &  -0.580 &   0.293 &   1.253 &   0.292 &   1 &   0 &   0.522 \\
4894 &   3 &  22.644 & -33.950 &  15.825 &   0.071 &  15.129 &   0.046 &2008.649 &  14.930 &   0.033 &2009.266 &  14.835 &   0.007 &2012.452 &  14.911 &   0.003 &  14.580 &   0.079 &   2.066 &   0.298 &  -3.504 &   0.273 &   0 &   0 &   0.000 \\
... &&&&&&&&&&&&&&&&&&&&&&&&& \\
\hline
\multicolumn{25}{c}{Field 4} \\
\hline
6566 &   4 &  22.048 &  10.509 &  15.729 &   0.049 &  12.466 &   0.074 &2009.274 &  11.066 &   0.006 &2012.452 &  11.077 &   0.001 & -99.000 & -99.000 & -99.000 &   9.889 &   0.024 &  -0.043 &   0.395 &  -0.068 &   0.576 &   1 &   0 &   0.706 \\
6567 &   4 &  38.605 &   8.312 &  15.828 &   0.080 &  11.931 &   0.094 &2009.274 &  10.379 &   0.003 &2012.452 &  10.317 &   0.001 & -99.000 & -99.000 & -99.000 &   9.138 &   0.018 &   2.154 &   0.396 &  -1.774 &   0.575 &   0 &   0 &   0.008 \\
... &&&&&&&&&&&&&&&&&&&&&&&&& \\
\hline
\multicolumn{25}{c}{Field 5} \\
\hline
7907 &   5 &  19.300 & -14.900 &  12.928 &   0.055 &   9.594 &   0.194 &2009.274 &   8.427 &   0.198 &2012.586 &   7.601 &   0.002 & -99.000 & -99.000 & -99.000 &   6.281 &   0.004 &   8.176 &   3.260 &  -0.836 &   0.720 &   0 &   1 &   0.004 \\
7908 &   5 &  35.105 & -11.798 &  13.951 &   0.066 &  10.202 &   0.097 &2009.274 &   8.915 &   0.097 &2012.586 &   9.089 &   0.001 & -99.000 & -99.000 & -99.000 &   6.697 &   0.074 &   5.465 &   1.121 &   0.976 &   1.001 &   0 &   1 &   0.000 \\
... &&&&&&&&&&&&&&&&&&&&&&&&& \\
\hline
\end{tabular}
\end{minipage}}
 \rotatebox{90}{
 \begin{minipage}{\textheight}
\tablefoot{Positional offsets in right ascension and declination are given in arcseconds, 
relative to the Wolf-Rayet member Q12 in the cluster core, RA 17:46:15.13, DEC -28:49:34.7.
Membership candidates are indicated, but membership selection is not absolute 
(1: cluster member candidate, 0: likely non-member, -1: membership unknown).
The first few bright sources shown for reference here are Wolf-Rayet cluster stars, yet their motions are 
uncertain owing to non-linearity effects in $K_s$.
The full version of the table will be available in the online version of the journal and from the CDS.}
 \end{minipage}}
\thispagestyle{empty}
\end{table*}


\begin{table*}[p]
\centering
 \rotatebox{90}{
 \begin{minipage}{\textheight}
\caption{\label{jhktab_arch} $JHK_s$ sources and $L'$ detections in the Arches cluster}
 \end{minipage}}
\tabcolsep1mm
\tiny
 \rotatebox{90}{
 \begin{minipage}{\textheight}
\begin{tabular}{rcccrrrrrrrrrrrrccccccccc}
\hline
  Seq & Field &  dRA   &   dDEC  &    J    & $\sigma_J$ & H & $\sigma_H$ & epoch1 & $K_s1$  & $\sigma_{Ks1}$ & epoch2 & $K_s2$ & $\sigma_{Ks2}$ & epoch3 & $K_s3$ & $\sigma_{Ks3}$ & $L'$ & $\sigma_{L}$ & $\mu_{\alpha cos\delta}$ & $\sigma_{\mu_{\alpha cos\delta}}$ & $\mu_\delta$ & $\sigma_{\mu_\delta}$ & mem & exc \\
      &       &  [``]  &   [``]  &  [mag]  &   [mag] &   [mag] &   [mag] & [year]  & [mag] &   [mag] & [year] &  [mag] & [mag] & [year] &[mag] & [mag] & [mag] & [mag] & [mas/yr] & [mas/yr] & [mas/yr] & [mas/yr] & & \\
\hline
\multicolumn{24}{c}{Field 1} \\
\hline
   1 &   1 &  -8.284 &  -0.125 &  11.730 &   0.117 &  10.961 &   0.001 &2002.246 &  10.353 &   0.005 &2010.605 &  10.345 &   0.006 &2011.246 &  10.376 &   0.001 &  10.280 &   0.012 &  -1.254 &   0.376 &   0.553 &   0.248 &   1 &   0 \\
   2 &   1 &   1.220 &  11.209 &  14.697 &   0.031 &  13.875 &   0.005 &2002.246 &  12.875 &   0.007 &2010.605 &  12.776 &   0.006 &2011.246 &  12.821 &   0.001 &  12.505 &   0.065 &  -2.176 &   0.376 &  -0.011 &   0.248 &   0 &   0 \\
   3 &   1 &   1.917 &   4.710 &  14.970 &   0.051 &  12.169 &   0.003 &2002.246 &  10.256 &   0.008 &2010.605 &  10.240 &   0.004 &2011.246 &  10.191 &   0.001 &   8.698 &   0.003 &  -0.211 &   0.376 &  -0.376 &   0.248 &   1 &   0 \\
   4 &   1 &  -0.000 &   0.000 &  15.051 &   0.052 &  12.009 &   0.004 &2002.246 &   9.998 &   0.008 &2010.605 &   9.919 &   0.004 &2011.246 &   9.933 &   0.001 &   8.383 &   0.003 &  -0.405 &   0.376 &   0.035 &   0.248 &   1 &   0 \\
... &&&&&&&&&&&&&&&&&&&&&&&& \\
\hline
\multicolumn{24}{c}{Field 2} \\
\hline
 468 &   2 & -32.961 & -22.767 &  16.094 &   0.040 &  11.797 &   0.064 &2008.430 &   9.329 &   0.008 &2010.605 &   9.419 &   0.005 & -99.000 & -99.000 & -99.000 &   7.430 &   0.005 &  -1.696 &   0.725 &  -0.433 &   0.931 &   0 &   0 \\
 469 &   2 & -42.392 & -28.073 &  16.902 &   0.079 &  16.117 &   0.011 &2008.430 &  15.444 &   0.015 &2010.605 &  15.431 &   0.016 & -99.000 & -99.000 & -99.000 &  15.305 &   0.072 &   0.709 &   0.766 &  -5.065 &   0.986 &   0 &   0 \\
... &&&&&&&&&&&&&&&&&&&&&&&& \\
\hline
\multicolumn{24}{c}{Field 3} \\
\hline
 707 &   3 &  29.488 &   7.715 &  16.704 &   0.127 &  13.222 &   0.009 &2008.430 &  11.355 &   0.001 &2011.712 &  11.370 &   0.001 &2012.446 &  11.327 &   0.003 &   9.975 &   0.010 &   2.503 &   0.480 &  -0.429 &   0.620 &   0 &   0 \\
 708 &   3 &  13.644 &  -3.423 &  17.080 &   0.093 &  13.569 &   0.018 &2008.430 &  11.403 &   0.004 &2011.712 &  11.373 &   0.002 &2012.446 &  11.319 &   0.002 &   9.629 &   0.022 &   0.111 &   0.489 &  -0.083 &   0.621 &   1 &   0 \\
... &&&&&&&&&&&&&&&&&&&&&&&& \\
\hline
\multicolumn{24}{c}{Field 4} \\
\hline
1342 &   4 & -13.063 &  28.697 &  19.279 &   0.080 &  16.183 &   0.028 &2008.430 &  14.335 &   0.008 &2011.715 &  14.304 &   0.005 &2012.446 &  14.371 &   0.009 &  13.106 &   0.014 &  -0.583 &   0.480 &   1.045 &   0.621 &   1 &   0 \\
1343 &   4 & -14.238 &  30.169 &  19.356 &   0.045 &  17.134 &   0.008 &2008.430 &  15.792 &   0.014 &2011.715 &  15.770 &   0.002 &2012.446 &  15.891 &   0.018 &  15.060 &   0.123 &  -0.109 &   0.492 &   0.537 &   0.624 &   1 &   0 \\
... &&&&&&&&&&&&&&&&&&&&&&&& \\
\hline
\multicolumn{24}{c}{Field 5} \\
\hline
1708 &   5 &   4.901 &  30.116 &  17.794 &   0.044 &  14.849 &   0.004 &2008.441 &  13.204 &   0.007 &2011.718 &  13.193 &   0.001 &2012.446 &  13.177 &   0.005 &  11.949 &   0.017 &  -0.462 &   0.490 &   0.899 &   0.620 &   1 &   0 \\
1709 &   5 &  23.376 &  20.347 &  17.925 &   0.013 &  14.763 &   0.007 &2008.441 &  12.974 &   0.002 &2011.718 &  12.992 &   0.001 &2012.446 &  12.917 &   0.003 &  11.871 &   0.007 &   1.070 &   0.484 &   0.147 &   0.619 &   1 &   0 \\
... &&&&&&&&&&&&&&&&&&&&&&&& \\
\hline
\end{tabular}
\end{minipage}}
 \rotatebox{90}{
 \begin{minipage}{\textheight}
\tablefoot{Positions are relative to the brightest $K_s$ cluster member: RA 17:45:50.42, DEC -28:49:21.84.
Proper motion uncertainties are dominated by the residual rms of the geometric transformation for bright stars, 
and hence are identical in the small selection shown here. Membership indicators follow the selections in the proper motion plane
(1: cluster member candidate, 0: likely non-member, -1: membership unknown).
The full version of the table will be available in the online version of the journal.}
\end{minipage}}
\end{table*}

\end{document}